\documentclass[useAMS]{mn2e}
\usepackage{graphicx}
\usepackage{txfonts}
\usepackage{aalongtable}
\usepackage{multirow}

\def\ha{\relax \ifmmode {\rm H}\alpha\else H$\alpha$\fi}
\def\pa{\relax \ifmmode {\rm Pa}\alpha\else Pa$\alpha$\fi}
\def\arcsec{\hbox{$^{\prime\prime}$}}
\def\nii{\relax \ifmmode {\rm N\,{\sc ii}}\else N\,{\sc ii}\fi}
\def\hii{\relax \ifmmode {\rm H\,{\sc ii}}\else H\,{\sc ii}\fi}
\def\hi{\relax \ifmmode {\rm H\,{\sc i}}\else H\,{\sc i}\fi}
\def\deg{\hbox{$^{\circ}$}}

\title[AINUR: Atlas of Images of NUclear Rings]{AINUR: Atlas of Images of NUclear Rings}
\author[S.~Comer\'on et al.]{S. Comer\'on$^{1}$\thanks{E-mail:
sebastien@iac.es}, J.~H.~Knapen$^{1,2}$, J.~E.~Beckman$^{1,2,3}$, E.~Laurikainen$^{4}$, H.~Salo$^{4}$, \newauthor
I.~Mart\'inez-Valpuesta$^{1,2}$, and R.~J. Buta$^{5}$\\
$^{1}$Instituto de Astrof\'isica de Canarias, E-38200 La Laguna, Tenerife, Spain\\
$^{2}$Departamento de Astrof\'isica, Universidad de La Laguna, E-38205 La Laguna, Tenerife, Spain\\
$^{3}$Consejo Superior de Investigaciones Cient\'ificas, Spain\\
$^{4}$Division of Astronomy, Department of Physical Sciences, University of Oulu, Oulu FIN-90014, Finland\\
$^{5}$Department of Physics and Astronomy, University of Alabama, Tuscaloosa, AL 35847, USA}
\begin{document}

\date{Accepted. Received; in original form }

\pagerange{\pageref{firstpage}--\pageref{lastpage}} \pubyear{2009}

\maketitle

\label{firstpage}

\begin{abstract}

We present the most complete atlas of nuclear rings to date. We include 113 rings found in 107 galaxies, six of which are elliptical galaxies, five are highly inclined disc galaxies, 18 are unbarred disc galaxies, and 78 are barred disc galaxies. Star-forming nuclear rings occur in $20\pm2\%$ of disc galaxies with types between $T=-3$ and $T=7$. We aim to explore possible relationships between the size and morphology of the rings and various galactic parameters. We also aim to establish whether ultra-compact nuclear rings are a distinct population of nuclear rings or if they are merely the low-end tail of the nuclear ring size distribution. We produce colour index and structure maps, as well as \ha{} and \pa{} continuum-subtracted images from $Hubble Space Telescope$ archival data. We derive ellipticity profiles from $H$-band Two-Micron All-Sky Srrvey images in order to detect bars and find their metric parameters. We measure the non-axisymmetric torque parameter, $Q_{\rm g}$, and search for correlations between bar and ring metric parameters, and $Q_{\rm g}$.

Our atlas of nuclear rings includes star-forming and dust rings. Nuclear rings span a range from a few tens of parsecs to a few kiloparsecs in radius. Star-forming nuclear rings can be found in a wide range of morphological types, from S0 to Sd, with a peak in the distribution between Sab and Sb and without strong preference for barred galaxies. The ellipticities of rings found in disc galaxies range from $\epsilon_{\rm r}=0$ to $\epsilon_{\rm r}=0.4$, assuming that nuclear rings lie in the galactic plane. Dust nuclear rings are found in elliptical and S0 galaxies. For barred galaxies, the maximum radius that a nuclear ring can reach is a quarter of the bar radius. We found a nearly random distribution of position angle offsets between nuclear rings and bars. There is some evidence that nuclear ring ellipticity is limited by bar ellipticity. We confirm that the maximum relative size of a star-forming nuclear ring is inversely proportional to the non-axisymmetric torque parameter, $Q_{\rm g}$ (``stronger 
bars host smaller rings'') and that the origin of nuclear rings, even the ones in non-barred hosts, is closely linked to the existence of dynamical resonances. Ultra-compact nuclear rings constitute the low-radius portion of the nuclear ring size distribution. We discuss implications for the lifetimes of nuclear rings and for their origin and evolution.

\end{abstract}

\section{Introduction}

Most nuclear rings (also known as circumnuclear rings in the literature) are thought to be produced as a result of an inflow of gaseous material that has been driven to the central parts of a galaxy after the loss of angular momentum (Combes \& Gerin 1985), which may be driven by non-axisymmetric features such as bars, ovals or strong spiral arms (see e.g.~Shlosman et al.~1990; Athanassoula 1994; Combes 2001). They are generally assumed to be related to the Inner Lindblad Resonances (ILRs) in a barred potential (Simkin et al.~1980; Combes \& Gerin 1985). Nuclear rings are thus tracers of recent gas inflow to the circumnuclear region (Knapen et al.~1995). Since they have star formation rates of up to 5\%\ of the total star formation in the galaxy (Kennicutt et al.~2005), they are part of a secular process that may help to build up pseudo-bulges (Kormendy \& Kennicutt 2004). Star-forming nuclear rings are statistically related to $\sigma$-drops (circumnuclear drops in the stellar velocity dispersion with a typical scalelength of one kpc) which are tracers of young circumnuclear populations and thus are another key piece in the study of secular evolution (Comer\'on et al.~2008b). As nuclear rings are relatively small they are one of the galactic morphological features that is most likely to be related to nuclear activity. In fact, evidence for such a link has been presented by Knapen (2005) and Knapen et al.~(2006). Nuclear rings have been related to nuclear bars because for both features, in many cases, their size is limited by the outer ILR (Laine et al.~2002; Englmaier \& Shlosman 2004).

Some new bars have been discovered recently, thanks to observations in new spectral ranges, notably the near-infrared (see e.g.~Scoville et al.~1988; Thronson et al.~1989; Eskridge et al.~2000; Knapen et al.~2000) and the millimetre domain (e.g.~in the study of NGC~6946 performed by Ball et al.~1985), but the bar fraction estimate has not significantly increased. The canonical estimate of the fraction of bars is around two-thirds (Mulchaey \& Regan 1997; Men\'endez-Delmestre et al.~2007; Marinova \& Jogee 2007) among luminous, normal galaxies. Fourier analysis techniques applied to bar detection have confirmed these results (Laurikainen et al.~2004). Whether the bar distribution shows some variation with Hubble morphological type is still an uncertain issue; some authors show that the bar distribution is nearly constant with the morphological type (Eskridge et al.~2000; Marinova \& Jogee 2007), other authors have shown that lenticular galaxies are less likely to be barred than spiral galaxies (Knapen et al.~2000; Aguerri et al.~2009; Laurikainen et al.~2009). It is, however, certain that bars, even if they are common, are not found in every spiral galaxy. Unbarred galaxies can host oval distortions which are deviations of the disc from a perfect axisymmetric shape. Ovals have lower ellipticities than bars and generally do not have Fourier terms with order higher than $m=2$ (Laurikainen et al.~2009). Their dynamical effects are very similar to those produced by bars (Kormendy \& Kennicutt 2004).

Detailed knowledge of whether a galaxy hosts a bar or not is necessary to explore the links between non-axisymmetric features and circumnuclear star-forming activity. The central regions of galaxies can maintain strong star-forming nuclear rings or Ultra-Compact Nuclear Rings (UCNRs), the latter defined by Comer\'on et al.~(2008a) as circumnuclear rings with a radius smaller than 200\,pc. Most nuclear rings have enhanced star formation rates, making them plausible tracers of recent inflow of material (Knapen et al.~2006).

In barred galaxies the gas is driven inward along shocks that may be delineated by dust lanes (Roberts et al.~1979; Athanassoula 1992; Shlosman 1999; Shlosman 2001). When the shocked gas crosses an ILR, the trajectory is gradually shifted from (nearly) radial to (nearly) circular (Knapen et al.~1995; Shlosman 2001). This shock-focused azimuthal flow creates a molecular nuclear ring between the two ILRs (when two ILRs are found) or just inside a single ILR (Shlosman 1999; P\'erez-Ram\'irez et al.~2000). The molecular gas may at some point become dense enough to initiate star formation. However, at exactly which radius (relative to the ILRs) a nuclear ring may appear is a subject that is still under discussion (see e.g.~the introduction of van de Ven \& Chang 2009). Some small nuclear rings could also be related to resonances that are even closer to the nucleus such as nuclear Lindblad Resonances (see e.g.~Fukuda et al.~1998; Shlosman 1999; Maciejewski 2003).

Even though nuclear rings have often been related to non-axisymmetric features and ILRs, a few have been found in galaxies with no known non-axisymmetric features. This is the case for NGC~473 (Heckman 1978), NGC~2985 (Comer\'on et al.~2008a), NGC~3011 (Gil de Paz et al.~2003), NGC~4826 (Pogge 1989), NGC~7217 (Pease 1917), and NGC~7742 (Morgan 1958).

Another phenomenon that may be called a nuclear ring is the presence of hundred pc-scale dust rings in elliptical and early lenticular galaxies (see e.g.~Lauer et al.~2005). These features have the size of nuclear rings but they show little or no star formation. Rings in elliptical galaxies may be caused by the depletion from inside out of dust nuclear discs (Lauer et al.~2005).

The aim of this paper is to present the most complete atlas of nuclear rings up to now and to study their links with other galactic features, notably bars. Samples of nuclear rings have already been published (S\'ersic \& Pastoriza 1965; Pogge 1989; Buta \& Crocker 1993; Knapen et al.~2002; Knapen 2005) but {\it Hubble Space Telescope} ($HST$) has provided much better resolution imaging that allows us to study the intermediate and small-sized nuclear rings. Since the launch of $HST$, tens of nuclear rings have been discovered but they have been studied individually or in small groups and no attempt has yet been made to make a uniform study of all of them. We also aim to explore possible relationships between the size and morphology of the rings and various galactic parameters such as the non-axisymmetries.

This paper is structured in the following fashion. In Sect.~2 we define nuclear rings and in Sect.~3 we describe our selection criteria. We then describe in Sect.~4 the reduction procedures of the data and data processing. In Sect.~5 we describe the bars that are found in the sample.  Sect.~6 is devoted to the description of how the ring parameters of the Atlas of Images of NUclear Rings (AINUR) were measured and we describe the ring size distribution. In Sect.~7 we search for links between the bar parameters and the nuclear rings. Sects.~8$-$9 we explore the links between nuclear rings and Hubble morphological type and nuclear activity. Sect.~10 contains comments about nuclear rings in unbarred galaxies. Sect.~11 has an individual discussion of all the rings included in AINUR. We discuss our results in Sect.~12 and present a summary of our conclusions in Sect.~13. Appendix~A contains tables with data about galaxy properties, and bar and nuclear ring and bar parameters. Appendix~B contains images of all the AINUR nuclear rings.

\section{Identification of nuclear rings}

\subsection{Definition of nuclear rings}

We define a star-forming nuclear ring as a ring of intense star formation in the proximity of the nucleus of a galaxy (see, e.~g., Buta et al.~2007). We will assume that nuclear rings lie in the galaxy plane. An inside-out surface brightness profile of them shows a rise and a fall in surface brightness with a maximum that we use to define the ring radius. We set the criterion that the width of a ring has to be at most half the ring radius; if not the nuclear ring is called a nuclear disc. These rings are usually located between the two ILRs where they exist and, if not, they are found just inside a single ILR (Knapen et al.~1995; Shlosman 1999). Nuclear rings are usually found inside the bar of a galaxy, but several nuclear rings are found in non-barred galaxies and may be related to ILRs due to strong spiral patterns, weak ovals or disturbances due to a companion galaxy. We will also include into this category purely stellar rings that once contained star formation and which have since faded.

In this paper we also include nuclear rings that are made of dust but have the size of star-forming nuclear rings. These dust nuclear rings are found in elliptical and early S0 galaxies. These features must not be confused with heavily dust-obscured star-forming rings which may need infrared imaging to directly see the star formation. The distinction between dust and star-forming nuclear rings is thus in their origin, not in their aspect.

NGC~2787 (Sil'chenko \& Afanasiev 2004) is a galaxy with a polar disc whose two nuclear rings do not lie in the main plane of the galaxy. As in the case of dust nuclear rings, the rings of this galaxy have been included in AINUR for completness (they have the size of UCNRs) but are not plotted nor used for statistics in which `normal' star-forming nuclear rings are involved.

Nuclear rings are in most cases easily distinguished from inner rings. In barred galaxies, inner rings are related to the inner 4:1 ultraharmonic resonance, found slightly inside the corotation radius of the bar (Schwarz 1984). Thus inner rings usually encircle main galactic bars (this is why they were called $\phi$-type spirals by Curtis in 1918) and nuclear rings are found well inside the bar (they may be encircling nuclear bars). When there is no obvious bar, distinguishing between nuclear and inner rings may be impossible if a galaxy does not have both a nuclear and inner ring.

The difference between pseudo-rings and rings is much fuzzier. According to Buta \& Crocker (1993) a pseudo-ring is a ring which is not well defined and probably represents a variation of the same phenomenon. A pseudo-ring may be seen as an intermediate state between a nuclear spiral and a well-defined nuclear ring. For practical purposes and due to the difficulty of quantifying how `ringy' a feature is, we are going to consider as a limiting case the nuclear ring in NGC~4321. We do not include pseudo-rings in this paper, but the distinction between the two types of features is not watertight and some of our nuclear rings may still be considered by some to have pseudo-ring characteristics. Previous catalogues of rings (not only nuclear, but also inner and outer) such as those by Buta \& Crocker (1993) and Buta (1995) include pseudo-rings. Thus, direct comparative statistics between these catalogues and AINUR may not be valid.

The nuclear rings in AINUR were detected by visual inspection of digital images (continuous {$R$-band}, $B$-band and UV images, continuum-subtracted \ha{} and \pa{} line emission images, colour-index maps and structure maps).

\subsection{Huge rings and tiny rings}

Prior to HST, the typical nuclear ring identified
on ground-based images had an average radius of
about 750\,pc (Buta \& Crocker 1993;
adjusted for current distance scale). Extensive
HST imaging has revealed a significant population
of much smaller rings, 200pc or less in radius,
called ultra-compact by Comer\'on et al. 2008a. It is not yet known whether UCNRs are the lower diameter tail of the nuclear ring size distribution or if they have a distinct origin. As one of the aims of this paper is to solve this problem we do not set a lower limit to the size of our rings.

The maximum radius of nuclear rings is uncertain. Laine et al.~(2002) relate the size distribution of nuclear rings to that of nuclear bars whose maximum radius is claimed by the authors to be around 1.6\,kpc. There are, nevertheless, some exceptionally big rings that have been defined as nuclear, as in ESO~565-11 (for a complete analysis of this galaxy see Buta et al.~1999). The innermost ring of this galaxy, whose radius is around 3.5\,kpc, is a {\it bona fide} nuclear ring because the galaxy has two more ring features that can be unambiguously identified as an inner ring and an outer pseudo-ring. In addition, the nuclear ring is inside the bar of the galaxy and cannot be an $x_{1}$ ring (a ring of gas at the largest non-looping orbit whose major axis is parallel to the major axis of the bar as defined by Regan \& Teuben 2004) because it is not aligned with the bar.

We decided to include in AINUR those nuclear rings with radii above 2.0\,kpc that were found inside a bar (ESO~565-11) or in unbarred galaxies that had outer, inner, and nuclear rings (ESO~198-13). If these criteria were not satisfied, we concluded that any ring feature with a radius above 2.0\,kpc could not be classified unambiguously and we thus did not include it in our sample. We have found three rings in the latter category, with radii between 2 and 3\,kpc in the unbarred galaxies NGC~3626, NGC~4750, and NGC~7213. These rings were not included in AINUR.

\section{Sample selection}

AINUR includes 113 nuclear rings distributed in 107 galaxies. The parameters of the host galaxies are found in Table~\ref{galax1} and the parameters of the nuclear rings and the host galaxy bar (if present) are found in Table~\ref{galax2}. Images of all the nuclear rings are presented in Appendix~B.

\subsection{Unbiased sample of nuclear rings}

Our galaxy sample has been based on the survey by Ho et al. (1995) aimed at `dwarf' Seyfert nuclei in nearby galaxies. They selected all those galaxies from the Revised Shapley-Ames Catalogue of Bright galaxies (RSA; Sandage \& Tammann 1981) and the Second Reference Catalogue of Bright Galaxies (RC2; de Vaucouleurs et al.~1976) with $B_{\rm T}\le12.5$\,mag and $\delta \ge 0\deg$. $B_{\rm T}$ is the apparent total $B$ magnitude reduced to the RC3 system (de Vaucouleurs et al.~1991). Twelve additional objects of special or historical interest were added to the sample by Ho et al.~(1995), reaching a total of 503 galaxies. To the 503 galaxies in this sample we added a similar selection from the southern celestial hemisphere: we selected all those galaxies with $\delta<0\deg$ and $B_{\rm T}\le12.5$\,mag in the Third Reference Catalogue of Bright Galaxies (RC3; de Vaucouleurs et al.~1991). We took into account the fact that some of the galaxies in RC3 just have a $V_{\rm T}$ measurement, and we corrected this effect by calculating the $B_{\rm T}$ magnitude as $B_{\rm T}=V_{\rm T}+(B-V)_{\rm T}$ in the cases where $(B-V)_{\rm T}$ is provided by RC3. The RC3 contains galaxies brighter that 15.5\,mag which implies that we have no problems with
completeness in the Southern hemisphere. Our combined sample consists of 888 galaxies with $B_{\rm T}\le12.5$\,mag in both Northern and Southern hemispheres.

As we were looking for nuclear rings that are assumed to lie in the galactic plane, we excluded from the list of 888 galaxies all those that were classified as edge-on in the RC3 catalogue (galaxies with a slash, `/',in the morphological type as reported by the RC3 catalogue). We also removed from our selection all the galaxies with an axis ratio smaller than 0.35 ($i>70\deg$) in HYPERLEDA (Paturel et al.~2003) because we considered that they are too highly inclined to give reliable results. We removed a total of 157 very inclined galaxies from the sample.

Next we retrieved those Advanced Camera for Surveys (ACS), Faint Object Camera (FOC), Near-Infrared Camera and Multi-Object Spectrometer (NICMOS), and Wide Field Planetary Camera 2 (WFPC2) images from the $HST$ archive which included the nuclear zones of our non edge-on galaxy sample. All the 248 galaxies that have not been observed by $HST$ were removed from the sample at this stage because the resolution reached by ground-based telescopes would not allow us to detect UCNRs at the distances of most of the galaxies. We also removed all the satellite galaxies of the Milky Way because these dwarf irregular galaxies, though very well studied, are known not to have rings. The final sample has 483 galaxies, with morphological types ranging from giant ellipticals to irregular dwarfs.

To summarize, our sample satisfies all of the following criteria:

\begin{itemize}

\item{$B_{\rm T} \le 12.5$ in the RC3 and/or they appear in the survey of galaxies by Ho et al.~(1995),}
\item{not catalogued as edge-on in the RC3,}
\item{axis ratio $d/D > 0.35$ in the HYPERLEDA database,}
\item{not a Milky Way satellite,}
\item{imaged by $HST$ with one or more of the following cameras: ACS, FOC, NICMOS, WFPC2.}

\end{itemize}

The 483 selected galaxies include 107 elliptical galaxies ($T\leq-3.0$), 363 disc galaxies ($-3.0<T\leq9.0$) and 13 irregular galaxies ($T>9.0$).

Among these 483 selected galaxies we have found 76 galaxies with at least one nuclear ring, which form the majority of the atlas of nuclear rings discussed in the present paper. However, for one of the galaxies, NGC~3414, the $HST$ image is not deep enough to detect the ring, which has been found in ground-based images by Buta et al.~(2007).

\subsection{Extended sample of nuclear rings in AINUR}

The survey of 483 galaxies searching for nuclear rings is used in Section~8, where an unbiased sample is necessary to investigate the nuclear ring proportion as a function of morphological type. In Section~9, where we study the nuclear activity, we have used the subset of galaxies which appears in the survey made by Ho et al.~(1995). For other purposes high statistics are preferred over an unbiased sample and that is why, except for Sections~8 and 9, we have used an extended sample, in which we include nuclear rings found in the literature.

Twenty-nine galaxies with a nuclear ring were added on the basis of the literature search. All these nuclear rings were confirmed by us using digital images. Section~3.4 discusses literature cases that we could not confirm. There are three reasons why these 29 nuclear ring host galaxies were not included in our 483 galaxy sample (galaxies can fall in more than one category), as follows.

\begin{itemize}

\item{Twenty-one galaxies are dimmer than $B_{\rm T}=12.5$ in the RC3. These galaxies are ESO~198-13, ESO~437-33, ESO~437-67, ESO~565-11, IC~1438, IC~4933, NGC~473, NGC~718, NGC~1241, NGC~1415, NGC~1819, NGC~2595, NGC~3011, NGC~3081, NGC~5135, NGC~5905, NGC~5135, NGC~5945, NGC~7469, NGC~7716, NGC~7570, and NGC~7771.}
\item{Three galaxies have $d/D<0.35$ in HYPERLEDA. Even though these galaxies are highly inclined, in a few cases it is possible to distinguish a nuclear ring. These galaxies are NGC~1808 ($d/D=0.34$), NGC~4100 ($d/D=0.28$), and NGC~6503 ($d/D=0.33$). In the case of NGC~1808 the value of the axis ratio given in HYPERLEDA is too low and we have used $d/D=0.74$ (Laurikainen et al.~2004).}
\item{Nineteen galaxies have not been imaged by $HST$: ESO~198-13, ESO~437-33, ESO~437-67, IC~1438, IC~4214, IC~4933, NGC~473, NGC~521, NGC~718, NGC~1343, NGC~1415, NGC~1819, NGC~2595, NGC~2935, NGC~3011, NGC~3313, NGC~5905, NGC~5945, and NGC~7570. Five of these galaxies (IC~4214, NGC~718, NGC~2935, NGC~3313, and NGC~5905) were bright enough to be in the 888 galaxy original sample.}

\end{itemize}

The literature sources of these 29 galaxies are: ESO~198-13, ESO~437-33, ESO~437-67, ESO~565-11, IC~1438, IC~4214, NGC~1343, NGC~1808, NGC~1819, NGC~2595, NGC~2935, NGC~3081, NGC~3313, NGC~4100, NGC~5945, and NGC~7469 from Buta \& Crocker (1993), IC~4933 from Ryder et al.~(2009), NGC~521 from Buta et al.~(2009), NGC~718 from Erwin \& Sparke (2003), NGC~1415 and NGC~5135 from Garc\'ia-Barreto et al.~(1996), NGC~3011 from Gil de Paz et al.~(2003), NGC~473, NGC~1241, NGC~5905, NGC~6503, NGC~7570, and NGC~7716 from Knapen et al.~(2006), and NGC~7771 from Smith et al.~(1999). Nine of these galaxies have been imaged by $HST$ and therefore they can be studied with the same detail as the other 76. These galaxies are ESO~565-11, NGC~1241, NGC~1808, NGC~3081, NGC~4100, NGC~5135, NGC~6503, NGC~7469, and NGC~7771. Seven galaxies, namely IC~1438, NGC~473, NGC~1343, NGC~5905, NGC~5945, NGC~7716, and NGC~7570 have been studied by us with the \ha{} images from Knapen et al.~(2006) and one more, NGC~1415, with an \ha{} image from Garc\'ia-Barreto et al.~(1996). NGC~2595 has been studied with Sloan Digital Sky Survey Data Release 7 (SDSS DR7; Abazajian et al.~2009) images. The images of the remaining galaxies have been provided by RJB.

Two more nuclear rings have been discovered by us in galaxies which do not meet the criteria to enter into the 483-galaxy sample and which do not appear in the literature. They are as follows. The first is a UCNR in an $HST$ \ha{} image of UGC~10445 in an unpublished study searching for nuclear clusters in late-type galaxies. The galaxy was too faint to be in our 483-galaxy sample. We also included the nuclear ring in NGC~5020, a galaxy not imaged by $HST$ and dimmer than $B_{\rm T}=12.5$, that was brought to our attention by a discussion on the forum pages of the galaxy zoo project (Lintott et al. 2008). This galaxy has been studied with SDSS DR7 (Abazajian et al.~2009) images.

\subsection{Incompleteness and biases}

A source of incompleteness in our catalogue is that we have not been able to access good quality images for all the galaxies that have nuclear rings in the literature, so we have not been able to confirm them. This is the case for ESO~219-37, NGC~3147, and NGC~4984 (Buta \& Crocker 1993). None of these galaxies has been included in our study.

\begin{figure}
\begin{center}
\begin{tabular}{c}
\includegraphics[width=0.45\textwidth]{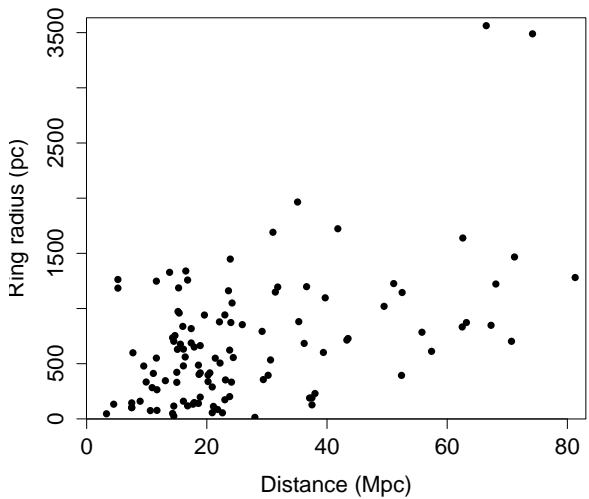}\\
\end{tabular}
\caption{\label{complet} Distribution of nuclear ring radius with distance to the host galaxy for the 113 nuclear rings of our sample. How the nuclear ring radius is measured is described in Sect.~5.}
\end{center}
\end{figure}

We can estimate the completeness of the Atlas from Fig.~\ref{complet}, where we plot the nuclear ring radius (Sect.~6) versus distance to the host galaxy. We can deduce empirically that we are able to detect UCNRs (defined as nuclear rings with a radius smaller than 200\,pc) using $HST$ at a distance of 40\,Mpc. This is confirmed when we consider that the largest UCNRs would span 2\arcsec\ at $D=40\,{\rm Mpc}$. Farther away, UCNRs are not detectable due to the lack of resolution.

AINUR is restricted to galaxies that have been imaged by $HST$ or that have been well-studied in the literature. It is possible that many close non-studied galaxies host nuclear rings. An SDSS-type survey in the whole sky would be very useful to detect these unknown nuclear rings.

Finally, it is possible that we are missing several nuclear rings, especially the smallest ones, due to dust obscuration in the inner parts of galaxies.

\subsection{Features called nuclear rings in the literature but not included in AINUR}

The Atlas of nuclear rings that we have produced does not include all galaxies that have been called `nuclear ring galaxies' in the literature.

Nuclear rings found in the literature for which we do not have access to digital images have not been included (see preceding subsection).

In three cases, what had been termed nuclear rings are most probably inner rings. NGC~3344 was considered to have a nuclear ring by Knapen (2005). We established that the bar of this galaxy is inside the ring and that there is no outer bar. Furthermore, this ring would be, compared to the extinction-corrected $D_{25}$ radius of the galaxy ($D_{\rm o}$, defined by Bottinelli et al.~1995), by far the biggest nuclear ring of the sample. NGC~3184 has a 2.2\,kpc ring that was considered a nuclear ring by Knapen (2005) but has a bar within it. Another case similar to NGC~3184 and NGC~3344 is NGC~253 (Engelbracht et al.~1998). Another ring that has been called a nuclear ring in the literature and that we now think is really an inner ring is the outermost ring-like feature in NGC~278. This galaxy has been studied in detail by Knapen et al.~(2004), who argue that this galaxy has two nuclear rings, one 450\,pc in diameter and another with a diameter of 2.2\,kpc. Knapen et al.~(2004) found no bar in this galaxy, but we found a small bar with a length of 0.8\,kpc using ellipticity profiles. This bar is evidence for the larger of the two rings being an inner ring. The relative radius (ring diameter divided by $D_{\rm o}$) of the larger ring is again much bigger than in any other nuclear ring in the sample. Even if this ring-like feature were nuclear, a close examination of $HST$ images leads to classify it as a pseudo-ring than a ring.

In various cases, features are called nuclear rings based on images that barely resolve the possible ring. As examples, we mention ESO~153-20, NGC~4254, and UGC~12646 (Buta \& Crocker 1993), NGC~5347 (Garc\'ia-Barreto et al.~1996), and NGC~1530 and NGC~5953 (Knapen et al.~2006), which nuclear rings have been reported, but which actually have nuclear spirals or pseudo-rings when observed with high resolution $HST$ images. We included nuclear ring galaxies without $HST$ images {\it only} when the images from literature were unambiguous.

There are some difficult cases that are in discussion in the community. NGC~1068 is reported to have a nuclear pseudo-ring by Buta \& Crocker (1993), but this feature is considered a nuclear ring by Knapen et al.~(2006). We included this feature as a nuclear ring because we consider it to be a complete ring which is partially covered by a dust lane. Another case under discussion, NGC~4321, has been included in the sample, albeit as the borderline case separating rings and pseudo-rings, as discussed in Sect.~2.1. We have not included the nuclear ring in NGC~2681 (Buta \& Crocker 1993) because it has been recognized as a pseudo-ring afterwards (Buta et al.~2007).

We have not included the possible nuclear ring in NGC~3489 (Erwin \& Sparke 2002) because it is too inclined for us to be completely sure about its nature.

We have not included features that are considered edge-on rings such as the one at 150\,pc radius in NGC~4570 (van den Bosch \& Emsellem 1998)  because we cannot accurately deproject the galaxies to confirm their ringed nature or to study the exact ring shape.

Finally, we have not included the nuclear ring-like features in the Circinus galaxy reported by Wilson et al.~(2000) because we consider that they are in fact pseudo-rings.

\subsection{AINUR}

The sample selection process has led us to create AINUR, an atlas with 113 nuclear rings in 107 galaxies. We found 82 nuclear rings in 76 galaxies in the 483 galaxy unbiased sample. Four galaxies (NGC~2787, NGC~4371, NGC~4459, and NGC~5248) are double-ringed and one (NGC~4800) is triple-ringed.

Most of the spiral and lenticular galaxies in AINUR are known to be barred. Eighteen galaxies (NGC~2985, NGC~2997, NGC~3011, NGC~3245, NGC~3593, NGC~4100, NGC~4138, NGC~4459, NGC~4571, NGC~4736, NGC~4826, NGC~5033, NGC~5247, NGC~6503, NGC~6753, NGC~6861, NGC~7217, and NGC~7742) are not known to be barred.

AINUR contains dust nuclear rings in six elliptical galaxies (NGC~3258, NGC~3268, NGC~3379, NGC~4494, NGC~5812, and NGC~6958). Two nuclear rings in lenticular galaxies (the inner one in NGC~4371 and the one in NGC~7049) may be either dust or star-forming rings. They have been considered as star-forming nuclear rings for the statistics in this paper.

The parameters of the galaxies in AINUR are found in Table~\ref{galax1}. In this table we use, if possible, the morphology given by Buta et al.~(2007). In other cases we used the morphology from NASA/IPAC Extragalactic Database (NED) which, in most cases, is the same as in the RC3. Some galaxies that are not listed as barred in Table~\ref{galax1} are barred following some literature sources. These galaxies are NGC~1068 (Scoville et al.~1988), NGC~4800 (Comer\'on et al.~2008a), NGC~5194 (Laurikainen \& Salo 2002), and NGC~5427 (Marinova \& Jogee 2007).

\section{Data processing}

We have used images from the $HST$ archive to identify nuclear rings and images from other archives only to confirm a nuclear ring in galaxies for which $HST$ images were not available. Two Micron All-Sky Survey $H$-band images have been used to determine bar parameters. Images of all the nuclear rings are presented in Appendix~B (Fig.~\ref{images}).

\subsection{Image processing of $HST$ images}

\subsubsection{Structure maps}

The structure map technique (Pogge \& Martini 2002) allows one to
observe the distribution of structure, and in particular of dust, on the scale of the point spread function (PSF) using only one image in one
filter. The structure map is mathematically defined as
\begin{equation}
S=\left(\frac{I}{I\otimes P}\right) \otimes P^{\rm t}
\end{equation}
where $S$ is the structure map, $I$ is the image, $P$ is the
PSF, $P^{\rm t}$ is the transform of the PSF, and
$\otimes$ is the convolution operator (Pogge \& Martini 2002).  We used
synthetic PSFs created with the Tiny Tim software
(Krist \& Hook 1999).

We used images taken through green or red optical filters to apply the structure map
operator, namely $F547M$, $F606W$, $F791W$, or $F814W$ for WFPC2, and $F625W$, $F606W$, or $F814W$ for ACS images.

For NGC~936 we produced a structure map from an $R$-band image taken with the William Herschel Telescope (WHT). The PSF was generated from field stars in the image.

\subsubsection{\ha\ images}

We used \ha\ ACS and WFPC2 images taken through the narrow-band filters
$F656N$ or $F658N$, and a continuum image taken through a red
broad-band filter. In the few cases where images taken through both
\ha{} filters were available in the archive, we chose the $F658N$
image, because the \ha\ line of our sample galaxies is better
centred in its passband. The images used for the continuum
subtraction were in most cases the same ones from which we derived the structure
maps.

The continuum subtraction procedure we used has been described in Knapen et al.~(2004), Knapen et al.~(2006), and Comer\'on et al.~(2008b).

\subsubsection{\pa{} images}

We downloaded relevant NICMOS Pa$\alpha$ narrow-band images from the archive. We used one of two techniques to remove the continuum component from the Pa$\alpha$ images, depending on which images are available in the archive. If there is a broad-band near-IR image at a wavelength close to that of Pa$\alpha$, the procedure is the same as for H$\alpha$. The wide band filter images used for continuum subtraction are taken through the $F110W$ ($\sim$ $J$) or the $F160W$ ($\sim$ $K$) filter. In some cases, the set of images in the archive is an $F187N$ and an $F190N$ image. At zero redshift, Pa$\alpha$, at a wavelength of 1874.5\,nm, is centred on $F187N$, and for galaxies with $v > 3000 \, \rm{km\,s^{-1}}$, on $F190N$. For galaxies with velocities lower than this threshold, we directly subtracted the $F190N$ image from the $F187N$ image after multiplying the $F190N$ image by a number ranging from 0.9 to 1.0 due to its slightly greater efficiency. This value was calculated in each 
case by plotting the counts from continuum emission areas in one filter against those in the other one, fitting a line and using its slope to find a correction factor. For galaxies with greater velocities we performed the inverse operation ($F190N$-$F187N$). No cases where the line is in the overlap region of filter transmission curves have been found.

\subsubsection{Colour-index images}

Most of the colour index images have been obtained using filters close to $V$ and $I$ with ACS and WFPC2. We first aligned the images and then we performed the operation $V-I=-2.5\log{V/I}$ to get the colour index image. In some cases where we had no $V$ or $I$ image, we produced $I-J$ images.

\subsubsection{UV images}

We also downloaded ACS, FOC, and WFPC2 UV images because it is possible to detect star-forming nuclear rings in the UV. No processing was needed for these images.

\subsection{Ground-based images}

Images that were obtained from authors of previous studies (see Sect.~3.2 for details) had already been processed by them. These images are mostly \ha\ and colour-index images. For IC~4933 we used a Pa$\beta$ image and for NGC~1819 we used a $V$-band image.

\subsection{2MASS image processing}

\subsubsection{Measuring bar length and ellipticity}

Bars have previously been detected in most of the galaxies in our sample. Unfortunately, the bar measurement procedures vary greatly from one paper to another (discussed by Athanassoula \& Misiriotis 2002). For example, bar length can be measured visually in $B$-band images (Martin 1995), as the radius at which is located a sharp rise in the amplitudes of the even Fourier components (e.g., Aguerri et al.~1998), as the location of the peak in ellipticity (e.g., Knapen et al.~2000) or as the length of the region with a constant $m=2$ phase in near-IR images (Laurikainen \& Salo 2002). Other authors, such as Erwin (2005), assign as a lower limit to the bar length the location of the maximum in ellipticity and as an upper limit the minimum of two ellipse fit measures: the first minimum in ellipticity outside the peak ellipticity of the bar, or the point at which the position angles (PAs) of the fitted ellipses differ by $\ge10\deg$ from the PA of the bar. Gadotti (2008) prefers to fit a S\'ersic profile to the bar with a truncation radius that corresponds to the bar length. This variety of criteria can cause the measured length of the bar to vary by a factor of up to two. Even though the length of a bar is a loosely defined concept, for practical purposes a consistency in the criteria would be useful. That is why we have performed our own measurements of bar lengths, as well as of bar maximum ellipticities, using $H$-band images from the 2MASS for all the galaxies hosting a nuclear ring.

First, we deprojected the 2MASS images, where possible using orientation parameters from deep optical and NIR images as reported by Laurikainen et al.~(2004), Laurikainen et al.~(2005), Buta et al.~(2006), and Laurikainen et al.~(2006). For other galaxies we used the axis ratio and PA values from the HYPERLEDA database, which were calculated using the $25\,{\rm mag}\,{\rm arcsec}^{-1}$ isophote in $B$-band plates. We used the deprojection parameters from these literature sources in all cases, except those ones listed below, where we could improve them using the data sources identified. In particular, we used data from Ondrechen et al.~(1989) for NGC~1097, Lindblad et al.~(1997) for NGC~1300, Ryder et al.~(1996) for NGC~1433, Hawarden et al.~(1979) for NGC~1512, Becker et al.~(1988) for NGC~1566, Mulder et al.~(1995) for NGC~3310, and Erwin (2005) for NGC~3945. For NGC~4448 we used our own determination of the orientation parameters from a 2MASS $H$-band image. For NGC~7570 we did the same as for NGC~4448 but using an $r$-band SDSS image. In the case of ESO~437-67 the orientation data from HYPERLEDA clearly correspond to the bar. Deeper images from the DSS show a roughly face-on outer pseudo-ring feature, which led us to consider this galaxy to be face-on. Finally, we preferred the HYPERLEDA orientation parameters for NGC~5248 to the one given by Laurikainen et al.~(2004) because it leads to a more realistic deprojection. The HYPERLEDA values we used for NGC~5248 are very close to the ones derived by Jogee et al.~(2002).

In the deprojected images, we masked all the stars and background or close-companion galaxies and then ran the ELLIPSE task from IRAF to fit the ellipticities and the PAs of the galaxies as a function of radius. We first fixed the centre of the galaxy to the position found using the centroid search option from the IMEXAM task. We plotted the data to visualize the variations of ellipticity and PA with radius. Then we followed the most restrictive criterion from Knapen et al.~(2000) to identify bars, namely that a bar is characterized by a significant rise in the radial ellipticity profile followed by a significant fall over a range in radius where the position angle of the major axis is roughly constant. The amplitude of the ellipticity variation must be at least 0.1. The highest values of bar ellipticity, $\epsilon_{\rm b}=0.8$, correspond to the more elliptical class of bars (the upper limit in bar ellipticity is set by the appearence of chaotic orbits in very elliptical bars; see Mart\'inez-Valpuesta \& Shlosman 2004).

We define the end of the bar as the radius where the ellipticity reaches its local maximum value in a region of roughly constant PA, following the approach of, among others, Knapen et al.~(2002). This length estimation is known to be a lower limit to the true bar length (Erwin 2005). The uncertainty in the measurement of the maximum in ellipticity and thus in the measurement of the bar length is of the order of 5-10 per cent. The bulge component usually has a radius of only one kpc or so and is not expected to affect the measured bar lengths to a significant degree. We used the galaxy distances as given in Table~\ref{galax1}, obtained from the references given bellow the table, to calculate absolute bar lengths. The basic bar parameters (length and maximum ellipticity) are tabulated in Table~\ref{galax2}.

Bar parameters such as bar length and bar maximum ellipticity are highly dependent on the orientation parameters which are the main error source. In particular, the characteristics of the weakest bars are highly dependent on the fine-tuning of the deprojection. For example, the bar that appears in the nearly face-on galaxy NGC~2985 (whose bar does not appear in any catalogue) is invisible in the non-deprojected image even though the ellipticity of the disc of the galaxy is only $\epsilon_{\rm{d}}=0.19$. Some weak bars can be hidden or fictiously created by an incorrect deprojection.

In two galaxies, NGC~473 and NGC~1386, we have not found any trace of a bar after the deprojection of the galaxy even though NED and HYPERLEDA state that they are barred. NGC~1386 is one of the most inclined galaxies in the sample, which makes bar identification difficult. NGC~5427, which is listed as barred only in HYPERLEDA, does not seem to be barred.

We also used the deprojected $H$-band 2MASS images to calculate the scalelength of the stellar discs of the galaxies. We performed a radial S\'ersic+exponential fit to the deprojected image of the galaxy which yields the disc scalelength, as tabulated in Table~\ref{galax1}. This disc scalelength will be used in the next subsection as a key element for the the measurement of $Q_{\rm g}$.

\subsubsection{Non-axisymmetric torque parameter ($Q_{\rm g}$)}

The non-axisymmetric torque parameter, $Q_{\rm g}$, quantifies the impact of non-axisymmetries in a galaxy (Combes \& Sanders 1981). It is defined as the highest value of tangential forces normalized by the axisymmetric force field. The mathematical expression is
\begin{equation}
Q_{\rm g}={\rm max}(F^{\rm max}_{\rm T}(r)/\langle F_{\rm R}(r)\rangle),
\end{equation}
where $F^{\rm max}_{\rm T}(r)$ is the maximum tangential force at a given radius and $\langle F_{\rm R}(r)\rangle$ is the average radial force at the same radius (Combes \& Sanders 1981). This maximum is located in the region of the bar or a strong arm perturbation. Spurious absolute $Q_{\rm g}$ maxima can appear in the outer parts of galaxies due to low signal-to-noise ratio. Higher values of $Q_{\rm g}$ are in a statistical sense related to strong bars and lower values to arm perturbations and oval distortions. However, exceptionally strong arms can cause high $Q_{\rm g}$ values in unbarred galaxies (e.g., in NGC~5427). For around half of the galaxies, we found non-axisymmetric torque parameter values in at least one of the following papers: Block et al.~(2001), Buta \& Block (2001), Laurikainen \& Salo (2002), Block et al.~(2004), Laurikainen et al.~(2004), Buta et al.~(2006), and Laurikainen et al.~(2006). Not all the previous papers report $Q_{\rm g}$; some list $Q_{\rm b}$, the perturbation related just to the bar (bar strength) which is very close to $Q_{\rm g}$ in most cases (see Durbala et al.~2009). We decided to determine our own set of $Q_{\rm g}$ values from $H$-band 2MASS images so that all galaxies have a $Q_{\rm g}$ value measured from an image of similar quality and in a consistent manner.

The method we followed is that of Laurikainen et al.~(2004), Buta et al.~(2006), and Laurikainen et al.~(2006). The basics of the method are described in Laurikainen \& Salo (2002) and the bulge subtraction procedure has been described in Laurikainen et al.~(2004). To properly perform a $Q_{\rm g}$ measurement following the steps explained in the literature, we need to know the orientation parameters of the galaxy and the scaleheight of the disc. We also need to model the bulge.

For deprojecting a galaxy, we used the same orientation parameters as when we performed the ellipticty profiles and the maximum bar ellipticity measurement. The deprojection of the whole galaxy causes the deformation of the bulge within it, and will cause the stretching of an intrinsically spherical component, and possibly a wrong $Q_{\rm g}$ value. To avoid this problem, we subtract a bulge model from the image before the deprojection and add it again once the deprojection has been performed. In that way, even though the disc has been deprojected prior to the measurement of $Q_{\rm g}$, the bulge remains unchanged. This procedure assumes that the bulge is intrinsically spherical. To subtract and then add a bulge model we performed 2D decompositions of the 2MASS $H$-band images of the sample galaxies. The software we used for the 2D decomposition is described in Laurikainen et al.~(2005). We fixed the disc PA and axis ratio as obtained from the literature and the disc scalelength from the 1D S\'ersic+exponential fit described in the previous subsection. Sometimes the nuclear ring was so intense that it strongly perturbed the bulge fit. In these cases, the nuclear rings were (crudely) modelled with a Ferrers function (Ferrers 1877) with an index $n=0$. This added component is not ring-like (actually it is lens-like), but the excess light it adds inside the nuclear ring is significantly fainter than the bulge contribution.

The disc scaleheight is an important parameter and one of the main sources of uncertainty in the $Q_{\rm g}$ measurement (see discussion in Laurikainen \& Salo 2002). De Grijs (1998) showed that the scaleheight of the disc is proportional to the scalelength and that the proportionality factor varies with morphological type. From his results we deduce that this factor is 1/4 for galaxies with $T<1.5$, 1/5 for galaxies with $1.5<T<4.5$, and 1/9 for galaxies with $T>4.5$. The disc scalelength was taken from the 1D luminosity profiles as derived from the $H$-band 2MASS images.

The same non-axisymmetric feature (e.g.~a bar) in two different galaxies may cause different $Q_{\rm g}$ values. On average, $Q_{\rm g}$ will be smaller in an earlier type galaxy. There are two reasons for this as follows. First, a higher central mass concentration affects the $Q_{\rm g}$ by increasing the $F_{\rm R}(r)$ factor in eq.~2 (by higher central mass concentration we are meaning a higher total mass of the bulge, not a higher S\'ersic index). Secondly, a bigger disc scaleheight makes the $Q_{\rm g}$ go down. This second effect is well seen in Fig.~6 from Laurikainen et al.~(2004) where it is shown that, on average, S0s have smaller $Q_{\rm g}$ than later-type galaxies.

We have been able to obtain a $Q_{\rm g}$ value for all the disc galaxies except for NGC~1386, NGC~3593, NGC~4100, NGC~4526, and NGC~6503 which are too inclined and to detect a bar or to produce $Q_{\rm g}$ profiles. The 2MASS images of ESO~198-13 and UGC~10445 are too shallow to measure $Q_{\rm g}$ reliably.

\begin{figure}
\begin{center}
\begin{tabular}{c}
\includegraphics[width=0.45\textwidth]{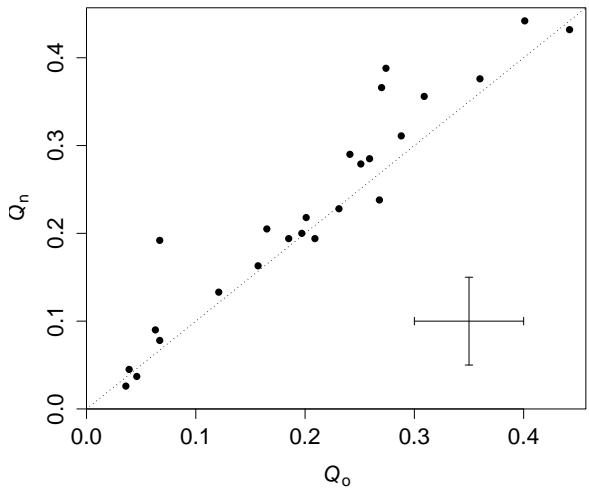}\\
\end{tabular}
\caption{\label{oldandnew} Comparison of $Q_{\rm g}$ values obtained by us ($Q_{\rm n}$) and in the literature ($Q_{\rm o})$ in those galaxies for which we have used the same orientation parameters as in the literature. Data used for this plot come from Laurikainen \& Salo (2002), Laurikainen et al.~(2004; 2006) and Buta et al.~(2006). The dashed line has a slope equal to one.}
\end{center}
\end{figure}

We check our results by comparing our $Q_{\rm g}$ values with those reported in the literature for 25 galaxies (Fig.~\ref{oldandnew}). The correlation factor between $Q_{\rm new}$ and $Q_{\rm old}$ is $\rho=0.95$ which corresponds to a scatter value lower than 0.03 in $Q_{\rm g}$. We found that only NGC~4571 has a significantly different $Q_{\rm g}$ value (Laurikainen et al.~2004), caused by these authors using a much lower S\'ersic index in their bulge model than we did. In general we found that even though the near-IR images we used are shallower, the results are very similar, and can thus conclude that 2MASS images are indeed good enough to measure $Q_{\rm g}$.

The $Q_{\rm g}$ values as well as the radii, $r_{\rm Q_{g}}$, at which they are measured are found in Table~\ref{galax2}.

\section{Comments on bar parameters}

\begin{figure*}
\begin{center}
\begin{tabular}{c}
\includegraphics[width=0.45\textwidth]{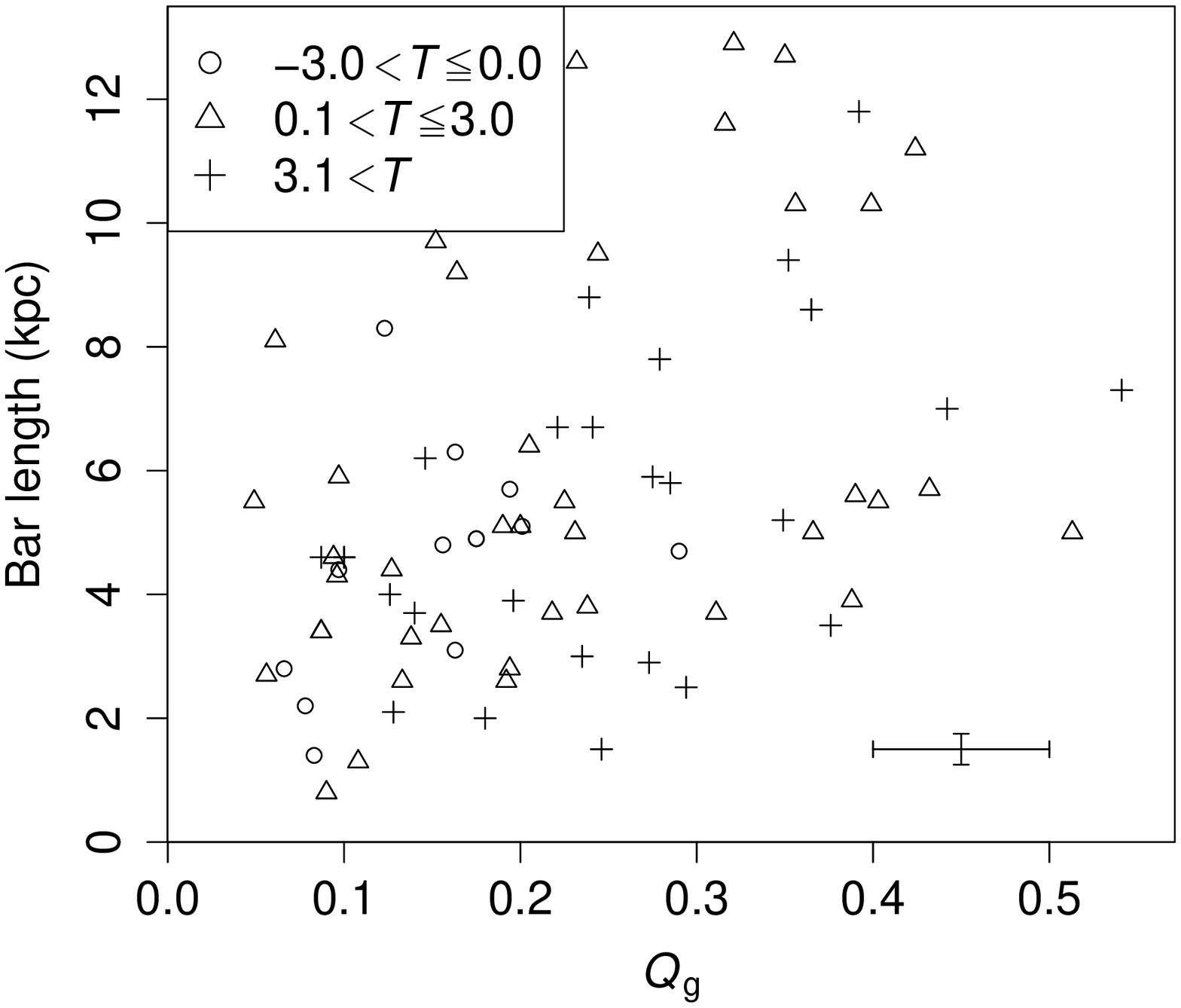}
\includegraphics[width=0.45\textwidth]{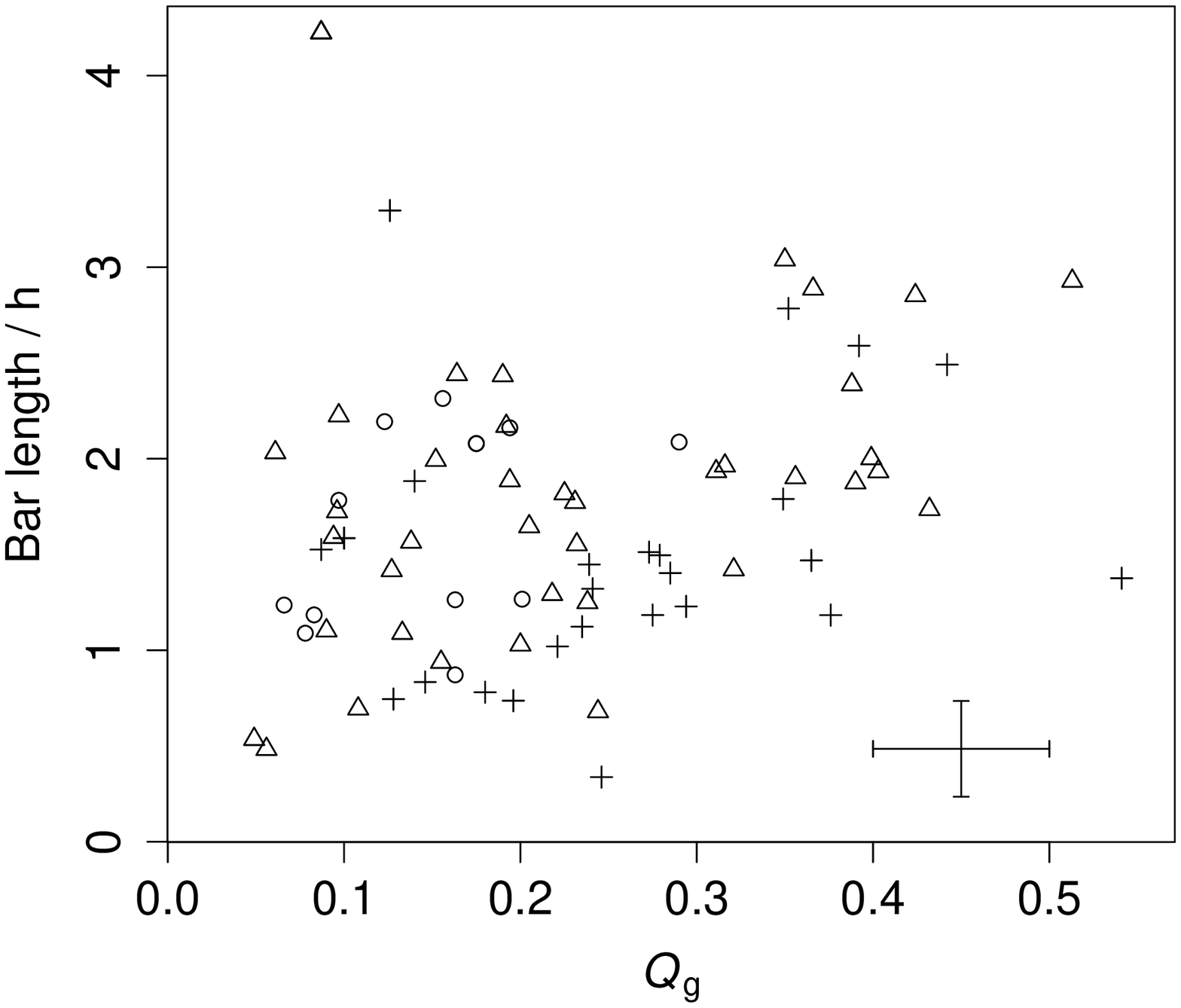}\\
\includegraphics[width=0.45\textwidth]{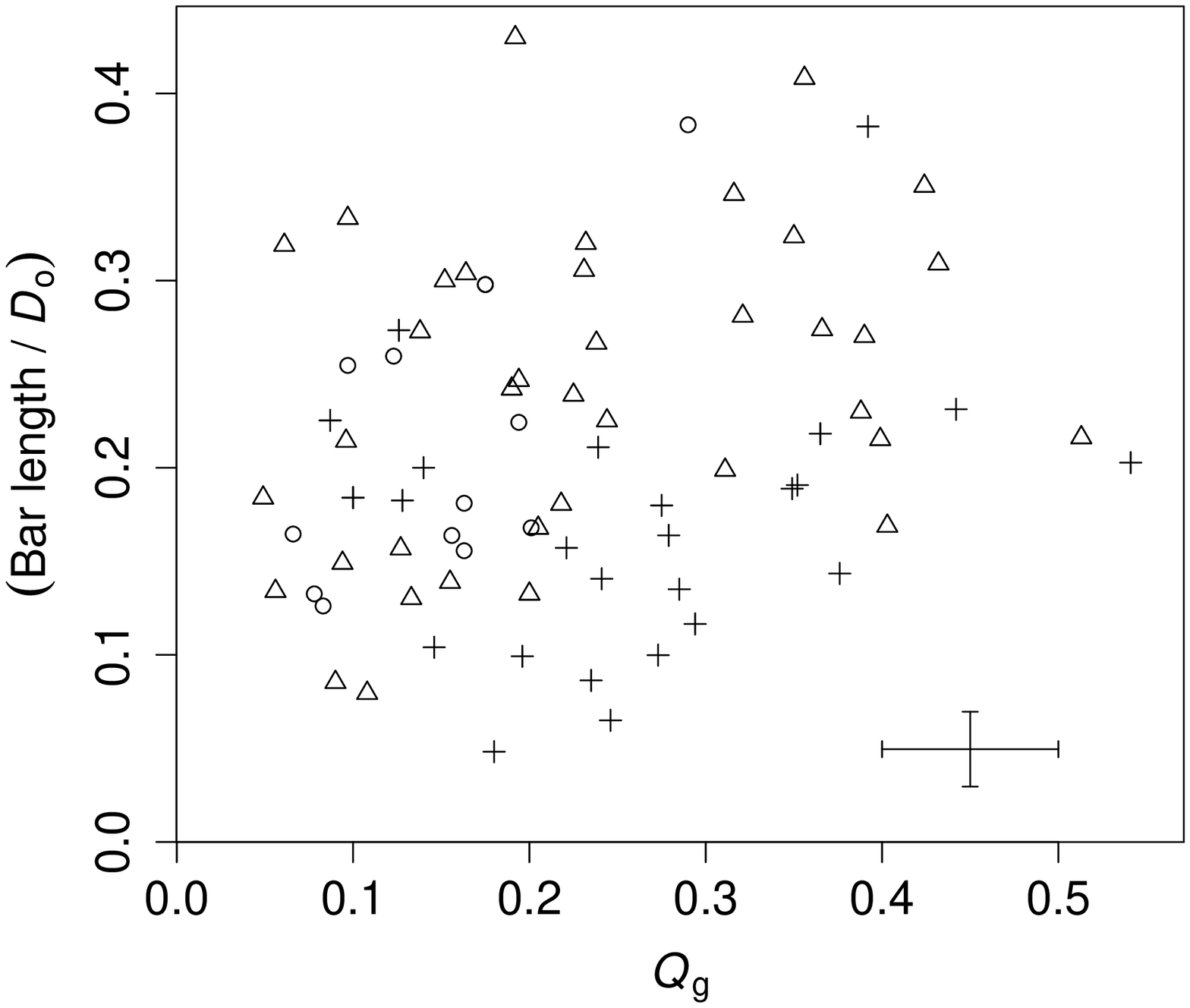}
\includegraphics[width=0.45\textwidth]{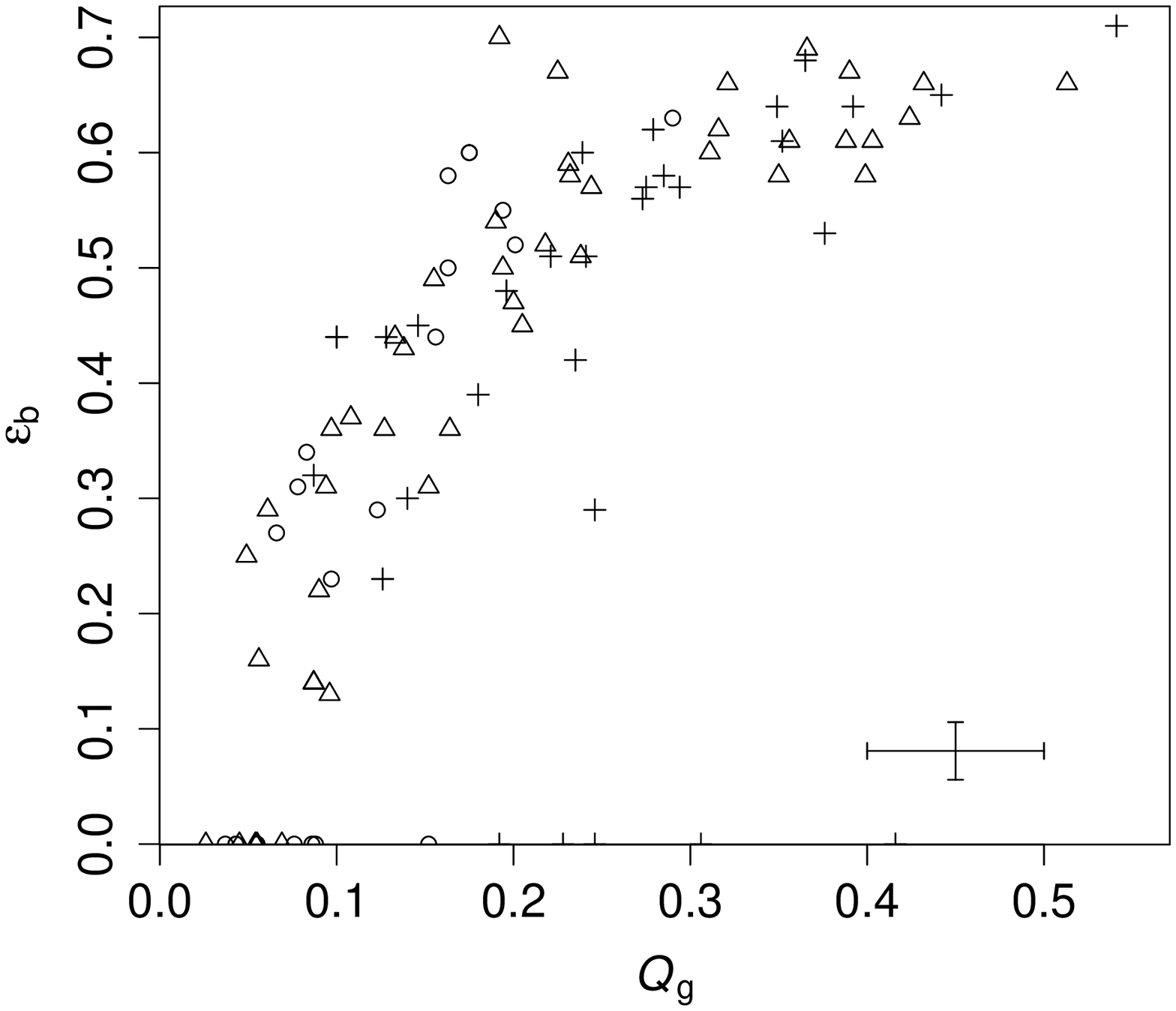}\\
\end{tabular}
\caption{\label{blothings} Bar length and strength parameters. In the top left-hand panel, the bar length as function of the non-axisymmetric torque parameter $Q_{\rm g}$ is given. In the top right-hand panel, the relative bar length, i.e.~the ratio of bar length and exponential disc scalelength, as a function of $Q_{\rm g}$ is given. In the bottom left-hand panel, the ratio of bar length to $D_{\rm o}$ as a function of $Q_{\rm g}$ is given. In the bottom right-hand panel, the ellipticity of the bar, $\epsilon_{\rm b}$, as a function of $Q_{\rm g}$ is given. In the bottom-right panel, $Q_{\rm g}$ value of galaxies with no bar are plotted at $\epsilon_{\rm b}=0$.}
\end{center}
\end{figure*}

With the information in Table~\ref{galax2} we can check the correlation between the bar strength and length reported by Martinet \& Friedli (1997), Knapen et al.~(2002), and also discussed by Elmegreen et al.~(2007). In the top-left panel of Fig.~\ref{blothings} we show how the length of the bar is correlated with $Q_{\rm g}$. We find that that weak bars can be both short and long; on the other hand, there are no short strong bars. The edge between the populated and non-populated area of the graph is sharp. If we plot (Fig.~\ref{blothings} top-right) the length of the bar relative to the scalelength of the disc, the trend previously noted remains. A similar trend but with a much larger scatter is found for the ratio between the bar length and $D_{\rm o}$ (Fig.~\ref{blothings}, bottom-left). It thus appears that galaxy size, disc scalelength and minimum allowed bar length for a given $Q_{\rm g}$ are closely related and that there is a roughly linear relationship between $Q_{\rm g}$ and the minimum allowed bar length. Our results are perfectly compatible with Fig.~2 in Elmegreen et al.~(2007) if we consider that $Q_{\rm g}$ roughly scales well with $A_{2}$ (as shown by Laurikainen et al.~2004). In their Letter, Elmegreen et al.~(2007) state that corotation should move outwards with time as the bar evolves, causing an increase in the semi-major axis of the orbits of the stars in the bar that is related to bar length (theoretically described by Mart\'inez-Valpuesta et al.~2006). This would explain why only the upper-left parts of the top-row plots in Fig.~\ref{blothings} are filled and would set the minimum allowed bar size as an initial bar length for a given $Q_{\rm g}$.

The top left-hand panel in Fig.~\ref{blothings} shows that spiral galaxies with early morphological types, between $0.1<T\leq3.0$, have longer bars, confirming results by Laurikainen et al.~(2002) and Erwin (2005). Bars in galaxies with types between $0.1<T\leq3.0$ are longer when scaled with $D_{\rm o}$ (as reported by Elmegreen \& Elmegreen 1985; Martin 1995; Regan \& Elmegreen 1997; Erwin 2005) although the effect is not as spectacular as when the absolute bar size is used. This effect is not seen when bar size is scaled with $h$ (but it is seen in Erwin 2005), probably because a lack of statistics for high bar size relative to the disc length values.

The bottom right-hand panel of Fig.~\ref{blothings} shows how the maximum bar ellipticity, $\epsilon_{\rm b}$, is an indicator of $Q_{\rm g}$. The correlation factor between these two parameters is $\rho=0.81$. For a given $\epsilon_{\rm b}$ early-type galaxies are found to have a slightly lower $Q_{\rm g}$, as found by Laurikainen et al.~(2002). Thus early-types galaxies are on the upper envelope of the distribution. This is at least partly because for early-type galaxies, we are using a lower value of $h/h_{\rm z}$ (the ratio between the disc scalelength and the disc scaleheight). A bigger disc scaleheight reduces the impact of high ellipticity bars in the galaxy.

\section{Comments on nuclear ring parameters}

From our images we have measured the sizes of the nuclear rings, the deprojected differences between the PAs of the major axes of the rings and those of the bar of their host galaxy, and the deprojected, or intrinsic, ellipticities of the rings.

The dimensions and the orientation of the nuclear rings were measured directly from the deprojected images (mostly from $HST$, but some from literature sources as explained in Section~3) and their semi-major axis measurements are in Table~\ref{galax2}. The size of the rings was measured by fitting an ellipse to the ridge line in \ha{} images, \pa{} images, colour-index images, UV images, or in structure maps. This fitted ellipse allowed us to make an estimate of the ellipticity and the orientation of the rings. The uncertainties in these measurements are highly dependent on the errors in the orientation parameters that we used. These orientation parameters are less likely to be accurate in strongly-barred galaxies. Another factor that adds some uncertainty is the assumption that the nuclear rings lie in the galactic plane of disc galaxies, but we have no evidence to the contrary (except for NGC~2787. See Sil'chenko \& Afanasiev 2004).

\begin{figure}
\begin{center}
\begin{tabular}{c}
\includegraphics[width=0.45\textwidth]{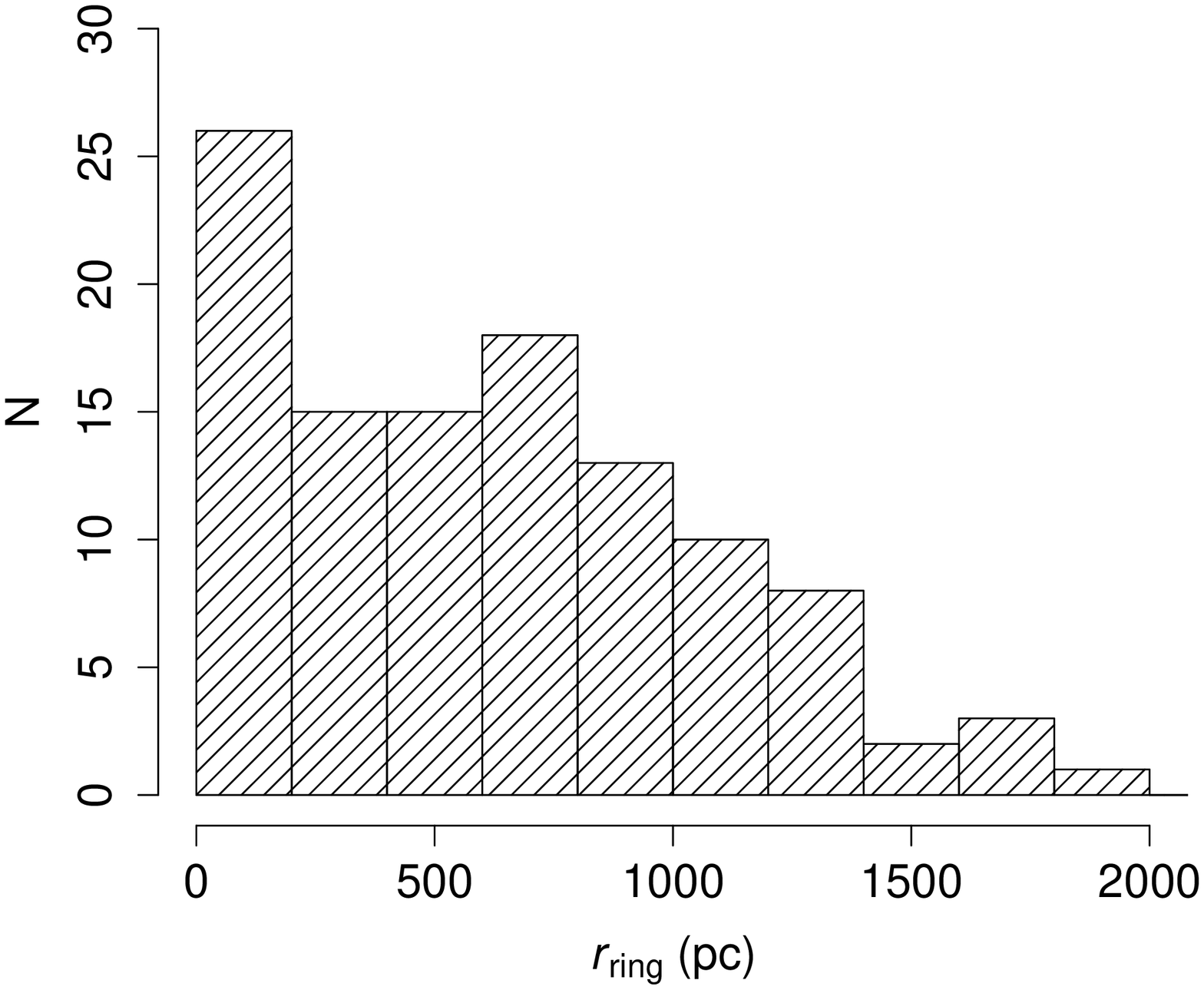}\\
\includegraphics[width=0.45\textwidth]{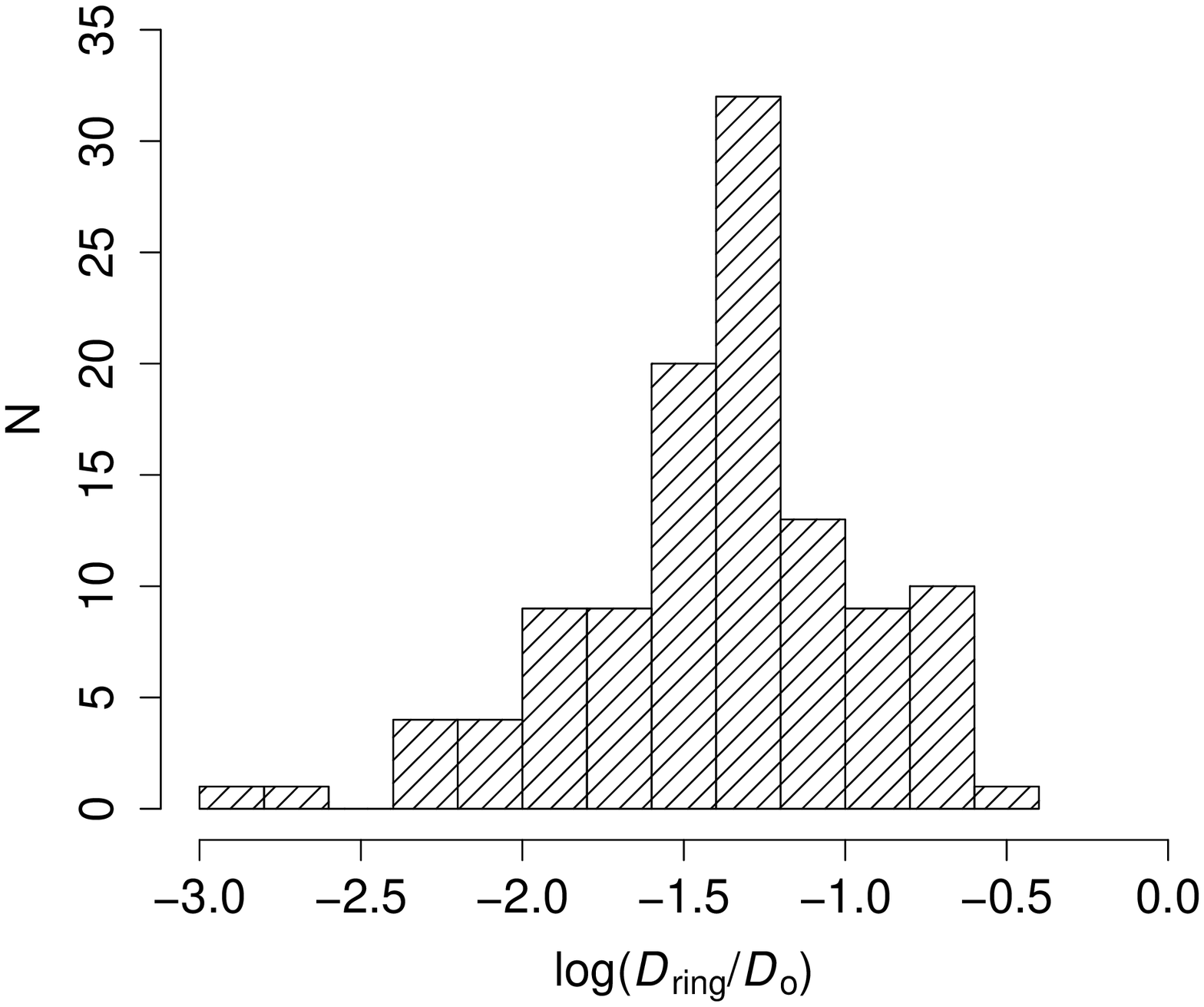}\\
\end{tabular}
\caption{\label{hist} Top panel: size distribution of the nuclear rings. ESO~198-13 and ESO~565-11 are too large to be in this plot. The bin size is 200\,pc. Lower panel: distribution of nuclear ring sizes relative to the size of the galaxy ($D_{\rm o}$).}
\end{center}
\end{figure}

\subsection{Ring size distribution}

Fig.~\ref{hist} shows the size distribution of the nuclear rings. It is evident that smaller rings are much more frequent than bigger rings, especially when considering resolution limitations and dust obscuration close to the centre. As an example, in our sample rings with sizes smaller than 200\,pc are four times more frequent than rings with size between 1000\,pc and 1200\,pc. The bottom panel of Fig.~\ref{hist} is very similar to the one presented by Laine et al.~(2002) who use $D_{25}$ instead of $D_{\rm o}$ for scaling the nuclear rings; the two values are almost identical for most of the galaxies and thus the plots can be safely compared. The only qualitative difference between the two plots is that our histogram shows a longer small-size tail. The reason is that Laine et al.~(2002) used mostly data from Buta \& Crocker (1993) which were collected before high-resolution images from $HST$ were available. Our plot peaks around $D_{\rm r}/D_{\rm o}=0.05$ (where $D_{\rm r}$ denotes the nuclear ring diameter), which is similar to the result from Laine et al.~(2002).

\section{Non-axisymmetries and nuclear rings}

\subsection{Ring sizes and $Q_{\rm g}$}

\begin{figure}
\begin{center}
\begin{tabular}{c}
a.\includegraphics[width=0.45\textwidth]{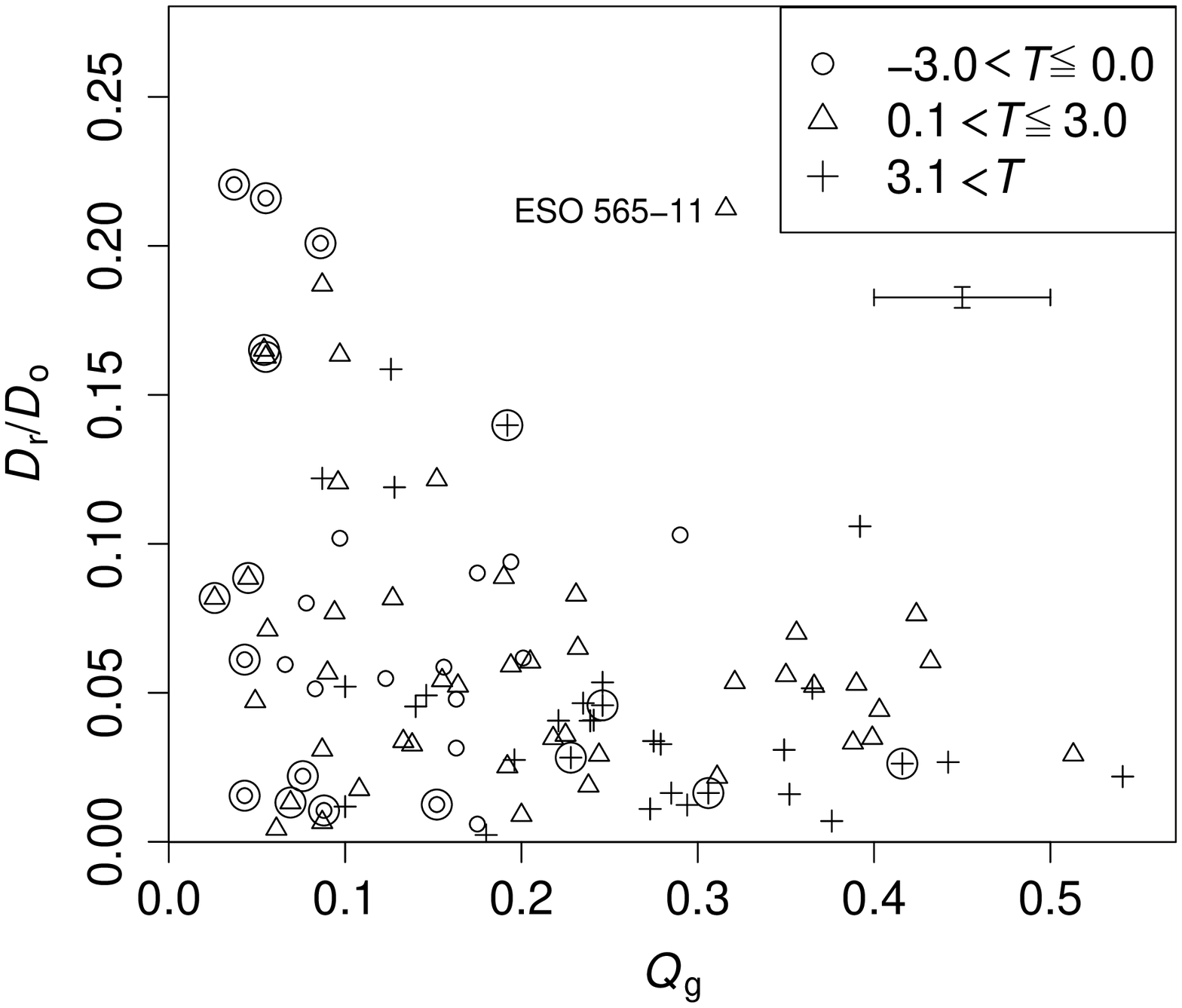}\\
b.\includegraphics[width=0.45\textwidth]{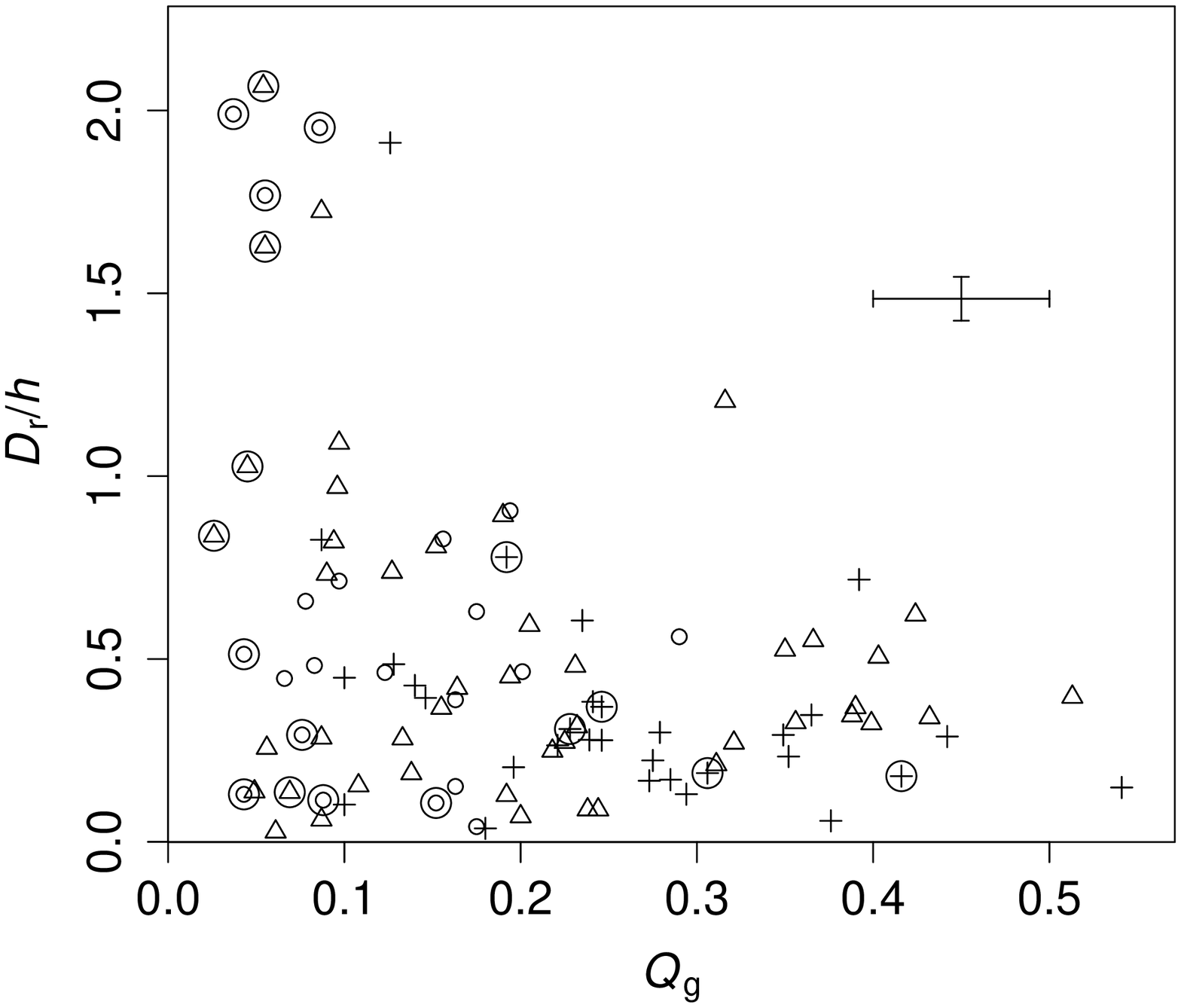}\\
c.\includegraphics[width=0.45\textwidth]{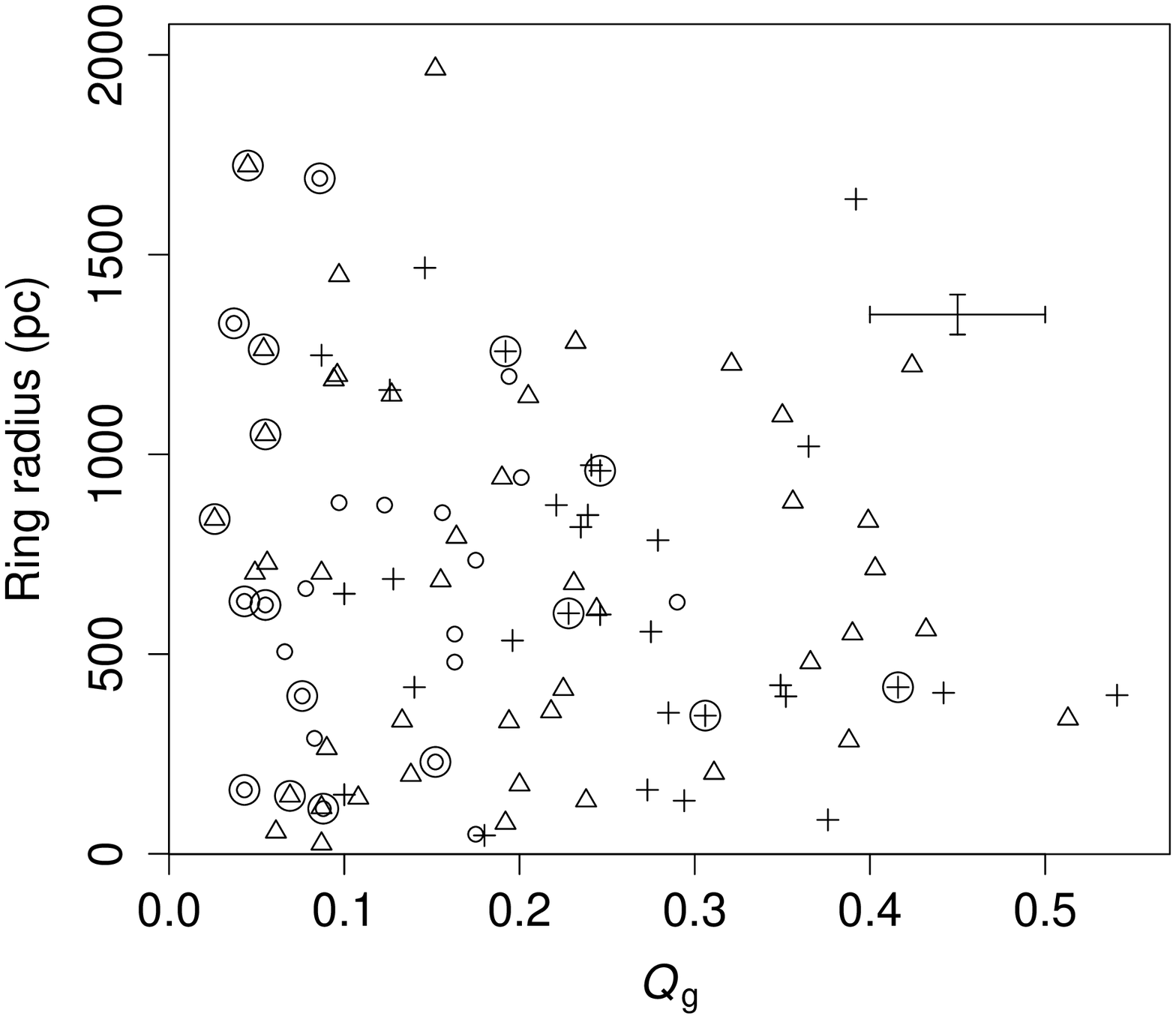}\\
\end{tabular}
\caption{\label{rdod25} Top panel: Nuclear ring size relative to $D_{\rm o}$ as a function of the non-axisymmetric torque parameter of the host galaxy $Q_{\rm g}$. Middle panel: same but using the disc scalelength to scale the ring size. Bottom panel: same but using absolute ring size. Two outlying points corresponding to ESO~198-13 and ESO~565-11 do not appear in the bottom panel. Encircled symbols denote unbarred galaxies.}
\end{center}
\end{figure}

Fig.~\ref{rdod25}(a) shows the relative size of the nuclear ring, defined as the ring diameter divided by the host galaxy diameter, $D_{\rm o}$, as a function of the non-axisymmetric torque, $Q_{\rm g}$, of the host. We see that, as a general trend, small $Q_{\rm g}$ values allow a wider range of relative sizes of nuclear rings and that for high $Q_{\rm g}$ values only small rings are allowed. Knapen (2005) found this from a sample of 16 nuclear rings: ``large nuclear rings occur in weak bars''. This effect has also been seen previously in simulations (Salo et al.~1999). The outlier in the $D_{\rm r}/D_{\rm o}$ plot is ESO~565-11, by far the biggest and most elliptical nuclear ring in the Atlas, which may be in an early stage of development (Buta et al.~1999). This ring, however, is not an outlier in the $D_{\rm r}/h$ plot (Fig.~\ref{rdod25}$b$). Fig.~\ref{rdod25}(c) shows that for absolute ring size, the trend is still marginally visible. The edge between the populated and unpopulated area in the third plot (Fig.~\ref{rdod25}$c$) is quite sharp, and almost linear with maximum ring radii of 2000\,pc at $Q_{\rm g}=0.1$ and around 500\,pc for $Q_{\rm g}=0.5$. The morphological type does not affect the relation between either absolute or relative ring size and $Q_{\rm g}$.

We explored possible relations between nuclear ring parameters and the local non-axisymmetric torque value. We define this local non-axisymetric torque as
\begin{equation}
Q_{\rm nr}=(F^{\rm max}_{\rm T}(r_{\rm nr})/\langle F_{\rm R}(r_{\rm nr})\rangle)
\end{equation}
where $F^{\rm max}_{\rm T}(r_{\rm nr})$ is the maximum tangential force at the deprojected major axis radius of the nuclear ring, $r_{\rm nr}$, and $\langle F_{\rm R}(r_{\rm nr})\rangle$ is the average radial force at the same radius. For small nuclear rings, the determination of $Q_{\rm nr}$ is harder because of a lack of resolution (PSF of 2\arcsec{} in 2MASS images) and because its behaviour becomes much more dependent on the model of the bulge that we subtract. We have found no trend when repeating the plots shown in Fig.~\ref{rdod25} using $Q_{\rm nr}$ in the horizontal axis. This may indicate that nuclear ring sizes do not depend on local conditions; higher resolution images and better 2D bulge modelling are necessary to confirm this.

\subsection{Ring and bar sizes}

\begin{figure}
\begin{center}
\begin{tabular}{c}
\includegraphics[width=0.45\textwidth]{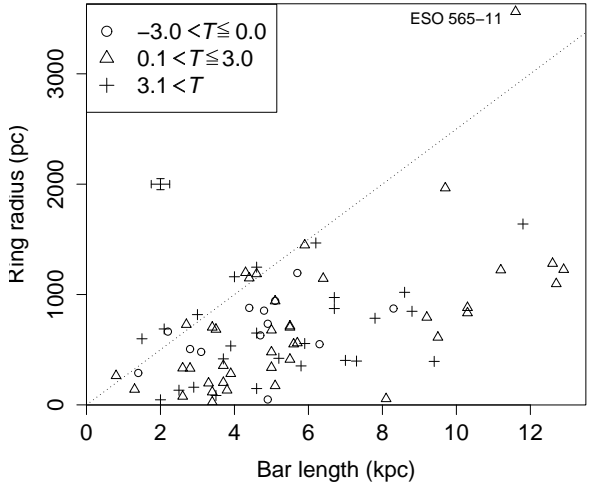}\\
\end{tabular}
\caption{\label{barsize} Nuclear ring size versus bar size as determined from fits to 2MASS data. The dashed line indicates $r_{\rm r}=r_{\rm b}/4$.}
\end{center}
\end{figure}

A relation between the bar length as derived from 2MASS images and the maximum nuclear ring radius can be seen in Fig.~\ref{barsize}, where only one galaxy is significantly above a sharp edge that is located at approximately $r_{\rm r}=r_{\rm b}/4$. The outlier galaxy is ESO~565-11. ESO~565-11 has a nuclear ring that is probably in a very quick initial phase of its evolution, and is, as we have already seen, an outlier in some of our plots (Buta et al.~1999). Fig.~\ref{barsize} also shows that the longest bars are found in galaxies of morphological type $0.1<T\leq3.0$, as already shown in Fig.~\ref{blothings}.

\subsection{PA offset between nuclear rings and bars}

\begin{figure}
\begin{center}
\begin{tabular}{c}
\includegraphics[width=0.45\textwidth]{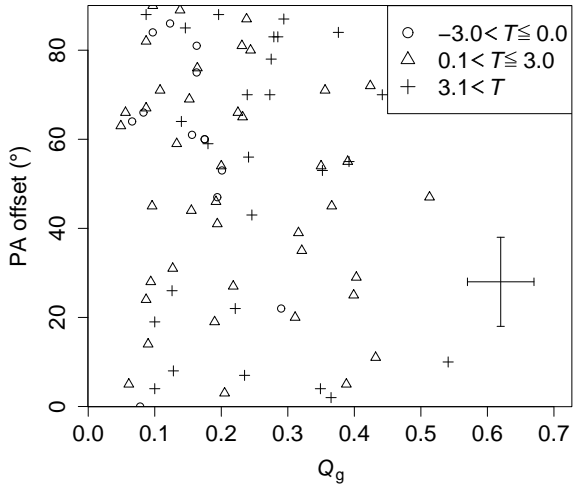}\\
\end{tabular}
\caption{\label{paoff} Difference between the major axis PA of the nuclear ring and that of the bar which hosts it, as a function of gravitational torque parameter $Q_{\rm g}$.}
\end{center}
\end{figure}

The PA offset between the bar major axes and the nuclear ring is the measure with the highest uncertainty in this paper. Many rings are intrinsically round which makes it difficult to measure the PA and makes the measurement very sensitive to small deprojection errors. Thus if any trend existed between PA offset and $Q_{\rm g}$, it would be much more evident for more elliptical rings. Knapen et al.~(2002) found no correlation between the PA offset and the $Q_{\rm b}$ of the galaxy. Our results are shown in Fig.~\ref{paoff} and they confirm the previously cited work. The morphological type of the host galaxy does not cause any segregation in the scatter plot. If instead of using $Q_{\rm g}$ on the horizontal axis we use the maximum bar ellipticity, $\epsilon_{b}$, the results are very similar.

When we compare the individual PA offset values from our survey with those from Knapen et al.~(2002), we find that the differences between the two surveys range between 3\deg{} and 32\deg{} (there is a bigger difference in NGC~2903 but the nuclear ring is defined differently), with an average difference of 18\deg. Uncertainties in the orientation parameters could explain these differences, since the differences in the inclination angle used for the deprojection are as large as 20\deg{}.

\subsection{Ring ellipticities and bar ellipticities}

\begin{figure}
\begin{center}
\begin{tabular}{c}
\includegraphics[width=0.45\textwidth]{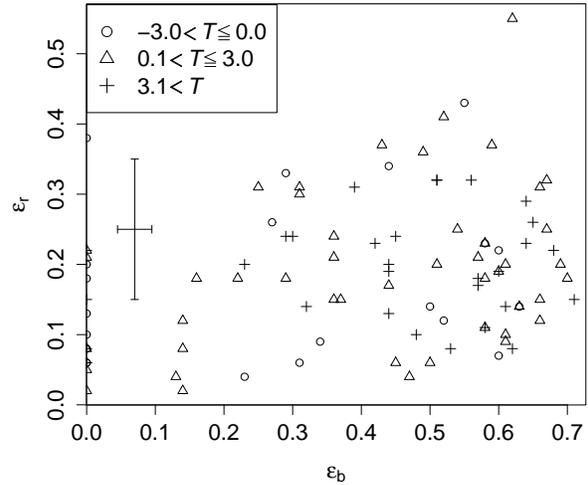}\\
\end{tabular}
\caption{\label{re} Deprojected ellipticity of star-forming nuclear rings as a function of bar ellipticity. Spiral galaxies with no noticeable bar are plotted at $\epsilon_{b}=0$. We do not include galaxies too edge-on to detect any bar and galaxies too faint to have a bar ellipticity reliably measured in 2MASS $H$-band images.}
\end{center}
\end{figure}

Figure~\ref{re} shows the ellipticities of the nuclear rings versus those of their host bars, and demonstrates how the most elliptical rings can only be found in the most elliptical bars. There is a maximum ring ellipticity for a given bar ellipticity and this limit is roughly linear as a first approximation. Galaxies with no bar have rings with ellipticities up to $\epsilon_{\rm r}\sim0.4$. This maximum value corresponds to NGC~6861 which is a very early-type galaxy with a dusty nuclear ring. The scatter plot does not show any hint of different behaviour for different morphological type bins. Plotting $Q_{\rm g}$ instead of the bar ellipticity causes a much more subtle effect.

We found no correlation between the ring ellipticity and the PA offset between bar and ring. There is a theoretical explanation for this. Knapen et al.~(1995) and Shlosman (1999) state that these parameters are regulated primarily by the gas inflow into the circumnuclear regions. Since gas inflow does not depend exclusively on the parameters of the bar but more generally on the available fuel, correlations of ring ellipticity and PA offset are not expected, and in fact not observed. The dynamical time scale in the circumnuclear region is around $10^{7}$\,years, and thus the availability of the fuel can be expected to be variable, reducing any possible correlation with the bar strength.

The two rings of NGC~2787 (not plotted in Fig.~\ref{re}) have ellipticities of $\epsilon_{\rm r}\sim0.7$ when projected into the main plane of the galaxy. This is because the rings are lying in a disc that is polar to the bar orientation of the galaxy (Sil'chenko \& Afanasiev 2004). The rings probably have a much smaller intrinsic elongation.

\section{Nuclear rings and morphological type}

\begin{figure}
\begin{center}
\begin{tabular}{c}
\includegraphics[width=0.45\textwidth]{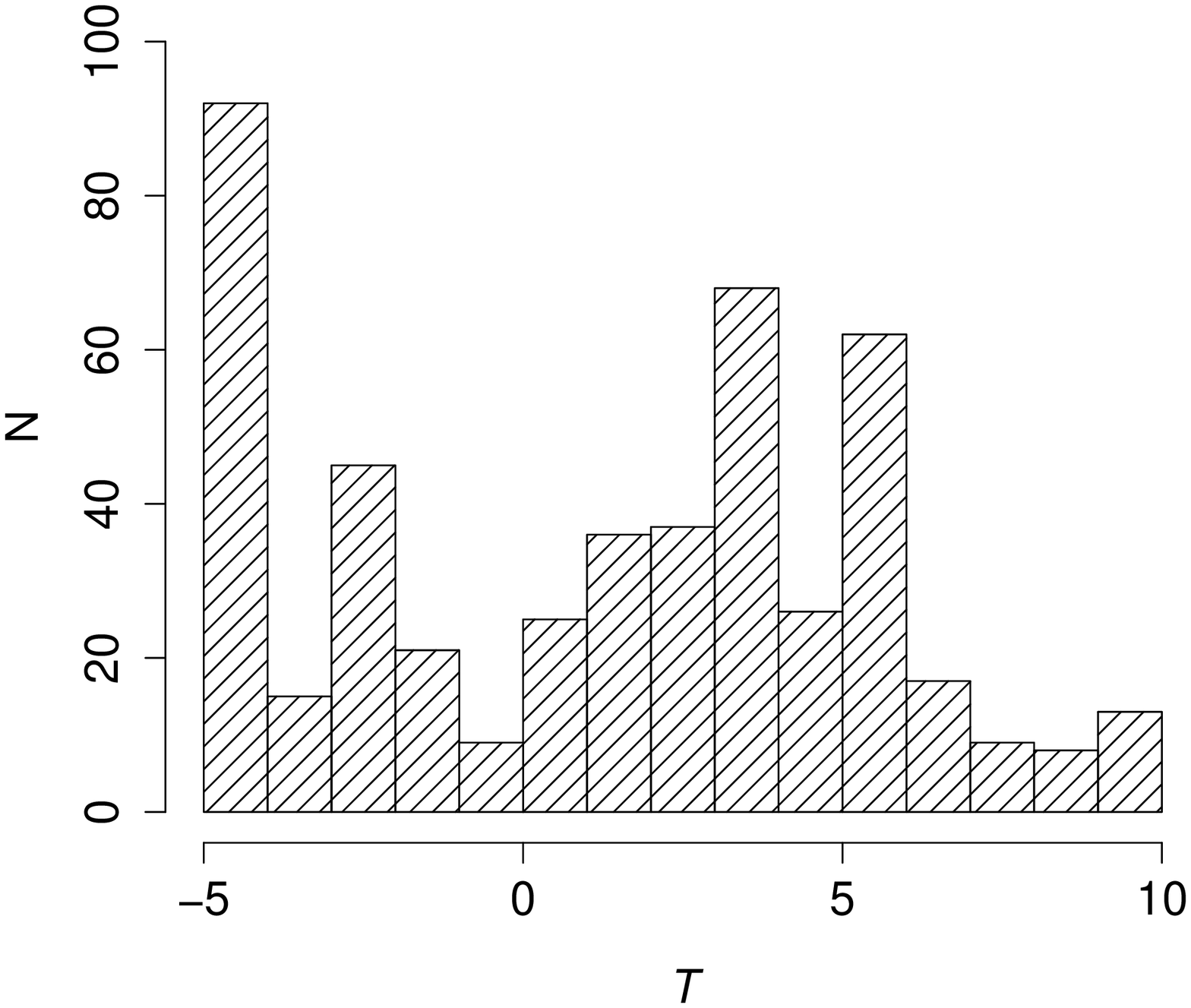}\\
\includegraphics[width=0.45\textwidth]{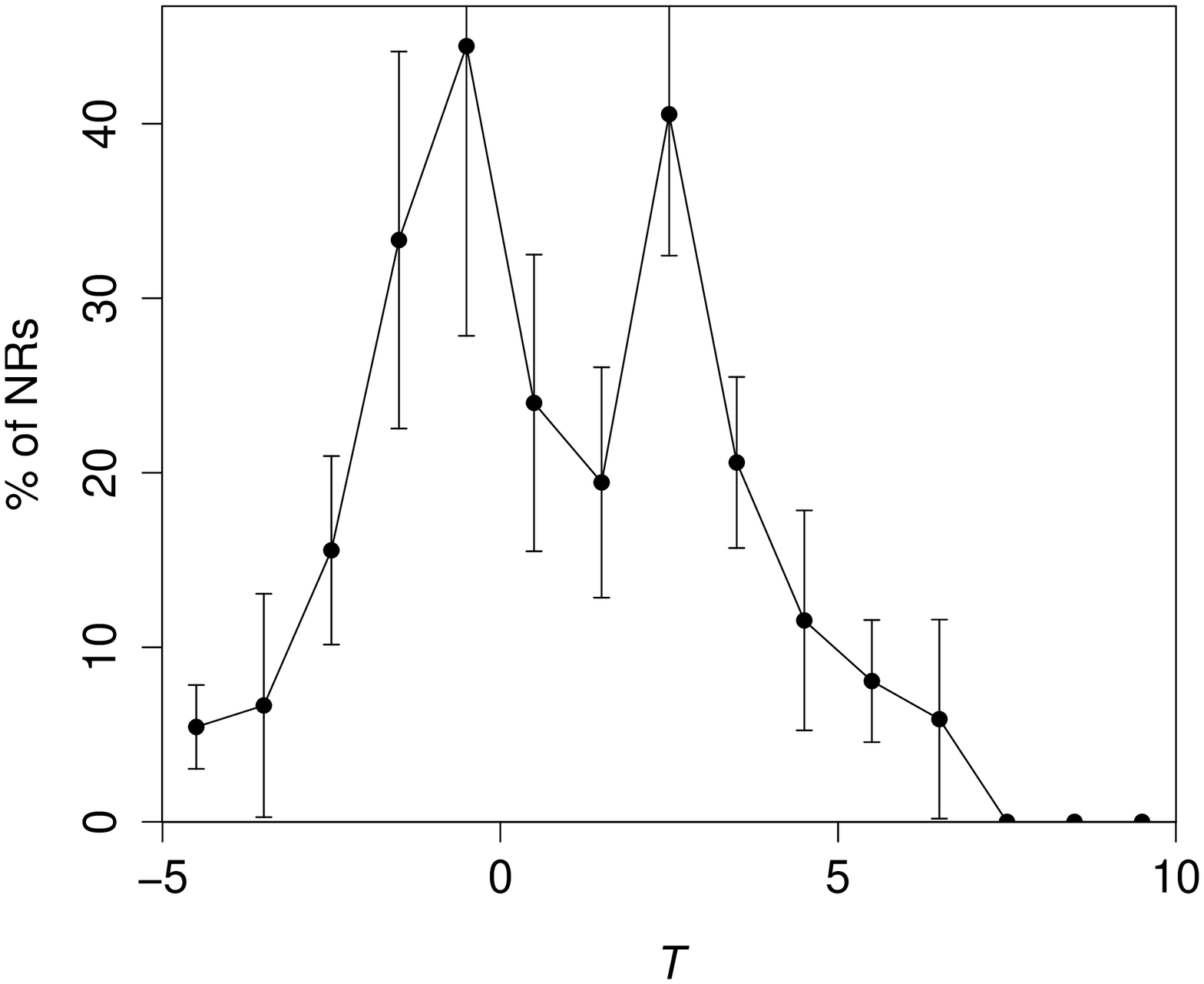}\\
\end{tabular}
\caption{\label{type} Top panel: histogram showing the distribution in the morphological type of the 483-galaxy sample. Overplotted (double-hatched areas) we show the galaxies with nuclear rings. Bottom panel: fraction of nuclear rings in each morphological type. Bin size $T=1$. Error bars calculated using Poisson binomial statistics.}
\end{center}
\end{figure}

Star-forming nuclear rings have been found to occur preferably in galaxies with morphological types between S0 and Sbc (Knapen et al.~2006).

To study the dependence of the fraction of nuclear rings on the morphological type of the host galaxy type we consider only the 76 nuclear-ringed galaxies that are found in our statistically meaningful sample of 483 galaxies. The other nuclear rings were found in a literature search and not in a survey and are thus less suitable for a reliable statistical analysis.

In Fig.~\ref{type} we show the absolute number of nuclear rings in each type bin (top panel, double-hatched) and the fraction of nuclear rings in each bin (bottom panel). Nuclear rings are mainly found in galaxies with $-2<T<5$ (S0 $-$ Sc). In $T\leq-3$ galaxies, rings usually are dusty and have probably little if anything to do with a resonance framework. For types later than Sd no nuclear rings have been found (see Knapen et al.~2006). Late-type  spirals (of types Sd to Sm) usually have
slowly rising, low-amplitude rotation curves which prevent ILRs from
forming (see the review by Sofue \& Rubin 2001 and a case study by Elmegreen et al.~1998). A galaxy without ILRs
would be very unlikely to host  a star-forming nuclear ring.

A possible alternative explanation for the lack of nuclear rings in
late-type galaxies can be deduced from the work of Dalcanton et
al.~(2004). They found from a study of edge-on late-type galaxies that
dust
lanes are found only in galaxies rotating faster than $V_{c}\sim120{\rm km
s^{-1}}$
because above this threshold the dust scaleheight drops drastically. As
cold molecular gas is linked to dust, Dalcanton et al.~(2004) speculate
that the ISM scaleheight should drop similarly at this threshold value of
$V_{c}\sim120{\rm km s^{-1}}$, which would
be the velocity at which the disc becomes gravitationally unstable. In a
stable disc,
turbulent velocities are much higher and the star formation rate drops
considerably (by a
factor of roughly 3; Dalcanton et al.~2004). In a more rapidly rotating,
earlier-type, gravitationally unstable disc the ISM
would be concentrated closer to the plane and would form molecular clouds
that would
interact with resonances and thus may form nuclear rings.

Knapen (2005) found that the maximum of the nuclear rings distribution occurs at $T=4$ (Sbc) but we find that it occurs in earlier-type galaxies ($T=2-3$ Sab $-$ Sb). There is another maximum around $T=-1$ (S0) but low statistics cause a big uncertainty in this particular measurement.

We find that $6\pm2\%$ of the elliptical galaxies ($-5\leq T\leq-3$) have dust rings. $24\pm5\%$ of lenticular galaxies ($-3<T\leq0$), and $19\pm2\%$ of non late-type spiral galaxies ($0<T\leq7$) have star-forming nuclear rings. We find that $20\pm2\%$ of the galaxies with types $-3<T\leq7$ (the range of disc galaxies where nuclear rings are found) host star-forming nuclear rings. Irregular galaxies never have a nuclear ring. These results confirm those of Knapen (2005) who concluded that a fifth of disc galaxies have star-forming nuclear rings.

\section{Nuclear activity}

In discussing nuclear activity we are excluding those galaxies classified as nuclear starbursts, which are naturally correlated with nuclear rings, as small nuclear rings may be confused with genuine nuclear star formation. As a consequence, many galaxies with a nuclear ring are listed in catalogues as starbursting galaxies because they have a nuclear ring.

The most extensive survey to date of nuclear activity in local galaxies has been published by Ho et al.~(1997a; 1997b). Forty-nine of our nuclear ring galaxies appear in their survey. We compared the proportion of each kind of nuclear activity (LINER, Seyfert, and Transition) in Ho et al.~(1997a; 1997b) with that in the 49 AINUR galaxies which matched with their catalogue, taking into account the morphological type of AINUR galaxies. We find no significant differences (larger than one sigma) between the proportion of active galaxies in AINUR and Ho et al.~(1997a; 1997b).
This is in contradiction both with Knapen et al.~(2006), who found a significantly higher proportion of active galaxies among galaxies with nuclear rings, and with Comer\'on et al.~(2008b), who found that nuclear rings are related with $\sigma$-drops, which are in turn related with nuclear activity. This may be because in Knapen et al.~(2006) and Comer\'on et al.~(2008b) the sample selection criteria favour `famous' galaxies, which tend to be active (Comer\'on et al.~2008b).

\section{Nuclear rings in unbarred disc galaxies}

Several of the nuclear rings we studied are found in unbarred galaxies. If we do not consider NGC~1386, NGC~3593, NGC~4100, NGC~4526, and NGC~6503 which are too highly inclined to tell whether they are barred or not, we have 18 {\it bona fide} unbarred spiral galaxies with nuclear rings. We explore the properties of these galaxies in some detail to find evidence that their nuclear rings are of resonance origin.

NGC~2997, NGC~5033, and NGC~5247 have very prominent arms that cause $Q_{\rm g}>0.2$ and are likely to generate resonance effects similar to those of a bar (see e.g.~Lindblad (1961); Shlosman (1999); Shlosman (2001); Rautiainen \& Salo 2000).

Several galaxies have ovals: NGC~4571 (Eskridge et al.~2002), NGC~4736 (Bosma et al.~1977; Trujillo et al.~2009), NGC~6753 (Laurikainen et al.~2004), and possibly NGC~7049 (Corwin et al.~1985). Ovals have dynamical effects very similar to those produced by bars (Kormendy \& Kennicutt 2004).

Other galaxies can have dynamical resonances due to a tidal interaction with a close companion (see e.g.~Buta et al.~1992). A good case is the ring in NGC~5427, a galaxy which may be interacting with NGC~5426 (even if according to Fuentes-Carrera et al.~2004, NGC~5427 has a bar of 32\arcsec\ in length that we have not detected). Other nuclear rings which may be caused by the same mechanism are the ones in NGC~3245 (this galaxy is close to NGC~3245A), NGC~6861 (near IC~4943), and NGC~4459 (which could be weakly interacting with NGC~4468 and NGC~4474). Simulations also show that an inflow of material to nuclear regions due to tidal interactions (Hernquist 1989; Hernquist \& Mihos 1995) may cause starbursts in the central areas of the galaxy.

Counterrotating components have been reported in NGC~4138 (Jore et al.~1996), NGC~4826 (Braun et al.~1992), NGC~7217 (Merrifield \& Kuijken 1994) and NGC~7742 (de Zeeuw et al.~2002; Mazzuca et al.~2006). In both these unbarred galaxies the presence of a nuclear ring and a counter-rotating component would be explained by a past minor merger with a gas-rich dwarf galaxy in a retrograde orbit (see Jore et al.~1996 for NGC~4138, Braun et al.~1994 for NGC~4826, Sil'chenko \& Moiseev 2006 for NGC~7217, and Mazzuca et al.~2006 for NGC~7742). Such a merger could have led to the counter-rotating component and to an oval distortion strong enough to create a nuclear ring.

We then still have three galaxies (ESO~198-13, NGC~473, and NGC~3011) that have nuclear rings for which we have no observational indications for an explanation in terms of a resonance framework.

We have found that nuclear rings are mostly found in barred galaxies. Most of the unbarred hosts have features that may be driving resonances. For the three remaining galaxies, the existence of an undetected weak oval distortion cannot be ruled out.

\section{Notes on individual galaxies}

For each galaxy we first list the original citation to the nuclear ring if this exists. We also describe the particularities of the ellipticity and the 2D bulge/disc/fit when necessary. Other interesting particularities have been described when necessary.

{\bf ESO~198-13}: Buta (1986). This galaxy is visually unbarred. We have not been able to perform an ellipticity profile to check it because the 2MASS $H$-band image is too shallow. This galaxy has three ring features which makes the identification of the innermost ring as a nuclear ring secure, although it is more than 3\,kpc in radius.

{\bf ESO~437-33}: Buta \& Crocker (1991). The non-axisymmetry profile is very flat which causes a large uncertainty in the radius at which $Q_{\rm g}$ is found.

{\bf ESO~437-67}: Buta \& Crocker (1991). We decided to adopt a face-on orientation for this galaxy because the HYPERLEDA values are clearly for the bar rather than the disc of the galaxy. Also, the galaxy has an $R'_{1}$ ring that looks nearly face-on. During the 2D decomposition we have not been able to fit a disc due to its shallowness in the 2MASS $H$-band image.

{\bf ESO~565-11}: Buta \& Crocker (1991), see Buta et al.~(1995; 1999) for detailed studies. This galaxy has the biggest and most intrinsically elliptical ring in our sample. It is unambiguously a nuclear ring because the galaxy has also inner and outer resonance features. The nuclear ring cannot be an $x_{1}$ ring because its major axis is not aligned with the bar. The ring may be at an early and fast stage of its formation (Buta et al.~1999).

{\bf IC~342}: Ishizuki et al.~(1990) in CO line emission. B\"oker et al.~(1997) in NIR. This nearby galaxy is very close to the Milky Way plane so it is highly extinguished. The 2MASS images we used for the ellipticity profile and the 2D decomposition had to be cleaned of many foreground stars (of the order of 1000). Apart from fitting the disc, the bulge, and the bar, the 2D decomposition also needs two circular Ferrers components to produce a reasonable fit. The smaller one has the size of the nuclear ring, while the other corresponds to a circular lens of 10\arcsec\ in radius. The nuclear ring is small and irregular.

{\bf IC~1438}: Buta \& Crocker (1993).

{\bf IC~4214}: Buta \& Crocker (1991).

{\bf IC~4933}: Ryder et al.~(2009).

{\bf NGC~278}: Pogge (1989). This galaxy has an unambiguous ring 9\arcsec{} in diameter with a somewhat octagonal shape. It has an other ring-like structure that we think is an inner pseudo-ring. Knapen et al.~(2004) called it a nuclear ring on the basis that this galaxy has no bar and that the feature is 2.2\,kpc in radius. We found, however, a bar inside this feature. This bar is doubtful and very weak, but it has also been found by P.~Erwin and L.~Guti\'errez (private comunication). Another argument against the 2.2\,kpc ring being a nuclear ring is that its size relative to the galaxy or to the disc scalelength is much larger than that of all the other nuclear rings in the sample.

{\bf NGC~473}: Heckman (1978). It is considered an inner ring in de Vaucouleurs et al.~(1980). The galaxy appears as barred in NED and RC3, but there is no sign of a bar or an interacting companion that could explain a nuclear ring. The nuclear ring has an ellipticity that is different from that of the disc of the galaxy, so a bar can be fitted to it, which may explain the false detection in the catalogues. The nuclear ring is one of the biggest in the sample.

{\bf NGC~521}: Buta et al.~(2009). This ring is made of dust and is visible as a dark annulus in $B$ images. Its morphology is coincident with that of nuclear star-forming rings; thus it is likely to be a star-forming nuclear ring deeply embedded in dust. The nucleus of the galaxy is off-centred. (Note: The ring radius of 7.7\arcsec\ given by Buta et al.~(2009) was based on an incorrect image scale. The radius given in Table.~\ref{galax2} uses the correct image scale.)

{\bf NGC~613}: Hummel et al.~(1987). The nuclear ring has also been considered a nuclear spiral (Buta \& Crocker 1993). The presence of the ring makes it difficult to fit the inner 10\arcsec\ in the 2D decomposition, but adding a round lens component with the size of the ring yielded a reasonable fit. Most of the ring surface is made up of dust which at some places has been torn away by intense star formation.

{\bf NGC~718}: Erwin \& Sparke (2002).

{\bf NGC~864}: Martini et al.~(2003). This is one of the smallest rings in the sample. It is composed of bright knots of star formation arranged in a {\it c} shape, with the `open' part of the ring located to the south-west.

{\bf NGC~936}: Erwin \& Sparke (2002). This ring is purely stellar with no traces of recent star formation. It encircles an inner bar. It appears as a brighter area in a structure map. The archive $HST$ images from this galaxy are too shallow to detect the nuclear ring; thus we used a WHT $r$-band image.

{\bf NGC~1068}: Scoville et al.~(1988). This ring is controversial and, sometimes interpreted as a nuclear pseudo-ring (de Vaucouleurs \& Buta 1980; Knapen et al.~2006). We consider that it is a complete ring, partially obscured by a dust lane that comes from the bar to the centre of the galaxy. This galaxy needed two bars to be properly fitted (with lengths of 8\arcsec{} and 80\arcsec{}). It also needed a central point-like component to take into account the bright Seyfert nucleus.

{\bf NGC~1079}: Maoz et al.~(1996). The ring looks smooth; there are few detectable clumps.

{\bf NGC~1097}: Burbidge \& Burbidge (1962). S\'ersic (1958) found a feature that he described as a `single spiral arm' with the same size as the nuclear ring. This galaxy appears strongly barred but $Q_{\rm g}$ is only $0.241$. The bar may look stronger than it really is due to the interaction with the dwarf elliptical galaxy NGC~1097A. The ring can be fitted approximately with a lens component. Buta \& Crocker (1993) state that the ring is a pseudo-ring. At high resolution, we see that the ring is made up of small spiral sections with a high pitch angle.

{\bf NGC~1241}: B\"oker et al.~(1999).

{\bf NGC~1300}: Pogge (1989). Suspected by Peterson \& Huntley (1980). This galaxy has the strongest bar in the sample ($Q_{\rm g}=0.541$). The north-west section of the ring is double.

{\bf NGC~1317}: Schweizer (1980). The bar is not very evident in the ellipticity profile and in the 2D decomposition, and the $Q_{\rm g}$ profile is not very peaked. The 2D fit of the galaxy is dominated by the disc component. The galaxy is double-barred and has a nuclear bar inside the nuclear ring that is clearly seen in the images and in the $Q_{\rm g}$ profile.

{\bf NGC~1326}: Garc\'ia-Barreto et al.~(1991a). The ring causes a poor fit in the inner parts of the 2D decomposition.

{\bf NGC~1343}: Zwicky \& Zwicky (1971). The ring again causes a poor fit in the inner parts of the 2D decomposition. The 2MASS image is so shallow that it has not been possible to fit the bar.

{\bf NGC~1386}: Tsvetanov \& Petrosian (1995). This galaxy is too highly inclined to measure the bar parameters. We have not detected a bar in the ellipticity profile but the NED reports it as barred. The ring is bright enough in \ha{} to be clearly traced even at the high inclination of the galaxy.

{\bf NGC~1387}: Laurikainen et al.~(2006) by subtracting a modelled galaxy from a 2.2\,micron image. The ring is seen in a colour-index image as blue patches distributed in a ring embedded in a dusty red disc.

{\bf NGC~1415}: Garc\'ia-Barreto et al.~(1996). The nuclear ring is well fitted with a lens component.

{\bf NGC~1433}: Buta (1983). The 2D decomposition yields a nuclear bar with the same length as the ring radius. This inner bar was discovered by Buta (1986). The ring looks a bit like a pseudo-ring but there are clear dust lanes crossing the ring, and covering part of it. The detection of the ring was performed using a colour-index image with exposures taken by $HST$ in 2001. The galaxy was previously imaged by $HST$ in the UV using the FOC camera (Maoz et al.~1996), leading to the identification of a feature defined as a nuclear ring by the authors. Their ring partly corresponds to the one presented here, but of half its diameter and off-centred. We think that this ring-like feature is an artefact and that it can be ruled out on the basis of more recent imaging in other bands.

{\bf NGC~1512}: Buta et al.~(1988). Hawarden et al.~(1979) described the nuclear ring as an `exquisite central spiral pattern' and showed a picture where the ring is clearly identifiable. The ellipticity of the nuclear ring is very different from that of the inner ring. This is probably due to the distortions caused by the bar ($Q_{\rm g}=0.366$).

{\bf NGC~1566}: Indirectly detected by Ag\"uero et al.~(2004), who added the radial fluxes along six spectrograph slits and discovered three radii where the \ha{} is enhanced, at the approximate locations where ILR, corrotation and outer Lindblad resonance should be. To our knowledge the present paper shows the first direct detection of the ring, which is seen as a dusty feature in the colour-index map. The paper of Ag\"uero et al.~(2004) suggests that it should have strong star-forming areas. Unfortunately, we do not have \ha{} imaging for this galaxy and the $HST$ UV image we used is too shallow to show any ring. The dusty feature we have found is distorted in the south-east.

{\bf NGC~1672}: D\'iaz et al.~(1999). The eastern part of the ring is fainter than the western part which may indicate that it is more heavily obscured by dust.

{\bf NGC~1808}: S\'ersic \& Pastoriza (1965) as a `hot spot' nucleus. The ring is nearly round after deprojection, except for its south-east part. The orientation parameters we used for this galaxy are the ones of the outer ring. The inner parts of the galaxy make it appear as a highly-inclined Sb in shallow images.

{\bf NGC~1819}: Buta \& Crocker (1993). The ring is so bright that a lens component has to be introduced in the 2D fit to take it into account. The nuclear ring is incomplete in the east. The western part is made up of two arcs of intense star formation.

{\bf NGC~2273}: Ferruit et al.~(2000). This nuclear ring is composed of two tightly wound star-forming arms.

{\bf NGC~2595}: Buta \& Crocker (1993).

{\bf NGC~2787}: Sarzi et al.~(2001) for the dust ring. The two rings are found in a polar disc perpendicular to the main disc of the galaxy and to the outer \hi{} ring of the galaxy (Sil'chenko \& Afanasiev 2004). The dust ring is the innermost part of a tightly wound dust spiral. Just inside the dust ring and with a slightly higher ellipticity is an \ha{} ring, which may be caused by the ionization of the inner part of the dust disc by the Seyfert nucleus. The two rings are probably polar and their deprojected ellipticity and PA values are thus not reliable. Because we round some of the values in Table~\ref{galax2}, they appear identical for the two rings, even though the ionization ring is slightly smaller than the other.

{\bf NGC~2859}: Erwin \& Sparke (2002). This star-forming nuclear ring is a dusty feature that is not visible in an \ha{} image and has an inner bar inside it (Erwin \& Sparke 2002).

{\bf NGC~2903}: Sakamoto et al.~(1999). The ring size is a matter of discussion. Alonso-Herrero et al.~(2001) consider the ring to be twice as large as we do, but their ring is extremely elliptical after deprojection while what we consider to be the ring is intrinsically much rounder. The 2D decomposition yields an almost bulgeless galaxy where the central regions are dominated by a lens that has the size of the intense star-forming area. This area of star formation causes an artificial peak in the $Q_{\rm g}$ profile.

{\bf NGC~2935}: Buta \& Crocker (1991).

{\bf NGC~2950}: Erwin \& Sparke (2002). This ring is purely stellar with no traces of recent star formation. It encircles an inner bar. It appears as a brighter area in a structure map. It has been sugested to be either a faded star-forming ring or a side-effect of the double-barred nature of NGC~2950 (Erwin \& Sparke 2002).

{\bf NGC~2985}: Comer\'on et al.~(2008a). This nuclear ring is one of the smallest in the sample. This galaxy is unbarred in the literature; however, we found that in the deprojected image a slight oval distortion is detectable even though its strength and even its existence are very dependent on the orientation parameters. We could not fit a bar in the 2D decomposition. A possible cause of a resonance ring may be the dwarf irregular satellite KDG~59 that is probably causing the disturbances seen in \hi{} maps (Comer\'on et al.~2008a). NGC~3027 may also cause these disturbances but there is no evidence for a tidal tail connecting the two galaxies (Noordermeer et al.~2005; Oosterloo, private communication).

{\bf NGC~2997}: S\'ersic \& Pastoriza (1965) as a `hot spot' nucleus. The northern part of the ring is obscured by dust. It consists of several spiral fragments made of \hii{} regions.

{\bf NGC~3011}: Gil de Paz et al.~(2003). The galaxy is a dwarf spiral with a 5\arcsec\ ($\sim600\,{\rm pc}$) nuclear ring and it also contains an outer ring of 19\arcsec\ ($\sim2\,{\rm kpc}$) in radius (Gil de Paz et al.~2003). These rings are the first rings discovered in a dwarf galaxy. As the galaxy has no obvious bar, Gil de Paz et al.~(2003) suggests that the nuclear ring may have been caused by the sweeping out of the interstellar medium by a nuclear episode of star formation. This gas may reach densities high enough to become molecular and to form \hii{} regions. The relative size of the nuclear ring is not incompatible with the `classical' rings in the sample, although it is found on the upper boundary.

{\bf NGC~3081}: Heckman (1978). There is an inner bar inside the ring (Erwin 2004) and no bright source at the centre of the ring. Several compact knots of emission could be candidates for the nucleus, but they are off-centred. A lens with the radius of the nuclear ring was needed for the 2D decomposition. Buta et al.~(2007) show a colour-index image of this galaxy with the outer, inner, and nuclear rings perfectly seen; in this image, the three rings are particularly sharp and are elongated with different PAs and ellipticities which is evidence for the rings' not being circular.

{\bf NGC~3185}: Comer\'on et al.~(2008c).

{\bf NGC~3245}: Barth et al.~(2001). They suggest that the ring may be the outer edge of a nuclear disc. The western part of the ring is probably covered by dust. The galaxy has no bar and the ring may be caused by the gravitational perturbations of NGC~3245A.

{\bf NGC~3258}: first reference as a ring in Comer\'on et al.~(2008c) but has been called disc by Capetti \& Balmaverde (2005).

{\bf NGC~3268}: first reference as a ring in Comer\'on et al.~(2008c) but has been referred to as a small disc of dust and ionized gas by Ferrari et al.~(1999) and as dust filaments by Capetti \& Balmaverde (2005).

{\bf NGC~3310}: Telesco \& Gatley (1984). Called an inner arm system by Walker \& Chincarini (1967). The nucleus is off-centred. This galaxy is the one least well fitted in the 2D decomposition, due to the intense star-formation at a radius of 15\arcsec. We fitted this activity with a lens but this did not yield a particularly good approximation to the morphology of the central region of the galaxy.

{\bf NGC~3313}: Buta \& Crocker (1993).

{\bf NGC~3351}: S\'ersic \& Pastoriza (1965) as a `hot spot' nucleus. A lens was needed to fit the nuclear ring in the 2D decomposition.

{\bf NGC~3379}: van Dokkum \& Franx (1995). This ring is clearly visible as a dust feature in structure maps and in colour-index images but also emits in \ha{}. Inside the ring, there is some \ha{} that could correspond to small arms of star formation spiralling from the ring to the AGN. Statler (2001) states that the ring is a decoupled nuclear component whose orientation is not linked to the main body of the galaxy.

{\bf NGC~3414}: Buta et al.~(2007). This galaxy has a nearly round component with an exponential profile and a thin bar-like component. Many explanations have been given for this peculiar morphology. The bar-like component could be either a real bar (Sil'chenko \& Afanasiev 2004) or a polar ring (Whitmore et al.~1990). The possible bar, although it is extremely thin in the 2D models of the galaxy, is very weak ($Q_{\rm g}=0.097$) and has no dust lanes.

{\bf NGC~3486}: Buta \& Crocker (1993). Appears as an inner ring in de Vaucouleurs \& Buta (1980). The ring appears to be a collection of spiral arms with high pitch angle. This galaxy is known to have only a nuclear bar (Martin 1995), but we have also found a very weak bar by looking the ellipticity profile.

{\bf NGC~3504}: Buta \& Crocker (1993). The northern part of the ring is probably hidden by dust. It is surprising that the size we measured for this ring is a half of the values that appear in literature (Buta \& Crocker 1993, Elmegreen et al.~1997).

{\bf NGC~3593}: Pogge (1989). The inner parts of the galaxy are too hidden by the disc to allow a proper 2D decomposition and thus to make a bulge model to measure $Q_{\rm g}$ with precision. The galaxy type could be either S0/a pec or I0 pec according to Buta et al.~(2007).

{\bf NGC~3945}: Erwin \& Sparke (2002). This ring is purely stellar with no traces of recent star formation. It encircles an inner bar. It appears as a brighter area in a structure map. The disc scalelength we have measured is the one from the outer disc, which is heavily affected by a strong outer ring. It has been sugested to be either a faded star-forming ring or a side-effect of the double-barred nature of NGC~3945 (Erwin \& Sparke 2002).

{\bf NGC~3982}: de Vaucouleurs \& Buta (1980) as an inner ring. The ring needed a lens in the 2D decomposition to be fitted. The nuclear ring consists of flocculent \ha{} areas and is not very well defined.

{\bf NGC~4100}: Buta \& Crocker (1993). The galaxy is too inclined to measure $Q_{\rm g}$. Eskridge et al.~(2002) state that this galaxy is barred.

{\bf NGC~4102}: Comer\'on et al.~(2008c).

{\bf NGC~4138}: de Vaucouleurs \& Buta (1980) as an inner ring. This large nuclear ring is in an unbarred galaxy, so it may be an inner ring. We do not have \ha{} or \pa{} $HST$ imaging for this galaxy and the UV exposures are too short to detect the ring. In the structure map, subtle dust structures tracing the ring can be seen. The $Q_{\rm g}$ profile is flat and close to zero which indicates a very symmetric structure. This galaxy contains a counter-rotating component (Jore et al.~1996).

{\bf NGC~4245}: Erwin \& Sparke (2002).

{\bf NGC~4262}: first reference to this dusty star-forming nuclear ring, which appears as featureless and wide in structure maps.

{\bf NGC~4274}: Falc\'on-Barroso et al.~(2006). The nucleus of the galaxy is off-centred.

{\bf NGC~4303}: M\"ollenhoff \& Heidt (2001), referred to as a nuclear spiral by Colina et al.~(1997) and Colina \& Wada (2000). The ring is wide and made up of small spiral bits. This galaxy has a nuclear bar with 175\,pc in radius (Colina \& Wada 2000, adjusted for distance).

{\bf NGC~4314}: Benedict et al.~(1977).

{\bf NGC~4321}: S\'ersic \& Pastoriza (1965) as a `hot spot' nucleus. This nuclear ring could be considered a pseudo-ring, but we find in the structure maps that the dust is distributed in arms that cross the areas where parts of the ring are missing. We used a round lens to fit the ring area in the 2D decomposition.

{\bf NGC~4340}: Erwin (2004). The nuclear ring appears sharp and thin, and it contains an inner bar (Erwin 2004).

{\bf NGC~4371}: Erwin \& Sparke (1999) for the outer nuclear ring. They consider the inner nuclear ring as a possible dust nuclear disc. The outer nuclear ring appears much bluer than the inner one in the colour-index maps. Thus, the outer nuclear ring is probably a resonance star-forming ring related with the bar and the inner one may have its origin in the depletion of inner parts of a dusty disc.

{\bf NGC~4448}: Comer\'on et al.~(2008c). The ring is made of several aligned knots of star formation which leave an empty space to the north-east.

{\bf NGC~4459}: de Vaucouleurs (1957) for the outer nuclear ring. AINUR contains the first known reference for the inner nuclear ring. The inner nuclear ring, unlike that in NGC~4371, looks thin and sharp; thus it is less likely to be an inner disc whose inner parts have been depleted. In the UV $HST$ image the innermost ring is clearly seen as emitting. The outer nuclear ring cannot be seen. The UV image shows that inside the innermost ring, there is a lens of diffuse star formation. NGC~4459 has no bar (the $Q_{\rm g}$ profile is completely flat) and the rings could be due to a weak interaction with NGC~4468 and/or NGC~4474.

{\bf NGC~4494}: Lauer et al.~(2005). Similar to NGC~3258.

{\bf NGC~4526}: Sandage (1961). Dust ring in a nearly edge-on galaxy where we have not been able to measure $Q_{\rm g}$.

{\bf NGC~4571}: de Vaucouleurs \& Buta (1980) as an inner pseudo-ring. This galaxy is not barred but it has a lens with the size of the nuclear  ring (Eskridge et al.~2002).

{\bf NGC~4579}: Terashima \& Wilson (2003). The nuclear ring is off-centred which can be explained in terms of a resonance ring in a mass distribution that is induced by a superposition of $m=1$ and $m=2$ perturbations (Comer\'on et al.~2008a). The inner arcs could be related to shocks caused by the interaction of a compact radio jet with the cloudy ambient gas (Contini 2004).

{\bf NGC~4593}: Gonz\'alez Delgado et al.~(1997).

{\bf NGC~4736}: Pease (1917) as the `periphery of an ellipse'. This SA galaxy has an oval disc too subtle to be detected in the 2MASS images (Erwin 2004; Trujillo et al.~2009). If we consider this oval as causing the nuclear ring, it has often been misclassified as an inner ring. This galaxy has also a nuclear bar, 20\arcsec{} in length (Erwin 2004).

{\bf NGC~4800}: de Vaucouleurs \& Buta (1980) for the outer nuclear ring, Comer\'on et al.~(2008a) for the two other nuclear rings. The innermost nuclear ring may be doubtful but structure maps make it appear clearly (Comer\'on et al.~2008a). The galaxy appears unbarred, but ellipse fitting in the $H$ band reveals a weak bar.

{\bf NGC~4826}: Pogge (1989). This galaxy is unbarred and has counter-rotating components (Braun et al.~1992).

{\bf NGC~5020}: first reference to this nuclear ring. The southern part of the nuclear ring is hidden by a dust lane.

{\bf NGC~5033}: Considered as a `spiral or ring-like pattern of \hii{} regions' in Mediavilla et al.~(2005). The galaxy is unbarred, but it has a strong arm pattern that could cause a nuclear ring.

{\bf NGC~5135}: Garc\'ia-Barreto et al.~(1996). What we see in the image is actually half of a nuclear ring. The other half is probably hidden in the dust patches that are seen in visible $HST$ images.

{\bf NGC~5194}: Worden (1974). The orientation parameters of this galaxy are very difficult to derive because of its interaction with NGC~5195. The galaxy has a weak bar but the peak in $Q_{\rm g}$ is located in the prominent arms caused or enhanced by the interaction. The ring is wide and not very well defined and it is made of small patches of star formation distributed in spiral arms.

{\bf NGC~5236}: Buta \& Crocker (1993). The nuclear region has such a huge star formation rate that the 2D decomposition yields uncertain bulge parameters. We used an inner lens to model this star formation region; this lens has a radius similar to the nuclear ring.

{\bf NGC~5247}: first reference to this nuclear ring. The UV $HST$ image we have is shallow, but it seems that the ring is not complete and some parts are missing in the south-west. This galaxy is non-barred but the prominent non-axisymmetric arms yield one of the highest $Q_{\rm g}$ values in the sample. The strong arms probably affect the determination of the orientation parameters.

{\bf NGC~5248}: S\'ersic \& Pastoriza (1965) as a `hot spot' nucleus for the outer nuclear ring and Laine et al.~(1999) for the inner nuclear ring. Both rings are patchy. The inner one appears as a double ring of \hii{} regions. The intense circumnuclear star formation makes a reliable determination of the bulge parameters in the 2D decomposition difficult.

{\bf NGC~5377}: Carollo et al.~(1998). This star-forming ring is dusty and wide, made up of tightly wound dust spirals.

{\bf NGC~5427}: Buta \& Crocker (1993). This circumnuclear feature is an ambiguous case between nuclear rings and nuclear pseudo-rings. Parts of it are missing in both north and south. This galaxy is unbarred, but its possible interaction with NGC~5426 may be the cause of the ring. Fuentes-Carrera et al.~2004, however, detected a bar in this galaxy. This bar is not detected by us because the peak in ellipticity is found in a radius of fast-changing PA, which is contradiction with the bar detection criteria we have used.

{\bf NGC~5728}: Buta \& Crocker (1993). The ring is highly off-centred.

{\bf NGC~5806}: first reference to this nuclear ring.

{\bf NGC~5812}: called dusty disc by Tran et al.~(2001) and nuclear ring by Comer\'on et al.~(2008c). This is a dust ring in an elliptical galaxy, probably the same kind of structure as in NGC~3258 and NGC~4494, but here the ring is much smaller. The ring is so perfectly round that it can lead to the false conclusion that it results from the saturation of the camera. We tested this and confirmed that even the brightest pixel of the nucleus has a count value well below the saturation level.

{\bf NGC~5905}: Friedli et al.~(1996). They say that the ring is just outside a secondary bar. The ring is brighter in its north-east sector.

{\bf NGC~5945}: Buta \& Crocker (1993). The ring is on the edge of an \ha-emitting lens.

{\bf NGC~6503}: Knapen et al.~(2006). This galaxy has been classified as unbarred, but its high inclination makes it difficult to detect a bar. The ring is 1.2\,kpc in radius, which is not exceptional, but the galaxy is quite small, making this one of biggest nuclear rings relative to $D_{\rm o}$.

{\bf NGC~6753}: de Vaucoulers \& Buta (1980) as an inner ring. This galaxy has no bar and a completely flat $Q_{\rm g}$ profile. The large size of the ring and the lack of a bar could give the impression it is an inner ring, but Laurikainen et al.~(2004) have found an oval with a size bigger than the ring.

{\bf NGC~6782}: Buta \& Crocker (1993).

{\bf NGC~6861}: First reference to this dust nuclear ring located in a very early-type unbarred S0 galaxy. This galaxy has a companion, IC~4943, which may cause ILRs. The other possibility in such an early-type galaxy is to have a ring formation mechanism similar to that found in elliptical galaxies.

{\bf NGC~6951}: Burbidge \& Burbidge (1962) as a `spiral ring'.

{\bf NGC~6958}: Comer\'on et al.~(2008c). A dust ring in an elliptical galaxy.

{\bf NGC~7049}: V\'eron-Cetty \& V\'eron (1988). A dust ring in a lenticular unbarred galaxy. The ring may be caused by $x_{2}$ orbits related to a possible oval distortion (Corwin et al.~1985). It also may have an origin similar to that of dust rings in elliptical galaxies (see e.~g. NGC~3258).

{\bf NGC~7217}: Pease (1917). This ring is in an unbarred galaxy with a completely flat non-axissymmetry profile. The ring could have been caused by the oval due to a minor merger with a dwarf gas-rich galaxy in a retrograde orbit that could also have caused the counter-rotating component of the galaxy (Sil'chenko \& Moiseev 2006).

{\bf NGC~7469}: Wilson et al.~(1991). As NGC~4321 and NGC~5427, this is an intermediate case between rings and pseudo-rings.

{\bf NGC~7552}: Forbes et al.~(1994) in radio. We see the ring as a dust feature in optical images, but radio observations show that it is actually a star-forming ring.

{\bf NGC~7570}: Knapen et al.~(2006). Although NED and HYPERLEDA state that it is a highly inclined SBa galaxy, an inspection of SDSS images shows that this galaxy has a nearly face-on outer ring.

{\bf NGC~7716}: de Vaucouleurs \& Buta (1980) as an inner ring. It probably is a nuclear ring because we found a bar that is bigger than the ring radius. The $Q_{\rm g}$ profile is rather flat and the peak caused by the bar is barely noticeable.

{\bf NGC~7742}: Morgan (1958) as a `brilliant ring'. This nuclear ring is double and is located in an unbarred galaxy with a completely flat $Q_{\rm g}$ profile. As in NGC~7217, the ring could have been caused by the oval due to a minor merger with a dwarf gas-rich galaxy in a retrograde orbit that could also have caused the counter-rotating component of the galaxy (Mazzuca et al.~2006).

{\bf NGC~7771}: Smith et al.~(1999). The eastern hole in the ring as seen in the \pa\ emission image is due to a dust lane. This galaxy is probably interacting with NGC~7770.

{\bf UGC~10445}: First reference to this nuclear ring. This Sd galaxy is rather exceptional because very few late-type galaxies host nuclear rings. It is a barred galaxy in the NED and HYPERLEDA catalogues but we have not been able to make ellipticity profiles or $Q_{\rm g}$ measurements because the galaxy is so faint that it is barely visible in 2MASS images.

\section{Discussion}

\subsection{Nuclear ring lifetimes}

In Sect.~8 we reported our result that $20\pm2\%$ of disc galaxies in the range of morphological types $-3<T\leq7$ host a star-forming nuclear ring.
This confirms, using a much larger and more robustly selected sample, the result of Knapen (2005) who found $21\%\pm5\%$ from a
sample of 57 galaxies, with 12 nuclear rings. Two effects may have influenced the result by Knapen (2005): namely ground-based images were used so small nuclear rings were not included, and the definition of a nuclear ring was slightly different, so the former study included some rings (those in NGC~3184 and NGC~3344) that we consider to be inner rather than nuclear rings in the present study. These two effects
go in opposite directions, which leads to the almost identical ring fractions in the two studies.

Two models for star-forming nuclear rings can be envisaged, each having different implications for
the nuclear ring life-time. In the first, star-forming nuclear rings are short-lived starburst-like events, with a 
lifetime of a few Myr, while in the second, the rings would be long-lived, but undergoing several distinct 
episodes of enhanced massive star formation. The evidence is overwhelmingly in favour of the second of these two options, on the basis of observations (e.g., Garc\'ia-Barreto et al.~1991b; Knapen et
al.~1995), simulations (Knapen et al.~1995; Heller \& Shlosman 1996; Salo et al.~1999;
Regan \& Teuben 2003), and stellar population synthesis (Allard et al.~2006; Sarzi et
al.~2007). In the case of the nuclear ring in NGC~4321, numerical modelling shows that the dynamical structure upholding the ring is stable (Knapen et al. 1995), the amount of molecular gas estimated to be present in the central region could be sustaining episodes
of star formation for $\sim6\times10^{8}$ years, even without inflow (e.g., Knapen et al. 1995), and stellar population modelling shows that short bursts of star formation have been going on periodically in this nuclear
ring for about 500\,Myr (Allard et al.~2006). The total lifetime for the nuclear ring in NGC~4321 is estimated to be at least
a Gyr and quite plausibly more if we consider that there is a bar-driven inflow of gas into
the circumnuclear region.

Making the safe assumption that nuclear rings are long-lived phenomena, we can use the derived fraction of occurrence of nuclear rings to place back-of-the-envelope limits on how long-lived they may be. To do this, we first assume that at a given moment in their existence all disc galaxies are able to create a
stable star-forming nuclear ring, and that they have been able to do so for a Hubble time. Thus, the observed present fraction of around 20\% of disc galaxies with nuclear rings 
would yield a
lifetime of order 2$-$3\,Gyr. If we then consider that only galaxies from $z=1$ have been able to create
star-forming nuclear rings like those studied here, because the physical conditions may well have been different at earlier times, we arrive at an estimate for a typical nuclear ring lifetime of around
1.5\,Gyr. This value increases if not all disc galaxies are able to
create a nuclear ring, but the typical lifetimes deduced are compatible with observational and numerical results in the literature.

\subsection{Are we missing small nuclear rings?}

One of the main results of this paper is the relative size distribution of
nuclear rings, which, among other things, seems to
show that there is no separate population of very small rings. The
existence of such a population of UCNRs was speculated on by Comer\'on et
al. (2008a), but our current systematic study using the highest spatial
resolution imaging available for a large sample of galaxies shows that the
size distribution of nuclear rings shows one peak, at ${\rm log}(D_{\rm r}/D_{\rm o})=-1.3$, accompanied by
a gradual rise at smaller sizes and a gradual fall at larger ones. The
latter is not affected by observational biases, because even if large
rings fell outside the field of view of our {\it HST} images such
large rings would have been noticed in ground-based imaging and reported
in the literature. Definition problems confusing inner with nuclear rings
can in principle occur for large nuclear rings, but only in very few
cases, as discussed in previous sections.

There are, however, two observational bias issues which could lead us to
miss small nuclear rings. First, the presence of significant quantities of
dust in the central regions of galaxies
could hide nuclear rings of small absolute size, and, second, insufficient
spatial resolution can
make us miss angularly small nuclear rings. The first effect, that of dust
extinction, is difficult to quantify and
model. But it seems extremely unlikely that a significant population of
galaxies exists in which large and uniformly distributed quantities of
dust would completely hide a star-forming nuclear ring from view, even in
the red H$\alpha$ band (if the dust were distributed preferentially in a
ring shape we would see it as such in our structure maps, and the galaxy
might well be classified as having a dust ring). Even if this were the
case, the quantities of dust needed (we estimate typical $A_V$ of some
5\,mag to hide a typical star-forming nuclear ring), over the scales, of
a few tens of parsec, needed to cover up a hypothetical population of
ultra-small rings would certainly show up in broad-band, colour index,
and structure images of galaxies. There is no evidence for this effect to
be present, and we conclude that dust is extremely unlikely to hide a
population of very small nuclear rings.

We can, on the other hand, make estimates on the second effect identified,
namely how many nuclear rings may be  missing
in our Atlas due to insufficient spatial resolution. The
maximum of  ${\rm log}(D_{\rm r}/D_{\rm o})=-1.3$
(bottom panel in Fig.~\ref{hist}) in the relative ring size distribution is unaffected by resolution issues
because even in the most distant galaxies in our sample this size
corresponds to an angular resolution of 5\,arcsec---such nuclear rings
can easily be distinguished in our {\it HST} data. In fact,
Laine et al.~(2002) found the same maximum using ground-based data.
However, the small-size tail in Laine et al.~(2002) is shorter than the
one found here
(Fig.~\ref{hist}), which implies that they were indeed missing rings due
to a lack of resolution.

\begin{figure}
\begin{center}
\begin{tabular}{c}
\includegraphics[width=0.45\textwidth]{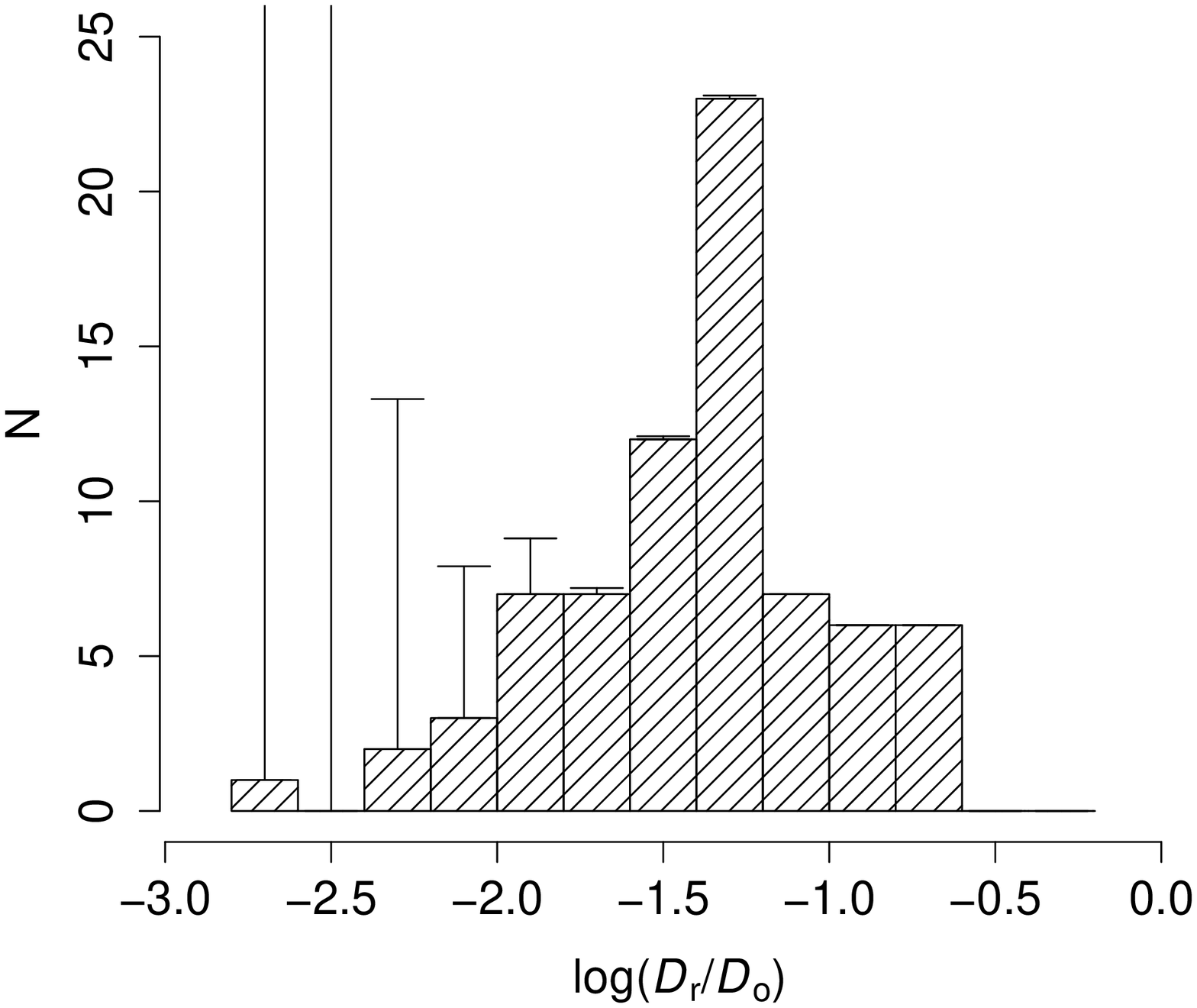}\\
\includegraphics[width=0.45\textwidth]{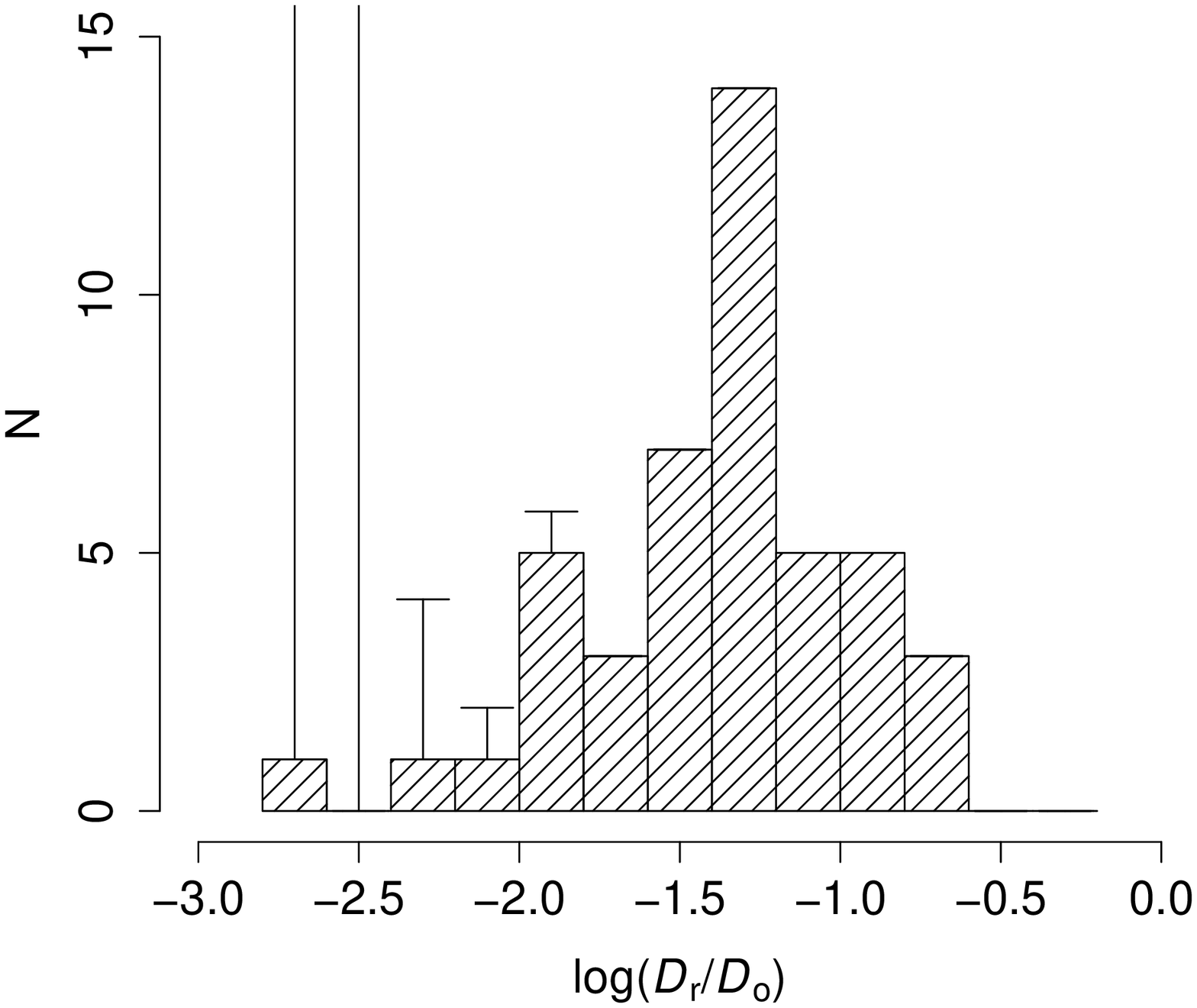}\\
\end{tabular}
\caption{\label{ansatz} Size distribution of star-forming nuclear rings
found in the sample of 347 disk galaxies with types $-3<T\leq7$ for which we have  {\it HST}
imaging (top panel), and of the subset of those which are at
a distance of less than 20\,Mpc (bottom panel). The error bars show an
estimate of the
nuclear rings that may be missing from our Atlas due to a lack of spatial
resolution in the imaging (see text for details).}
\end{center}
\end{figure}

For galaxies imaged by {\it HST} we estimate that the threshold above
which we can find a
nuclear ring is
2\,arcsec in diameter (see Fig.~\ref{complet} and Sect.~3.3), although
this may be slightly smaller (about 1.5\,arcsec) for galaxies imaged with
an {\it HST} camera with small pixels and with a high signal-to-noise
ratio. To check what the likelihood is that we have missed a
significant population of very small nuclear rings, we started by
considering each of the
347 disc galaxies with types $-3<T\leq7$ (the range of types in which star-forming nuclear rings are found) of the sample that we studied with {\it HST}
images. For
each galaxy we determine the relative
size, $D_{\rm r}/D_{\rm o}$, of a nuclear ring that corresponds to a
diameter of
2\,arcsec, which is the
detection threshold for $HST$ images. We then count in how many galaxies
a star-forming nuclear ring of a given relative size $D_{\rm r}/D_{\rm o}$
would be detectable, and in how many
such a ring would be smaller than the detection limit. We then estimate
the uncertainty in the number of star-forming nuclear rings counted in a
given
$D_{\rm r}/D_{\rm o}$ bin by assuming a constant nuclear ring occurrence
frequency, in other words, that the proportion of star-forming nuclear
rings
in galaxies where the ring is, in principle, detectable is the same as
that in
galaxies where they are not detectable. The result of this test has been
represented in the top panel of
Fig.~\ref{ansatz}, where the error bars indicate our calculated estimates
of the numbers of undetected
rings. This figure shows that the observed peak in relative nuclear ring
size is true and that a possible second peak indicating a separate
population of smaller nuclear rings can in principle be located at
relative sizes ${\rm log}(D_{\rm
ring}/D_{\rm o})<-1.9$, corresponding to a radius of just under 200\,pc for a
galaxy of average size at
30\,Mpc.

If we repeat the experiment described above but reducing our sample to the
nearby galaxies,
those at $D<20\,{\rm Mpc}$, we find similar results (bottom panel of
Fig.~\ref{ansatz}). In this test, we have a sample of 45 galaxies, and
considering a sample of even closer galaxies is not feasible (selecting
only those at $D<15\,{\rm Mpc}$, for instance, would reduce the sample
size to 21 which is insufficient to draw statistically valid conclusions).
The maximum relative size of a possible second peak in the distribution,
which would correspond to a population of nuclear rings small enough to be
unresolved in our {\it HST} imaging, is now located at ${\rm log}(D_{\rm
ring}/D_{\rm o})<-2.1$. This corresponds to a maximum radius of 80\,pc for a
galaxy of average size at 20\,Mpc. The conclusion from this series of
tests are that the peak in the size distribution is real, although we cannot formally rule out the
possibility of a second population of smaller nuclear rings, with a peak
in radius smaller than 80\,pc. The implications of this will be further
discussed in the next section.

\subsection{Are star-forming UCNRs normal nuclear rings?}

The occurrence and properties of the smallest nuclear rings, with radii
smaller than a few hundred parsec, have so far not been explored
systematically. Such small nuclear rings were termed UCNRs by Comer\'on et al.~(2008a), and defined as those with a
radius smaller
than 200\,pc. Although this 200\,pc size limit is arbitrary, it is useful
in allowing us to define small rings as a class, and subsequently study
their properties. A handful of UCNRs were known from the literature, and
the study by Comer\'on et al.~(2008a) increased the known sample,
to 16 UCNRs. In the present Atlas, the total number of nuclear rings
smaller than 200\,pc in radius has been increased to 28, which allows us
to make more definite statements on their properties. In particular, it
allows it to ascertain that UCNRs do not, in fact, form a separate class
of nuclear rings, but instead merely constitute the small-size part of the
overall nuclear ring size distribution.

The justification for defining UCNRs as such (by Comer\'on et al.~2008a)
was based in part on the theoretical possibility of the existence of a
bimodal distribution of nuclear ring sizes. This would be related to the
smaller class of rings (UCNRs, presumably) forming near nuclear Lindblad
resonances (as described by e.g.~Fukuda
et al.~1998; Shlosman 1999; Maciejewski 2003). The size distribution of the nuclear rings
in our Atlas, of which 28 (out of 113 rings in total) can be classed
UCNR, however, is not bimodal (Fig.~\ref{hist}, and discussion on possible biases
in Sect.~12.2) when the peak of a nuclear ring distribution related to nuclear Lindblad resonances is expected to peak around ${\rm log}(D_{\rm r}/D_{\rm o})=-1.7$ (from ring expected size in Fukuda et al.~1998).

Considering their much smaller sample of three new and 13 literature UCNRs, Comer\'on et al.~(2008a) suggested that UCNRs may be related
to ILRs. The existence of an ILR at the
small radius implied by the UCNRs would imply a shallow radial mass
distribution in
the central kpc of the host galaxy, which in turn would imply that UCNR
host galaxies should have a small bulge. As a consequence, a deficiency
of UCNRs in early-type galaxies would be observed. In contrast, we find
here that star-forming UCNRs are, without preference, found in
host galaxies
ranging from morphological types S0 to Sd.  which makes this possibility
unlikely.

UCNRs may also be related to nuclear bars (Comer\'on et al.~2008a), implying they
would
be found near the corotation radius of the nuclear bar. To dismiss this
possibility for each individual UCNR we would need near-IR
images at significantly higher resolution than the 2MASS images to perform
detailed ellipticity fits of the inner parts of the galaxies.
A literature search, however, indicates that of the 27 UCNR
host galaxies in the Atlas, only NGC~4371 and NGC~5236 have been reported
as having nuclear bars (by Wozniak et
al.~1995; Elmegreen et al.~1998, respectively). For NGC~4371, Erwin \& Sparke 1999 argued that the nuclear bar from Wozniak et
al.~(1995) was actually the outer nuclear ring reported in the present paper. We conclude that a
direct and ubiquitous link between UCNRs and nuclear bars is unlikely.

The conclusion from the current work is that UCNRs merely constitute the
low-radius tail of the nuclear ring size distribution and that there is
no evidence that they are a manifestation of a different physical
phenomenon. The AINUR images show
that the morphology of the UCNRs is very similar to that of larger nuclear
rings. In addition, the UCNRs do not populate preferential regions in any
of the diagrams presented in this paper, which implies that the properties
of these rings and of their host bars and galaxies are not systematically
different from those of larger nuclear rings. We thus conclude that UCNRs
are not a separate class of nuclear ring phenomenon.

\subsection{What determines nuclear ring size?}

Our analysis of nuclear ring sizes confirms that stronger bars host
smaller nuclear rings, and that weaker bars can host larger nuclear rings,
but also smaller ones. The maximum allowed nuclear ring radius is thus
inversely proportional to the torque parameter $Q_{\rm g}$
(Fig.~\ref{rdod25}), while for a given $Q_{\rm g}$, the allowed range in
nuclear ring radius spans from this
maximum down to the smallest measured radius of a few tens of parsecs.
This is seen particularly clearly when plotting relative nuclear ring size
($D_{\rm r}/D_{\rm o}$; Fig.~\ref{rdod25}a), but is also recognisable in
absolute nuclear ring size (Fig.~\ref{rdod25}c). These results confirm those
published by Knapen et al. (2002) and Knapen (2005), but with more nuclear
rings due to the comprehensive nature of our new Atlas, and extending to
much smaller nuclear ring sizes due to the use of archival {\it HST}
imaging. Knapen (2005) noted that the findings are in agreement with
predictions from numerical and theoretical work on the evolution of
nuclear rings in a framework of resonances and orbit families in bars
(Knapen et al. 1995; Heller \& Shlosman 1996), something that we will
expand on below.

Two scenarios can help explain why the complete parameter space below the
limiting maximum ring radius at
a given $Q_{\rm g}$ is populated by nuclear rings. The first of these is
that nuclear rings do not necessarily form very close to the maximum
radius, most probably related to the outer ILR as explained above. In
terms of resonances and using a linear approximation, thus considering
that the gravitational potential of
the bar is axi-symmetric, it is quite intuitive that the nuclear ring will
form close to the outer ILR. But although the linear approximation may be
useful for weak bars, it is not correct in the presence of a strong bar
(Sellwood \& Wilkinson 1993). Avoiding the linear approximation, Shlosman (2001)
finds that no nuclear ring is found close to an outer ILR; rather, rings
are formed  halfway between the two ILRs when both of these resonances are
present.
Some simulations show that in the case of a weak bar, the nuclear ring is
created close to
the outer ILR (Salo et al.~1999).

When considering a formation scenario related to the $x_{2}$ orbits in a
barred potential, it is unclear where a nuclear ring will form. Although is it
known that $x_{2}$ orbits are all resonant (Regan \& Teuben 2003), and that they are located
between a pair of ILRs (Contopoulos \& Papayannopoulos 1980), there is no quantitative study of which of
the orbits are populated with gas or stars (or even particles in
theoretical studies). It is thus impossible to know whether the stars will
at a given time be at the right position and velocity to be trapped by an
$x_{2}$ orbit, and as a consequence, where exactly an $x_{2}$-orbit-induced
nuclear ring may form. This could thus, presumably, happen anywhere
between the ILRs, and it is unclear what physical parameters influence the
exact radius of the nuclear ring.

In addition, the maximum possible
radius is not a strictly defined quantity and there may well be some
scatter due mainly to the size and
mass of the bulge---changes in the central mass concentration can move the
ILR
position (see e.g.~Hasan \& Norman 1990; Thakur et
al. 2009).
The central mass concentration is not completely decoupled from the
$Q_{\rm g}$
measurement, though, and a dense large bulge results in both the ILRs
moving outwards and
$Q_{\rm g}$ decreasing (with differences up to 50\% in bar strength in the
case of
extremely elliptical bars).

The second scenario to explain the observations is an evolution of the
nuclear ring size.  It has been well established from numerical modelling
that nuclear rings
can shrink, mainly due to the
loss of angular momentum in shocks (Knapen et al.~1995; Heller \& Shlosman
1996; Regan \& Teuben 2003), possibly also by an effect called
gas shepherding (Chang 2008; van de Ven \& Chang 2009). In this scenario,
nuclear rings would form near the maximum radius allowed by their host
galaxy's $Q_{\rm g}$ and then evolve down to lower size, thus populating
the $Q_{\rm g}$-size diagram. It is not entirely clear, however, whether
the speed of this evolution can lead to the uniform spread in size shown
in Fig.~\ref{rdod25}, and by how much rings can and do shrink. For
instance, the nuclear rings in the paper of Regan \& Teuben (2003) evolve
so fast (compared to our nuclear ring timescales as discussed in
Sect.~12.1) that all but the very youngest would be observed with very
small sizes, and thus at the very lower part of our diagram, something
which is clearly not the case. A slow evolution would cause rings to
linger in the higher regions of the diagram, which is also not the
observed case. On the other hand, and causing further uncertainty, Heller
\& Shlosman (1996) have described the dynamical effects of a gaseous
nuclear ring on the stellar orbits in a barred potential, finding that the
rings, especially if they are massive, can weaken the bars which host
them. It is thus not clear whether time evolution of the nuclear rings
leads to systematic changes in their position in the $D_{\rm r}/D_{\rm o}$
vs $Q_{\rm g}$ diagram.

Whatever the exact initial position is of a nuclear ring
relative to the ILRs or the extent of the $x_{2}$ orbits in a barred galaxy,
it is clear that the maximum nuclear ring size is limited by either the
outer ILR, or by the extent of the $x_{2}$ orbits. Both of these will be
related to the bar strength, with stronger bars leaving less physical
space for the nuclear ring to form in, and thus hosting smaller rings.
This effect has been shown in a number of simulations which aim to
reproduce the formation of nuclear rings in different host potentials
(e.g.~Knapen et al. 1995; Heller \& Shlosman 1996; Salo et al.~1999; Rautiainen \& Salo 2000;
Regan \& Teuben 2003). This would lead to a  maximum radius for nuclear
rings, with stronger bars allowing only smaller rings, as seen in the edge
of the distribution of nuclear rings in Fig.\ref{rdod25}.

In conclusion, the existence of a maximum allowed nuclear ring size can be
explained by the fact that the radius of the largest $x_{2}$-orbits and the
outer ILR are larger in weaker bars. The fact that all the space below this maximum
is populated with nuclear
rings suggests that nuclear rings can form at a variety of radii, limited
primarily by the maximum identified above.
An additional effect is that the nuclear rings, once formed, evolve with
time to become smaller, although evolution of the ring may also stimulate
an evolution of the host bar properties. A combination of these effects
constitutes the most likely explanation for the observed distribution in
Fig.~\ref{rdod25}.

\subsection{Star-forming nuclear rings in unbarred galaxies: are they different?}

We find that 18 not very inclined galaxies with a star-forming nuclear ring do not host a
bar.
This constitutes a fraction of $19\pm4\%$ of the disc galaxies in this sample ($-3\leq T$), a
fraction that is not very
different from that of unbarred disc galaxies in the local Universe
($30\%-40\%$; Sandage 1961; de Vaucouleurs et al. 1991; Sellwood \&
Wilkinson 1993; Knapen et al. 2000; Marinova \& Jogee 2007). However, all
but three of these unbarred galaxies have evidence for some
non-axisymmetry, which should be able to generate ILRs
(strong spiral arms as suggested by simulations from Rautiainen \& Salo in 2000, oval distortions, counter-rotating components, or
interactions with
close companions; see Sect.~10 for details). One of the three remaining
galaxies is a blue compact dwarf (NGC~3011), for which Gil de Paz
et al. (2003) speculate an origin in terms of a starburst-driven shock
interacting with the surrounding interstellar medium.

Nuclear rings in unbarred galaxies usually look indistinguishable from
their counterparts in barred hosts, except those found in
galaxies with counter-rotating components which tend to be more regular
than average. For most of the
unbarred host galaxies we could determine (low) $Q_{\rm g}$ values, but a
few with strong arm patterns have $Q_{\rm
g}$ values that are comparable to those determined for barred galaxies
(e.g., $Q_{\rm g}=0.416$ for
NGC~5247, which is the most extreme case). Our sample of unbarred galaxies
thus covers a wide range of $Q_{\rm g}$. None of the nuclear rings in
unbarred disc galaxies is an outlier
in the diagram relating relative ring size to  Fig.~\ref{rdod25}, which
strongly indicates that their formation mechanism is the same as, or strongly
related, to that in barred galaxies, quite possibly involving
non-axisymmetry-induced resonances.

In some cases, nuclear rings found in a galaxies without a bar, or with a
weak
oval, might indicate that a previously existing bar has been destroyed.
Simulations have in fact shown that
bars can be weakened and/or destroyed by a central mass concentration
(e.g.~Friedli \& Benz
1993), by the transfer of angular momentum from the infalling gas to the
stellar bar
(e.g.~Bournaud et al.~2005), by the buckling instability (e.g., Raha et
al.~1991;
Mart\'inez-Valpuesta et al.~2004), or a combination of several factors
(Berentzen et
al.~2007). The mass of a growing nuclear ring itself can weaken the bar
which hosts it by increasing the extent of the
$x_{2}$ orbits and then depopulating the $x_{1}$ bar-supporting orbits
(Heller \&
Shlosman 1996). As nuclear rings can last for around a Gyr without gas
inflow (Knapen et
al.~1995), the nuclear ring could continue to exist even after the bar
destruction. We should note, however, that according to Shen \& Sellwood (2004), only very centrally concentrated masses could efficiently destroy bars, which suggests that nuclear rings would not be very efficient as bars destructors.

For the few galaxies where no non-axisymmetry has been detected at all,
there is the possibility
that they have not been studied in enough detail and that further scrutiny
may still reveal weak ovals,
counter-rotating components or small companions that are disturbing the
disc by close
flybys, so a resonant origin can hardly be excluded for any of the nuclear
rings in our sample.

It is worth noting that the non-axisymmetries in the unbarred galaxies
discussed here have been discovered by studying them in above-average
detail. If non-barred galaxies in general were subject to the same
level of examination, it is likely that many of those would also be found
to contain similar non-axisymmetries. Given the substantial number of
non-barred nuclear ring hosts found in our Atlas (18, or $19\pm4\%$) we are thus
led to the conclusion that unbarred nuclear ring hosts are not exceptional
cases, as they have often been considered in the past, but instead rather
normal. In fact, considering the similar fractions of unbarred galaxies
among the nuclear ring host galaxies and among the general disc galaxy
population, it is interesting to speculate whether the presence of a bar
is at all necessary for the presence of a nuclear ring, especially when
realising that nuclear ring hosts are subject to a more detailed search
for bars than their counterparts in the wider galaxy population (see
Sect.~4.3.1).

We postulate that nuclear rings are induced by non-axisymmetries in the
gravitational potential of the host galaxy, in most but not all cases
showing up in the form of a bar. The fact that only very few disc galaxies
will {\it not} have such a non-axisymmetry leads to the question of why
certain galaxies have nuclear rings, whereas most do not. It is possible
(as we discussed in Sect.~12.1 about nuclear ring lifetimes) that all
discs at some stage form a nuclear ring, but it is also possible that
specific combinations of physical parameters in the host potential and
interstellar medium are prerequisite to the formation of a nuclear ring.

\section{Summary}

This paper presents the most extensive atlas of nuclear rings so far published. It includes 113 rings distributed in 107 galaxies within a distance of 80\,Mpc. It includes all {\it bona fide} nuclear rings that appear in the literature, as well as seven new ones. A total of 17 of the 113 nuclear rings included have been discovered in the framework of the AINUR project (Comer\'on et al.~2008a; Comer\'on et al.~2008c; and the present paper).

We investigate how the general structure of non-axisymmetries in galaxies is related to rings, mainly by increasing the statistics and improving the methodology as compared to previous works. Our study is mostly based on the $HST$ archive from which we obtain colour-index maps, structure maps, \ha{}, \pa{}, and/or UV images. Additional images have been collected from the literature or sent by astronomers who have observed those nuclear rings which have not been imaged by $HST$. We have used a complete set of deprojected $H$-band 2MASS images to perform ellipse fits and to derive the non-axisymmetric torque parameter $Q_{\rm g}$, which were used to study the length, the ellipticity, and the strength of galactic bars and other non-axisymmetric features.

Our main results can be summarized as follows:

\begin{itemize}

\item{Star-forming nuclear rings are found in $\sim20\pm2\%$ of disc galaxies in the range of morphological types $-3<T\leq7$. Under the assumption that all spiral galaxies at some stage host a nuclear ring, this number would imply an effective nuclear ring lifetime of 2$-$3\,Gyr, in broad agreement with a large range of other observational and numerical studies.}

\item{Dust nuclear rings are found in $\sim6\pm2\%$ of elliptical galaxies.}

\item{Star-forming rings are predominantly found in galaxies with morphological types ranging from S0 to Sc and dust nuclear rings are found in ellipticals and early lenticulars. Only one nuclear ring in an Sd-type galaxy was found, and none in later types, because the shallow mass distribution of such galaxies is unfavourable to the existence of ILRs, and thus nuclear rings.}

\item{We establish that nuclear rings do not have a strong preference for barred host galaxies. Some $19\pm4\%$ of them (18 nuclear rings) occur in non-barred galaxies, though some evidence for non-axisymmetry in the potential can be discerned in most of these hosts. These non-axisymmetries can in principle cause resonances which can lead
to ring formation, but we speculate that many non-barred galaxies outside
our nuclear ring host sample will also host such non-axisymmetries, and
that some additional mechanism is apparently needed to stimulate the
formation of a nuclear ring.}

\item{We explore relations between $Q_{\rm g}$ and the geometric parameters of the rings, namely their size relative to the galaxy disc, intrinsic ellipticity, and PA offset between the ring and the bar that hosts it (if present). There is an upper limit to the relative size of a nuclear ring, and this upper limit decreases uniformly with increasing $Q_{\rm g}$. In barred galaxies, this trend can be described as stronger bars hosting smaller rings. Any relative size below the upper limit can and does occur in galaxies. This behaviour is exhibited by both barred and unbarred galaxies, and can be explained in a framework of nuclear rings forming between a set of dynamical resonances. Evolution with time, causing rings to shrink, may play an additional role.}

\item{Nuclear rings are not affected by {\it local} non-axisymmetric torques, which we measure by the amplitude of the torque parameter at the radius of the nuclear ring (the maximum amplitude of this parameter is $Q_{\rm g}$, and this maximum always occurs outside the nuclear ring). This confirms that the nuclear ring is affected by the non-axisymmetries on larger scales than the ring itself.}

\item{There is a weak correlation between the length and the strength of the bar among the barred host galaxies of the nuclear rings in our Atlas, where bar strength is measured through the non-axisymmetric gravitational torque parameter $Q_{\rm g}$. The bar length relative to the host galaxy disc scale length correlates somewhat better with $Q_{\rm g}$. These results confirm those by Elmegreen et al.~(2007) that bars have a specific range of lengths for a given strength when they form, but then evolve into longer bars.}

\item{Nuclear rings in barred galaxies are limited in size to one quarter of the length of the bar.}

\item{There is no strong evidence of any preferred PA offset between rings and bars.}

\item{There is some evidence that the maximum possible ellipticity for a nuclear ring is limited by the bar ellipticity. This conclusion, however, depends on the accuracy of the deprojection of images of strongly barred galaxies, and must thus be taken with caution.}

\item{We have found no correlation between the bulge shape and the properties of nuclear rings.}

\item{We cannot confirm the relation stated in the literature between the presence of nuclear activity and that of nuclear rings.}

\item{Star-forming UCNRs most probably constitute the low-size tail of the nuclear size ring distribution. We have no evidence for the existence of a separate population of such ultra-compact ($<200$\,pc radius) rings, although the spatial resolution of the {\it HST} images employed here does not allow us to exclude with certainty the existence of a population of much smaller rings, with radii smaller than 80\,pc.}

\item{Star-forming nuclear rings are closely related to ILRs and, with very few possible exceptions indeed, are related to dynamical resonances in their host galaxies. Given that they occur rather commonly in galaxies and that they constitute bright, thus well observable, tracers, this study thus confirms the value of nuclear rings, alongside inner and outer rings, in the observational study of the dynamics and evolution of galaxies.}

\end{itemize}

\section*{acknowledgements}

We thank our anonymous referee for comments that helped to improve the paper. We thank Emmanuel Davoust, Peter Erwin, Antonio Garc\'ia-Barreto, Armando Gil de Paz, Richard Pogge, Stuart Ryder, and Denise Smith for sending useful images for AINUR. We thank Kambiz Fathi for providing us a copy of the seminal paper by Bertil Lindblad on dynamical resonances in galaxies. Many thanks to Nick Scoville for suggesting the format of the images in the atlas. Support by the Ministerio de Educaci\'on y Ciencia (AYA 2004-08251-CO2-01 and AYA 2007-CO2-01), and the Instituto de Astrof\'isica de Canarias (P3/86 and 3I2407) is gratefully acknowledged. RJB acknowledges the support of NSF grant AST 050-7140.

Based on observations made with the NASA/ESA $HST$, obtained from the data archive at the STScI, which is operated by AURA under NASA contract NAS 5-26555. This research has made use of the NASA/IPAC Extragalactic Database (NED) which is operated by the Jet Propulsion Laboratory, California Institute of Technology, under contract with the National Aeronautics and Space Administration. We acknowledge the usage of the HyperLeda database (http://leda.univ-lyon1.fr). This publication makes use of data products from the Two Micron All Sky Survey, which is a joint project of the
University of Massachusetts and the Infrared Processing and Analysis Center/California Institute of Technology, funded by the National Aeronautics and Space Administration and the National Science Foundation.

Funding for the SDSS and SDSS-II has been provided by the Alfred P. Sloan Foundation, the Participating Institutions, the National Science Foundation, the U.S. Department of Energy, the National Aeronautics and Space Administration, the Japanese Monbukagakusho, the Max Planck Society, and the Higher Education Funding Council for England. The SDSS Web Site is http://www.sdss.org/. The SDSS is managed by the Astrophysical Research Consortium for the Participating Institutions. The Participating Institutions are the American Museum of Natural History, Astrophysical Institute Potsdam, University of Basel, University of Cambridge, Case Western Reserve University, University of Chicago, Drexel University, Fermilab, the Institute for Advanced Study, the Japan Participation Group, Johns Hopkins University, the Joint Institute for Nuclear Astrophysics, the Kavli Institute for Particle Astrophysics and Cosmology, the Korean Scientist Group, the Chinese Academy of Sciences (LAMOST), Los
Alamos National Laboratory, the Max-Planck-Institute for Astronomy (MPIA), the Max-Planck-Institute for Astrophysics (MPA), New Mexico State University, Ohio State University, University of Pittsburgh, University of Portsmouth, Princeton University, the United States Naval Observatory, and the University of Washington.

This research has made use of SAOImage DS9, developed by Smithsonian Astrophysical Observatory.

\clearpage

\appendix

\section{Tables}

\begin{longtable}{l c c c c l c c c c}

\caption{\label{galax1}Properties of the nuclear ring host galaxies.}\\

\hline
Galaxy  & RA    &   dec   &  Distance & Refs. & Morph. type &Nuclear    & Disc scale- & Disc PA & $\epsilon_{\rm d}$       \\
        &(hh mm)& (\deg{}')& (Mpc)  & &  &     activity  & length (kpc)& (\deg)  & \\
(1)     &  (2)  &   (3)    & (4)    & (5) & (6)          & (7)         & (8)     & (9) & (10)\\
\hline
\endfirsthead
\caption{continued.}\\
\hline
Galaxy  & RA    &   dec   &  Distance& Refs. & Morph. type &Nuclear    & Disc scale- & Disc PA & $\epsilon_{\rm d}$       \\
        &(hh mm)& (\deg{}')& (Mpc)    &    & & activity  & length (kpc)& (\deg)  & \\
(1)     &  (2)  &   (3)    & (4)    & (5) & (6)          & (7)         & (8)     & (9) & (10)\\
\hline
\endhead
\hline
\endfoot
ESO~198-13&02 29&$-$48 29  & 74.2 & (1) & (R)SA(r)ab              & $-$ & $-$       & 119.7 &  0.26 \\
ESO~437-33&10 40&$-$30 12  & 43.5 & (1) & (R$_1^{\prime}$)SAB(rl)a &$-$&  5.7  &  15.6   &    0.32    \\
ESO~437-67&10 52&$-$32 40  & 43.3 & (1) & (R$_1^{\prime}$)SB(r)ab & $-$&  2.8  & $-$   &    0.00    \\
ESO~565-11&09 29&$-$20 23  & 66.5 & (1) & (R$_1^{\prime}$)SB($\underline{\rm r}$s)a    & $-$        &   5.9       & 156.1   &    0.28    \\
IC~342  &  03 47& $+$68 06  &  3.3 & (2) & S$\underline{\rm A}$B(r$\underline{\rm s}$)cd      &\hii    &   2.5       &   $-$   &    0.05    \\
IC~1438 &  22 16& $-$21 26  & 36.2 & (1) & (R$_1$R$_2^{\prime}$)SAB(r)a& $-$ & 3.7 & $-$ & 0.07 \\
IC~4214 &  13 18& $-$32 06  & 31.4 & (1) & (R$_1^{\prime}$)SAB(r)a & $-$ &  3.1  & 165.4   &    0.38    \\
IC~4933 &  20 03& $-$54 59  & 67.3 & (1) & SB(rs)bc       & $-$         &   6.1  & 9.0     & 0.38\\
NGC~278 &  00 52& $+$47 33  & 11.7 & (1) & SAB(rs)b       & \hii        &   0.7       &  65.8   &    0.07    \\
NGC~473 &  01 20& $+$16 33  & 31.0 & (1) & SAB(r)0/a      & $-$        &  1.7      &  153.3   & 0.42 \\
NGC~521 &  01 25& $+$01 44  & 71.2 & (1) & SB(r)bc        & Transition 2 & 7.5     & $-$ & 0.07 \\
NGC~613 &  01 34& $-$29 25  & 18.7 & (1) & SB(rs)bc       & $-$   &   2.8       &  121.6  &    0.23    \\
NGC~718 &  01 53& $+$04 12  & 24.1 & (1) & (R$^{\prime}$)SA$\underline{\rm B}$(rs)a        & LINER 2        &   2.4       &    7.4  &    0.15    \\
NGC~864 &  02 15& $+$06 00  & 21.8 & (1) & S$\underline{\rm A}$B(rs)c       & \hii        &   2.9       &   28.7  &    0.16    \\
NGC~936 &  02 28& $-$01 09& 23.0   & (3) & SB($\underline{\rm r}$s)0$^+$ & $-$ & 4.0 & 122.7 & 0.26\\
NGC~1068&  02 43& $-$00 01& 15.3   & (1) & (R)SA(rs)b     & Sy 1.8    &   2.9       &   72.5  &    0.13    \\
NGC~1079&  02 44& $-$29 00& 17.7   & (1) & (R$_1$R$_2^{\prime}$)S$\underline{\rm A}$B($\underline{\rm r}$s)a  & $-$&   6.&   78.0  &    0.41    \\
NGC~1097&  02 46& $-$30 17& 15.2   & (1) & (R$_1^{\prime}$)SB(rs)b pec         & $-$        &   5.1       &  134.0  &    0.31    \\
NGC~1241&  03 11& $-$08 55& 55.8   & (1) & SAB(rs)b        & $-$        &   5.2       &  147.5  &    0.42    \\
NGC~1300&  03 20& $-$19 25& 20.2   & (1) & SB(rs)b       & $-$        &   5.3       &   87.0  &    0.18    \\
NGC~1317&  03 23& $-$37 06& 23.9   & (1) & (R$^{\prime}$)SAB(rl)a        & $-$        &   2.7       &   64.5  &    0.11    \\
NGC~1326&  03 24& $-$36 28& 16.1   & (1) & (R$_1$)SAB(r)0/a      & $-$      &   2.5       &   73.4  &    0.25    \\
NGC~1343&  03 38& $+$72 34& 35.1   & (1) & SAB(s)b pec    & $-$        &   4.9       &   77.5  &    0.45    \\
NGC~1386&  03 37& $-$36 00& 16.5   & (3) & SB(s)0$^+$        & $-$        &   1.8       &   24.6  &    0.65    \\
NGC~1387&  03 37& $-$35 30& 18.9   & (4) & SB0$^-$      & $-$        &  2.0         &$-$  & 0.00    \\
NGC~1415&  03 41& $-$22 34& 19.6   & (1) & (R)SAB(s)a   & $-$       &   2.1       &  151.7  &    0.61    \\
NGC~1433&  03 42& $-$47 13& 11.6   & (1) & (R$_1^{\prime}$)SB(r)ab & $-$        &   3.0   &  159.0  &    0.16    \\
NGC~1512&  04 04& $-$43 21&  9.5   & (5) & (R$^{\prime}$)SB(r)ab pec         & $-$        &   1.7       &   83.0  &    0.40    \\
NGC~1566&  04 20& $-$54 56& 17.4   & (1) & (R$_1^{\prime}$)SAB(s)bc& $-$        &   2.7       &   36.0  &    0.19    \\
NGC~1672&  04 46& $-$59 15& 15.0   & (1) & (R$_1^{\prime}$:)SB(r)bc & $-$        &   2.9       &   $-$   &    0.13    \\
NGC~1808&  05 08& $-$37 31& 10.9   & (1) & (R$_1$)SAB(s)b pec& $-$        &   1.6       &  119.5  &    0.26    \\
NGC~1819&  05 12& $+$05 12& 63.2   & (1) & SB0            & $-$         &   3.8       &   116.2    &  0.32  \\
NGC~2273&  06 50& $+$60 51& 29.4   & (1) & (RR)SAB(rs)a       & Sy 2        &   2.9       &   59.3  &    0.40    \\
NGC~2595&  08 28& $+$21 29& 62.6   & (1) & SAB(rs)c	      & $-$         &   4.6           &  11.8   &   0.32  \\
NGC~2787&  09 19& $+$69 12&  7.5   & (3) & SB(r)0$^+$        & LINER 1.9      &   0.8       &  113.5  &    0.45    \\
NGC~2859&  09 24& $+$34 31& 25.9   & (1) & (R)SB(rl)0$^+$      & Transition 2         &   2.1       &   85.0  &    0.13    \\
NGC~2903&  09 32& $+$21 30&  8.9   & (6) & SAB(rs)bc         & \hii       &   1.9       &   23.0  &    0.54    \\
NGC~2935&  09 37& $-$21 08& 30.6   & (1) & (R$_2^{\prime}$)SAB(s)b& $-$& 5.2   &  171.0  &    0.24    \\
NGC~2950&  09 43& $+$58 51& 22.2   & (1) & (R)SB(r)0$^{\rm o}$ & no activity & 2.3 & 140.0 & 0.37 \\
NGC~2985&  09 50& $+$72 17& 22.6   & (1) & (R$^{\prime}$)SA(rs)ab   & Transition 1.9      &   4.0       &  175.5  &    0.19    \\
NGC~2997&  09 46& $-$31 11& 13.1   & (1) & S$\underline {\rm A}$B(s)c         & $-$        &   3.7       &   96.6  &    0.40    \\
NGC~3011&  09 50& $+$32 13& 23.8   & (1) & S0/a           & $-$        &   0.7       &   59.5  &    0.17    \\
NGC~3081&  09 59& $-$22 50& 31.8   & (1) & (R$_1$R$_2^{\prime}$)SAB(r)0/a & $-$        &   2.6       &   97.0  &    0.16    \\
NGC~3185&  10 18& $+$21 41& 18.9   & (1) & (R)SB(r)a      & Sy 2        &   2.1       &  130.2  &    0.42    \\
NGC~3245&  10 27& $+$28 30& 21.1   & (1) & SA(l)0$^{\rm o}$   & Transition 2 &   2.0       &  176.0  &    0.44    \\
NGC~3258&  10 29& $-$35 36& 37.5   & (1) & E1             & $-$        &   $-$       &   $-$   &    $-$     \\
NGC~3268&  10 30& $-$35 20& 37.5   & (1) & E2             & $-$        &   $-$       &   $-$   &    $-$     \\
NGC~3310&  10 39& $+$53 30& 17.4   & (1) & SA(rs)bc pec       & \hii       &   2.8       & 167.0   &    0.38    \\
NGC~3313&  10 37& $-$25 19& 51.1   & (1) & SB(r)b         & $-$       &   9.1       &  55.0   &    0.19    \\
NGC~3351&  10 44& $+$11 42& 11.1   & (7) & SB(r)b         & \hii       &   3.0       &    9.9  &    0.40    \\
NGC~3379&  10 47& $+$12 35& 10.6   & (3) & E1             & LINER 2      &   $-$       &   $-$   &    $-$     \\
NGC~3414&  10 51& $+$27 59& 22.1   & (1) & S0$^{\rm o}$ pec   & LINER 2     &  2.5 &  22.7 & 0.14\\
NGC~3486&  11 00& $+$28 58& 11.6   & (1) & SAB(r)c        & Sy 2  &   3.0       &   80.0  &    0.29    \\
NGC~3504&  11 03& $+$27 58& 23.8   & (1) & (R$_1^{\prime}$)SA$\underline{\rm B}$(rs)ab    & \hii       &   1.9       &   $-$   &    0.02    \\
NGC~3593&  11 15& $+$12 49&  9.9   & (1) & S0/a pec        &\hii  &   2.0       &   86.2  &    0.51    \\
NGC~3945&  11 53& $+$60 41& 21.4   & (1) & (R)SB(rl)0$^+$ & LINER 2 & 7.3 & 158 & 0.43 \\
NGC~3982&  11 56& $+$55 08& 23.6   & (1) & SAB(r)b        & Sy 1.9        &   1.2       &    $-$  &    0.13    \\
NGC~4100&  12 06& $+$49 35& 18.6   & (1) & (R$^{\prime}$)SA(s)bc & \hii & 2.9 &  165.0  &    0.72    \\
NGC~4102&  12 06& $+$52 43& 18.6   & (5) & SAB(s)b?       &\hii&   1.8       &   38.0  &    0.44    \\
NGC~4138&  12 09& $+$43 41& 13.8   & (3) & SA(r)a        & Sy 1.9      &   1.3       &  148.2  &    0.40    \\
NGC~4245&  12 18& $+$29 36& 15.0   & (1) & SB(r)0/a         & \hii       &   1.5       &  174.1  &    0.18    \\
NGC~4262&  12 20& $+$14 53& 20.9   & (1) & SB(s)0$^-$?       & no activity        &   1.2       &  145.0  &    0.11    \\
NGC~4274&  12 19& $+$29 37& 15.6   & (1) & (R$^{\prime}$)SB(r)ab     & \hii      &   2.8       &  102.5  &    0.53    \\
NGC~4303&  12 22& $+$04 28& 23.1   & (1) & SAB(rs)bc      &\hii  &   4.2       &  146.9  &    0.14    \\
NGC~4314&  12 23& $+$29 54& 16.4   & (1) & (R$_1^{\prime}$)SB(r$^{\prime}$l)a        & LINER 2      &   3.3       &   61.8   &   0.04    \\
NGC~4321&  12 23& $+$15 49& 24.0   & (1) & SAB(s)bc       &Transition 2  &   6.6       &   $-$   &    0.09    \\
NGC~4340&  12 24& $+$16 43& 15.1   & (1) & SB(r)0$^+$        & no activity        &   2.2       &   98.6  &    0.44    \\
NGC~4371&  12 25& $+$11 42& 14.3   & (8) & SB(r)0$^{\rm o}$        & no activity        &   2.3       &   96.5  &    0.53    \\
NGC~4448&  12 28& $+$28 37& 11.7   & (1) & (R)SB(r)ab        & \hii        &   1.2       &   91.8  &    0.69    \\
NGC~4459&  12 29& $+$13 59& 16.1   & (8) & SA(r)0$^+$        &Transition 2&   2.5       &  108.3  &    0.24    \\
NGC~4494&  12 31& $+$25 46& 20.9   & (1) & E1-2           &LINER 2     & $-$ &$-$ & $-$\\
NGC~4526&  12 34& $+$07 42& 14.7   & (3) & SAB0$^{\rm o}$ sp& \hii        &   3.1       &  113.0  &    0.65    \\
NGC~4571&  12 37& $+$14 13& 16.8   & (5) & SA(r)c         & $-$        &   3.2       &   34.9  &    0.18    \\
NGC~4579&  12 38& $+$11 49& 23.0   & (1) & SA$\underline{\rm B}$(r$\underline{\rm s}$)ab       &Sy 1.9 &   5.0       &   94.8  &    0.22    \\
NGC~4593&  12 40& $-$05 21& 35.3   & (1) & (R$^{\prime}$)SB(rs)ab     & $-$        &   5.4       &   98.1  &    0.26    \\
NGC~4736&  12 51& $+$41 07&  5.2   & (3) & (R)S$\underline {\rm A}$B($\underline{\rm r}$s)ab     &LINER 2  &   1.2       &  105.0  &    0.13    \\
NGC~4800&  12 55& $+$46 32& 14.5   & (1) & SA(rs)b        & \hii       &   0.8       &   19.4  &    0.29    \\
NGC~4826&  12 57& $+$21 41&  7.5   & (3) & (R$^{\prime}$)SA(r)$\underline{\rm a}$b pec& Transition 2        &   2.1       &  115.0  &    0.50    \\
NGC~5020&  13 13& $+$12 36& 49.5   & (1) & SAB(rs)bc      & $-$ & 5.9 & 84.3 & 0.11 \\
NGC~5033&  13 13& $+$36 36& 15.4   & (1) & SA(s)c         & Sy 1.5      &   5.2       &  171.0  &    0.55    \\
NGC~5135&  13 26& $-$29 50& 57.4   & (1) & SB(l)ab        & $-$   &  13.9       &   $-$   &    0.09    \\
NGC~5194&  13 30& $+$47 12&  7.7   & (3) & SA(s)bc        & Sy 2&   4.3       &  163.0  &    0.21    \\
NGC~5236&  13 37& $-$29 52&  4.5   & (7) & SAB(s)c        &$-$&   2.0       &   $-$   &    0.07    \\
NGC~5247&  13 38& $-$17 53& 18.9   & (1) & SA(s)c        & $-$        &   4.6       &   40.0  &    0.17    \\
NGC~5248&  13 38& $+$08 53& 17.9   & (1) & (R$^{\prime}$)SAB(rs)bc      & \hii   &   2.9       &  125.0  &    0.41    \\
NGC~5377&  13 56& $+$47 14& 29.2   & (1) & (R$_1^{\prime}$)SA$\underline{\rm B}$(rs)a      & LINER 2        &   3.8       &   37.6  &    0.54    \\
NGC~5427&  14 03& $-$06 02& 39.4   & (1) & SA(rs)b$\underline{\rm c}$         & $-$   &   3.9       &  154.0   &    0.07    \\
NGC~5728&  14 42& $-$17 11& 39.7   & (1) & (R$_1$)SA$\underline{\rm B}$(r)a   & $-$ &   4.2       &   14.5  &    0.41    \\
NGC~5806&  15 00& $+$01 53& 20.5   & (1) & SAB(s)b        & \hii      &   2.0       &  174.3  &    0.48    \\
NGC~5812&  15 01& $-$07 27& 28.0   & (1) & E0             & $-$        &   $-$       &   $-$   &    $-$     \\
NGC~5905&  15 15& $+$55 31& 52.4   & (1) & SB($\underline{\rm r}$s)bc         & \hii   &   3.4       &  128.9  &    0.19    \\
NGC~5945&  15 30& $+$42 55& 81.3   & (1) & (R$_1$)SB(rs)ab       & $-$      &   8.1       &  115.0  &    0.07    \\
NGC~6503&  17 49& $+$70 09&  5.2   & (9) & SA(s)cd        & Transition 2 &   0.8       &  122.3  &    0.67    \\
NGC~6753&  19 11& $-$57 93& 41.8   & (1) & (R)SA(r)b     & $-$        &   3.4       &   29.3  &    0.15    \\
NGC~6782&  19 24& $-$59 55& 52.5   & (1) & (R$_1$R$_2^{\prime}$)SB(r)a     & $-$        &   3.9       &   34.9  &    0.10    \\
NGC~6861&  20 07& $-$48 22& 38.0   & (1) & SA(s)0$^-$        & $-$        &   4.3       &  133.1  &    0.36    \\
NGC~6951&  20 37& $+$66 06& 24.4   & (1) & SA$\underline{\rm B}$(rs)bc      & Sy 2  &   5.0       &  132.2  &    0.29    \\
NGC~6958&  20 49& $-$38 00& 37.1   & (1) & E$^+$             & $-$        &   $-$       &   $-$   &    $-$     \\
NGC~7049&  21 19& $-$48 33& 30.2   & (1) & SA(s)0$^{\rm o}$  & $-$        &   2.7       &   63.7  &    0.32    \\
NGC~7217&  22 08& $+$31 22& 16.0   & (1) & (R)SA(r)ab     & LINER 2   &   2.0       &   93.0  &    0.14    \\
NGC~7469&  23 03& $+$08 52& 70.7   & (1) & (R$^{\prime}$)SAB(rs)a   & $-$      &  10.2       &  126.5  &    0.13    \\
NGC~7552&  23 16& $-$42 35& 20.2   & (1) & (R$_1^{\prime}$)SB(s)ab    &$-$&   1.7       &  169.9  &    0.13    \\
NGC~7570&  23 17& $+$13 29& 68.1   & (1) & SBa            & $-$ &    3.9      & 152.0 & 0.12 \\
NGC~7716&  23 37& $+$00 18& 36.6   & (1) & SAB(r)b        & $-$        &   2.5       &   30.4  &    0.31    \\
NGC~7742&  23 44& $+$10 46& 24.2   & (1) & SA(r)ab         & Transition 2 &   1.3       &   $-$   &    0.05    \\
NGC~7771&  23 51& $+$20 07& 62.5   & (7) & SB(s)a pec         & $-$        &   5.1        &  67.0    &   0.50 \\
UGC~10445& 16 34& $+$28 59& 16.8   & (1) & SBc            & $-$        &   $-$       &  143.1  &    0.29    \\
\end{longtable}
Identification (col.~1), position on the sky in J2000.0 coordinates (cols.~2 and 3), distance (col.~4, from the literature for galaxies with radial velocity $v \le 1400$ where possible; In other cases from the radial velocity from HYPERLEDA, with $H_{0}=70 {\rm km}/{\rm s}\,{\rm Mpc}$), reference of the distance measurement (col.~5), morphological type (col.~6, from Buta et al.~2007 when available and from NED for the other galaxies), presence and type of nuclear activity (col.~7, from Ho et al.~1997b), scalelength of the exponential disc in kpc (col.~8), PA of the major axis of the outer disc (col.~9), and outer disc ellipticity (col.~10). Orientation parameters come different sources described in Sect.~3.3.1. Distance references: (1) Paturel et al.~(2003) (HYPERLEDA), (2) Karachentsev et al.~(2003), (3) Tonry et al.~(2001), (4) Jensen et al.~(2003), (5) Tully et al.~(1988), (6) Drozdovsky \& Karachentsev (2000), (7) Saha et al.~(2006), (8) Mei et al.~(2007), and (9) Karachentsev \& Sharina (1997).

\onecolumn

\begin{longtable}{l c c c c c c c c c c c}
\caption{\label{galax2} Properties of the nuclear rings and the bars in our sample galaxies.}\\
\hline
Galaxy  & \multicolumn{2}{c}{Ring semi-major axis}& Rel.~Ring     & $\epsilon_{\rm r}$ & PA offset (\deg)  & Bar length & $\epsilon_{\rm b}$ & $Q_{\rm g}$ & \multicolumn{2}{c}{$r_{Q_{\rm g}}$} & Type of\\
        & (\arcsec) &  (pc)      & size     &   Deproj.       &  Deproj.          & (kpc)      & Deproj.         &  & (\arcsec) & (kpc) & ring\\
(1)     &  (2)  &   (3)    & (4)    & (5) & (6)          & (7)         & (8)     & (9) & (10)  & (11) & (12)\\
\hline
\endfirsthead
\caption{continued.}\\
\hline
Galaxy  & \multicolumn{2}{c}{Ring semi-major axis}& Rel.~Ring     & $\epsilon_{\rm r}$ & PA offset (\deg)  & Bar length & $\epsilon_{\rm b}$ & $Q_{\rm g}$ & \multicolumn{2}{c}{$r_{Q_{\rm g}}$} & Type of\\
        & (\arcsec) &  (pc)      & size     &   Deproj.       &  Deproj.          & (kpc)      & Deproj.         &  & (\arcsec) & (kpc) & ring\\
(1)     &  (2)  &   (3)    & (4)    & (5) & (6)          & (7)         & (8)     & (9) & (10)  & (11) & (12)\\
\hline
\endhead
\hline
\endfoot
ESO~198-13&  9.7 & 3490       & 0.234    &   0.05          &    $-$               &  $-$          & $-$        &  $-$       & $-$    &  $-$ & sf     \\
ESO~437-33&  3.5 & 730        & 0.071    &   0.18          &    66             &   2.7      &   0.16          &   0.056 & 15  & 3.2 & sf   \\
ESO~437-67&  3.4 & 710        & 0.044    &   0.10          &    29             &   5.5      &   0.61          &   0.403 & 23  & 4.8 & sf  \\
ESO~565-11& 11.1 &3560        & 0.212    &   0.55          &    39             &  11.6      &   0.62          &   0.316 & 29  & 9.3 & sf  \\
IC~342  &     2.9 & 50           & 0.002    &   0.31          &    59             &   2.0      &   0.39          &   0.177 & 225 & 3.6 & sf   \\
IC~1438 &     3.9 & 680           & 0.054    &   0.36          &    44             &   3.5      &   0.49          &   0.155 & 17  & 3.0 & sf  \\
IC~4214 &     7.6 & 1150           & 0.081    &   0.21          &    31             &   4.4      &   0.36          &   0.127 & 27  & 4.1 & sf  \\
IC~4933 &     2.6 &  850          &  0.041   &    0.19         &     70            &    8.8     &    0.60         &    0.239 & 21  & 6.9 & sf\\
NGC~278 &     4.7 & 260           & 0.057    &   0.18          &    14             &   0.8      &   0.22          &   0.090 & 27  & 1.5 & sf  \\
NGC~473 &    11.3 & 1690           & 0.201    &   0.06          &    $-$            &   $-$      &   $-$           &   0.119 & 25  & 3.8 & sf  \\
NGC~521 &     4.3 & 1470           & 0.049    &    0.24         &    85              &   6.2     &   0.45           &   0.146 & 13 &  4.5 & sf \\
NGC~613 &     4.5 & 400           & 0.027    &   0.26          &    70             &   7.0      &   0.65          &   0.442 & 47  & 4.3 & sf  \\
NGC~718 &     2.9 & 330           & 0.034    &   0.17          &    59             &   2.6      &   0.44          &   0.133 & 21  & 2.5 & sf  \\
NGC~864 &     0.8 & 90           & 0.007    &   0.08          &    84             &   3.5      &   0.53          &   0.376 & 19  & 2.0  & sf \\
NGC~936 &     8.5 & 940          & 0.062     &   0.12         &    53             &    5.1     &   0.52          &   0.201  & 39 & 4.3  & sf \\
NGC~1068&    16.0 & 1190           & 0.077    &   0.31          &    28             &   4.6      &   0.31          &   0.094 & 55  & 4.1 & sf  \\
NGC~1079&     1.6 & 130           & 0.019    &   0.20          &    87             &   3.8      &   0.51          &   0.238 & 37  & 3.2 & sf  \\
NGC~1097&    13.2 & 970           & 0.041    &   0.32          &    56             &   6.7      &   0.51          &   0.241 & 75  & 5.5 & sf  \\
NGC~1241&     2.9 & 790           & 0.033    &   0.08          &    83             &   7.8      &   0.62          &   0.279 & 21  & 5.7 & sf  \\
NGC~1300&     4.1 & 400           & 0.022    &   0.15          &    10             &   7.3      &   0.71          &   0.541 & 63  & 6.2 & sf  \\
NGC~1317&    12.5 & 1450           & 0.163    &   0.15          &    90             &   5.9      &   0.36          &   0.097 & 35  & 4.1 & sf  \\
NGC~1326&     6.2 & 480           & 0.048    &   0.14          &    75             &   3.1      &   0.50          &   0.163 & 39  & 3.0 & sf  \\
NGC~1343&    11.6 & 1970           & 0.122    &   0.30          &    69             &   9.7      &   0.31          &   0.152 & 47  & 8.0 & sf  \\
NGC~1386&    16.8 & 1340           & 0.138    &   0.18          &    $-$            &   $-$      &   $-$           &   $-$   & $-$ & $-$ & sf   \\
NGC~1387&     7.3 & 660           & 0.080    &   0.06          &    0              &   2.2      &   0.31          &   0.078 & 19  & 1.7 & sf   \\
NGC~1415&     9.9 & 940           & 0.089    &   0.25          &    19             &   5.1      &   0.54          &   0.190 & 47  & 4.5 & sf  \\
NGC~1433&     9.8 & 550           & 0.053    &   0.32          &    55             &   5.6      &   0.67          &   0.390 & 67  & 3.8 & sf  \\
NGC~1512&    10.4 & 480           & 0.052    &   0.20          &    45             &   5.0      &   0.69          &   0.366 & 79  & 3.6 & sf  \\
NGC~1566&     9.7 & 820           & 0.047    &   0.23          &     7             &   3.0      &   0.42          &   0.235 & 71  & 6.0 & sf  \\
NGC~1672&     5.8 & 420           & 0.031    &   0.23          &     4             &   5.2      &   0.64          &   0.349 & 59  & 4.3 & sf  \\
NGC~1808&     5.4 & 280           & 0.033    &   0.20          &     5             &   3.9      &   0.61          &   0.388 & 73  & 3.9 & sf  \\
NGC~1819&     2.9 & 870           & 0.055    &   0.33          &    86             &   8.3      &   0.29          &   0.123 & 23  & 7.0 & sf  \\
NGC~2273&     2.5 & 360           & 0.035    &   0.41          &    27             &   3.7      &   0.52          &   0.218 & 21  & 3.0 & sf  \\
NGC~2595&     5.4 & 1640          & 0.106    &   0.29          &    55             &  11.8      &   0.64          &   0.392 & 29  & 8.8 & sf  \\
\multirow{2}{*}{NGC~2787$\Big\{$}
        &     2.8 & 100           & 0.026    &   $-$          &     $-$             &   1.7      &   0.47          &   0.137 & 37  & 1.3 & p  \\
        &     2.8 & 100           & 0.026    &   $-$          &     $-$             &   1.7      &   0.47          &   0.137 & 37  & 1.3 & p  \\
NGC~2859&     6.8 & 850           & 0.059    &   0.34          &    61             &   4.8      &   0.44          &   0.156 & 33  & 4.1 & sf  \\
NGC~2903&     3.7 & 160           & 0.011    &   0.32          &    70             &   2.9      &   0.56          &   0.273 & 61  & 2.6 & sf  \\
NGC~2935&     3.6 & 530           & 0.027    &   0.10          &    88             &   3.9      &   0.48          &   0.196 & 25  & 3.7 & sf  \\
NGC~2950&     4.7 & 510          &  0.059   &    0.26         &     64             &   2.8      &  0.27          &   0.066  & 23 & 2.5  & sf \\
NGC~2985&     0.5 & 60           & 0.004    &   0.18          &     5             &   8.1      &   0.29          &   0.061 & 23  & 2.5 & sf  \\
NGC~2997&     5.5 & 350           & 0.016    &   0.15          &    $-$            &   $-$      &   $-$           &   0.306 & 99  & 6.3 & sf  \\
NGC~3011&     5.4 & 620           & 0.216    &   0.08          &    $-$            &   $-$      &   $-$           &   0.055 & 15  & 1.7 & sf  \\
NGC~3081&     7.8 & 1200           & 0.094    &   0.43          &    47             &   5.7      &   0.55          &   0.194 & 35  & 5.4 & sf  \\
NGC~3185&     2.2 & 200           & 0.033    &   0.37          &    89             &   3.3      &   0.43          &   0.138 & 27  & 2.5 & sf  \\
NGC~3245&     1.1 & 110           & 0.011    &   0.10          &    $-$            &   $-$      &   $-$           &   0.088 & 13  & 1.3 & sf  \\
NGC~3258&     1.1 & 190           & 0.011    &   $-$           &    $-$            &   $-$      &   $-$           &   $-$   & $-$ & $-$ & d   \\
NGC~3268&     0.7 & 130           & 0.006    &   $-$           &    $-$            &   $-$      &   $-$           &   $-$   & $-$ & $-$ & d   \\
NGC~3310&     8.2 & 690           & 0.119    &   0.13          &     8             &   2.1      &   0.44          &   0.128 & 19  & 1.6 & sf  \\
NGC~3313&     5.0 & 1230           & 0.053    &   0.12          &    35             &  12.9      &   0.66          &   0.321 & 33  & 8.2 & sf  \\
NGC~3351&     7.7 & 410           & 0.036    &   0.25          &    66             &   5.5      &   0.67          &   0.225 & 63  & 3.4 & sf  \\
NGC~3379&     1.5 & 80           & 0.009    &   $-$           &    $-$            &   $-$      &   $-$           &   $-$   & $-$ & $-$ & d   \\
NGC~3414&     8.2 & 880          & 0.102     &    0.04        &      84           &   4.4      &    0.23         &   0.097  & 33 &  3.5 & sf  \\
NGC~3486&    22.2 & 1250           & 0.122    &   0.14          &    88             &   4.6      &   0.32          &   0.087 & 81  & 4.6 & sf  \\
NGC~3504&     1.8 & 200           & 0.022    &   0.19          &    20             &   3.7      &   0.60          &   0.311 & 31  & 3.6 & sf  \\
NGC~3593&     7.0 & 330           & 0.048    &   0.21          &    $-$            &   $-$      &   $-$           &   $-$   & $-$ & $-$ & sf   \\
NGC~3945&     5.3 & 550           & 0.031    &    0.23         &    81             &   6.3      &   0.58          &   0.163  & 53 & 5.5 & sf   \\
NGC~3982&    10.2 & 1160           & 0.159    &   0.20          &    26             &   4.0      &   0.23          &   0.126 & 13  & 1.5 & sf  \\
NGC~4100&     5.4 & 490           & 0.040    &   0.33          &    $-$            &   $-$      &   $-$           &   $-$   & $-$ & $-$ & sf   \\
NGC~4102&     1.6 & 140           & 0.018    &   0.15          &    71             &   1.3      &   0.37          &   0.108 & 13  & 1.2 & sf  \\
NGC~4138&    19.9 & 1330           & 0.221    &   0.20          &    $-$            &   $-$      &   $-$           &   0.037 & 35  & 2.3 & sf  \\
NGC~4245&     4.6 & 330           & 0.059    &   0.06          &    41             &   2.8      &   0.50          &   0.194 & 29  & 2.1 & sf  \\
NGC~4262&     2.9 & 290           & 0.051    &   0.09          &    66             &   1.4      &   0.34          &   0.083 & 11  & 1.1 & sf  \\
NGC~4274&     9.0 & 680           & 0.083    &   0.37          &    81             &   5.0      &   0.59          &   0.231 & 49  & 3.7 & sf  \\
NGC~4303&     3.2 & 350           & 0.016    &   0.11          &    83             &   5.8      &   0.11          &   0.285 & 41  & 4.6 & sf  \\
NGC~4314&     7.1 & 560           & 0.061    &   0.31          &    11             &   5.7      &   0.66          &   0.432 & 49  & 3.9 & sf \\
NGC~4321&     7.5 & 870           & 0.041    &   0.32          &    22             &   6.7      &   0.51          &   0.221 & 85  & 9.9 & sf  \\
NGC~4340&     8.6 & 630           & 0.103    &   0.14          &    22             &   4.7      &   0.63          &   0.290 & 57  & 4.2 & sf  \\
\multirow{2}{*}{NGC~4371$\Big\{$}
        &     0.7 & 50           & 0.006    &   0.07          &    55             &   4.9      &   0.60          &   0.175 & 63  & 4.4  & d/sf? \\
        &    10.6 & 740           & 0.090    &   0.22          &    55             &   4.9      &   0.60          &   0.175 & 63  & 4.4 & sf  \\
NGC~4448&     1.4 & 80           & 0.025    &   0.18          &    46             &   2.6      &   0.70          &   0.192 & 29  & 1.6 & sf  \\
\multirow{2}{*}{NGC~4459$\Big\{$}
        &     2.1 & 160           & 0.015    &   0.18          &    $-$            &   $-$      &   $-$           &   0.043 & 21  & 1.6 & sf  \\
        &     8.1 & 630           & 0.061    &   0.06          &    $-$            &   $-$      &   $-$           &   0.043 & 21  & 1.6 & sf  \\
NGC~4494&     0.6 & 60           & 0.004    &   $-$           &    $-$            &   $-$      &   $-$           &   $-$   & $-$ & $-$  & d  \\
NGC~4526&    10.6 & 760           & 0.052    &   0.29          &    $-$            &   $-$      &   $-$           &   $-$   & $-$ & $-$ & sf   \\
NGC~4571&    15.5 & 1260           & 0.140    &   0.06          &    $-$            &   $-$      &   $-$           &   0.192 & 37  & 3.0 & sf  \\
NGC~4579&     1.6 & 170           & 0.009    &   0.04          &    54             &   5.1      &   0.47          &   0.200 & 33  & 3.7 & sf  \\
NGC~4593&     5.2 & 880           & 0.070    &   0.09          &    71             &  10.3      &   0.61          &   0.356 & 43  & 7.4 & sf  \\
NGC~4736&    50.1 & 1260           & 0.165    &   0.22          &    $-$             &  $-$      &   $-$           &   0.054 & 139 & 3.5 & sf   \\
\multirow{3}{*}{NGC~4800$\Bigg\{$}
        &     0.4 & 30           & 0.007    &   0.12          &    67             &   3.4      &   0.14          &   0.087 & 19  & 1.3 & sf  \\
        &     1.7 & 120           & 0.031    &   0.08          &    24             &   3.4      &   0.14          &   0.087 & 19  & 1.3 & sf  \\
        &    10.0 & 700           & 0.187    &   0.02          &    82             &   3.4      &   0.14          &   0.087 & 19  & 1.3 & sf  \\
NGC~4826&     4.0 & 150           & 0.013    &   0.21          &    $-$            &   $-$      &   $-$           &   0.069 & 25  & 0.9 & sf  \\
NGC~5020&     4.3 & 1020          & 0.051    &   0.22          &    2              &   8.6      &   0.68          &   0.365 & 31  & 7.4 & sf  \\
NGC~5033&    12.9 & 960           & 0.046    &   0.06          &    $-$            &   $-$      &   $-$           &   0.246 & 29  & 2.2 & sf  \\
NGC~5135&     2.2 & 610           & 0.029    &   0.21          &    80             &   9.5      &   0.57          &   0.244 & 27  & 7.5 & sf \\
NGC~5194&    16.1 & 600           & 0.054    &   0.24          &    43             &   1.5      &   0.29          &   0.246 & 135 & 5.0 & sf   \\
NGC~5236&     6.1 & 130           & 0.012    &   0.18          &    87             &   2.5      &   0.57          &   0.294 & 81  & 1.8 & sf  \\
NGC~5247&     4.6 & 420           & 0.026    &   0.22          &    $-$            &   $-$      &   $-$           &   0.416 & 75  & 6.9 & sf  \\
\multirow{2}{*}{NGC~5248$\Big\{$}
        &     1.7 & 150           & 0.012    &   0.19          &    19             &   4.6      &   0.44          &   0.100 & 43  & 3.7 & sf  \\
        &     7.5 & 650           & 0.052    &   0.20          &     4             &   4.6      &   0.44          &   0.100 & 43  & 3.7 & sf  \\
NGC~5377&     5.6 & 790           & 0.052    &   0.24          &    76             &   9.2      &   0.36          &   0.164 & 47  & 6.7 & sf  \\
NGC~5427&     3.2 & 600           & 0.028    &   0.08          &    $-$            &   $-$      &   $-$           &   0.228 & 37  & 7.1 & sf  \\
NGC~5728&     5.7 & 1100           & 0.056    &   0.23          &    54             &  12.7      &   0.58          &   0.350 & 57  & 11.0 & sf  \\
NGC~5806&     4.2 & 420           & 0.045    &   0.24          &    64             &   3.7      &   0.30          &   0.140 & 41  & 4.1 & sf  \\
NGC~5812&     0.1 & 10           & 0.001    &   $-$           &    $-$            &   $-$      &   $-$           &   $-$   & $-$ & $-$ & d   \\
NGC~5905&     1.6 & 390           & 0.016    &   0.14          &    53             &   9.4      &   0.61          &   0.352 & 23  & 5.8 & sf  \\
NGC~5945&     3.3 & 1280           & 0.065    &   0.18          &    65             &  12.6      &   0.58          &   0.232 & 23  & 9.1 & sf  \\
NGC~6503&    47.0 & 1190           & 0.266    &   0.09          &    $-$            &   $-$      &   $-$           &   $-$   & $-$ & $-$ & sf   \\
NGC~6753&     8.5 & 1720           & 0.089    &   0.05          &    $-$            &   $-$      &   $-$           &   0.045 & 13  & 2.6 & sf  \\
NGC~6782&     4.5 & 1150           & 0.060    &   0.06          &     3             &   6.4      &   0.45          &   0.205 & 23  & 5.9 & sf  \\
NGC~6861&     1.3 & 230           & 0.013    &   0.36          &    $-$            &   $-$      &   $-$           &   0.152 & 15  & 2.8 & sf  \\
NGC~6951&     4.7 & 560           & 0.034    &   0.17          &    78             &   5.9      &   0.57          &   0.275 & 39  & 4.6 & sf  \\
NGC~6958&     1.1 & 190           & 0.014    &   $-$           &    $-$            &   $-$      &   $-$           &   $-$   & $-$ & $-$ & d   \\
NGC~7049&     2.7 & 400           & 0.022    &   0.13          &    $-$            &   $-$      &   $-$           &   0.076 & 15  & 2.2 & d/sf?  \\
NGC~7217&    10.8 & 840           & 0.082    &   0.08          &    $-$            &   $-$      &   $-$           &   0.026 & 41  & 3.2 & sf  \\
NGC~7469&     2.1 & 700           & 0.047    &   0.31          &    63             &   5.5      &   0.25          &   0.049 & 13  & 4.5 & sf  \\
NGC~7552&     3.5 & 340           & 0.029    &   0.15          &    47             &   5.0      &   0.66          &   0.513 & 47  & 4.6 & sf  \\
NGC~7570&     3.7 & 1220          & 0.076    &   0.14          &    72             &   11.2      &  0.63          &   0.424 & 29  & 9.6 & sf \\
NGC~7716&     6.8 & 1200           & 0.121    &   0.04          &    45             &   4.3      &   0.13          &   0.096 & 33  & 5.9 & sf \\
NGC~7742&     9.0 & 1050           & 0.163    &   0.05          &    $-$            &   $-$      &   $-$           &   0.055 & 21  & 2.5 & sf  \\
NGC~7771&     2.8 & 830           & 0.035    &   0.11          &    25             &  10.3      &   0.58          &   0.399 & 27  & 8.2 & sf  \\
UGC~10445&    1.5 & 120           & 0.025    &   0.04          &    $-$            &   $-$      &   $-$           &   $-$   & $-$ & $-$ & sf   \\
\end{longtable}

Identification (col.~1), nuclear ring semi-major axis in arcsecs and in pc (cols.~2 and 3), relative size of the nuclear ring, ring diameter divided by $D_{\rm o}$ (col.~4), nuclear ring ellipticity (col.~5), PA offset between the nuclear ring major axis and the bar major axis  (col.~6), bar length in kpc (col.~7), bar maximum ellipticity (col.~8), $Q_{\rm g}$ measurement (col.~9), radius at which the maximum non-axisymmetric torque is found in arcsec and in pc (col.~10 and 11), type of nuclear ring, where `d' stands for dust nuclear ring, `p' stands for polar nuclear ring, and `sf' stands for star-forming nuclear ring (col.~12). All the measurements have been done after deprojecting the galaxy using parameters found in Table~\ref{galax1}.

\clearpage

\section{Nuclear ring images}

\begin{figure*}
\begin{center}
\begin{tabular}{c}
\includegraphics[width=0.48\textwidth]{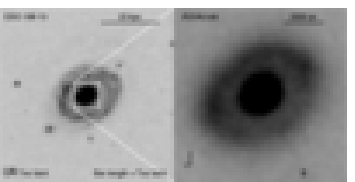}
\includegraphics[width=0.48\textwidth]{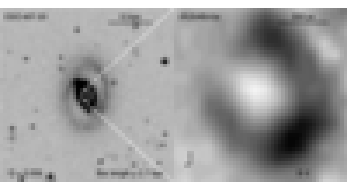}\\
\includegraphics[width=0.48\textwidth]{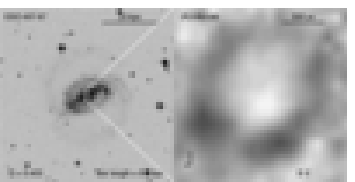}
\includegraphics[width=0.48\textwidth]{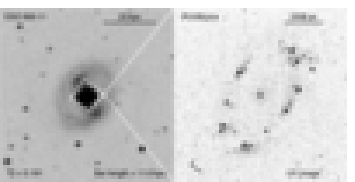}\\
\includegraphics[width=0.48\textwidth]{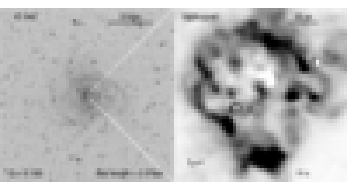}
\includegraphics[width=0.48\textwidth]{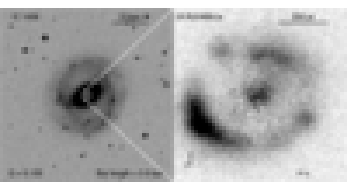}\\
\includegraphics[width=0.48\textwidth]{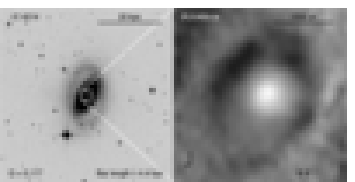}
\includegraphics[width=0.48\textwidth]{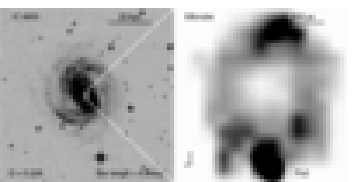}\\
\includegraphics[width=0.48\textwidth]{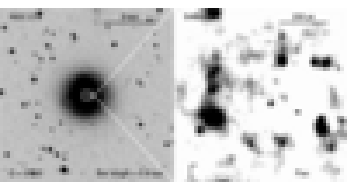}
\includegraphics[width=0.48\textwidth]{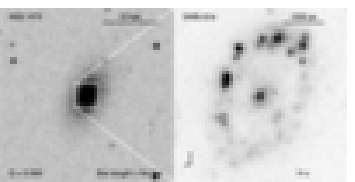}\\
\end{tabular}
\end{center}
\caption{\label{images} At the left of each pair of images, we present a DSS images whose size is twice $D_{25}$. At the right we present ring images, where the ring diameter is scaled to be $\sim80\%$ of the image side and where darker areas indicate star formation. From left to right and top to bottom, the labels correspond to: galaxy name, DSS image scale, galaxy classification from NED, scale of the nuclear ring image, value of the bar  non-axisymmetric torque parameter $Q_{\rm g}$ of the galaxy, bar length after galaxy deprojection, orientation of the nuclear ring image with the large arrow indicating north and the small one east, and kind of nuclear ring image.}
\end{figure*}

\begin{figure*}
\begin{center}
\begin{tabular}{c}

\includegraphics[width=0.48\textwidth]{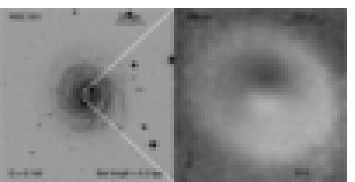}
\includegraphics[width=0.48\textwidth]{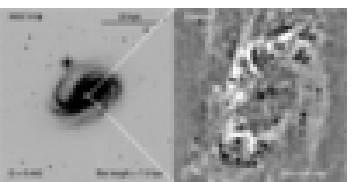}\\
\includegraphics[width=0.48\textwidth]{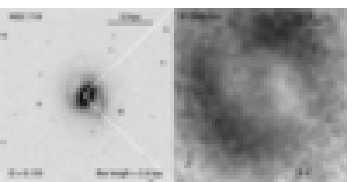}
\includegraphics[width=0.48\textwidth]{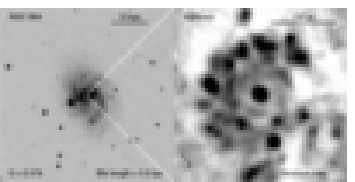}\\
\includegraphics[width=0.48\textwidth]{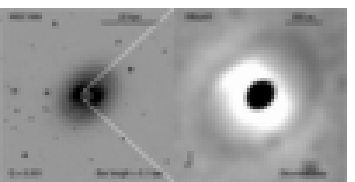}
\includegraphics[width=0.48\textwidth]{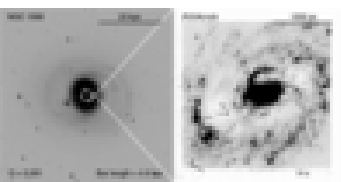}\\
\includegraphics[width=0.48\textwidth]{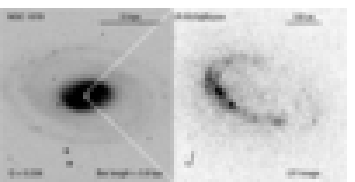}
\includegraphics[width=0.48\textwidth]{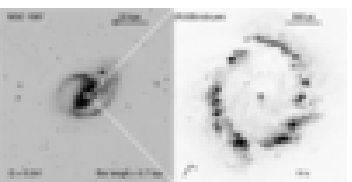}\\
\includegraphics[width=0.48\textwidth]{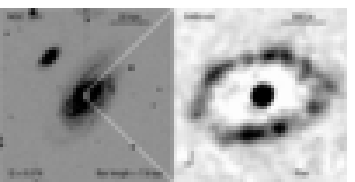}
\includegraphics[width=0.48\textwidth]{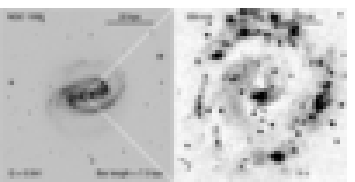}\\
\end{tabular}
\end{center}
{\bf Fig.~\ref{images}.} (continued)
\end{figure*}

\begin{figure*}
\begin{center}
\begin{tabular}{c}
\includegraphics[width=0.48\textwidth]{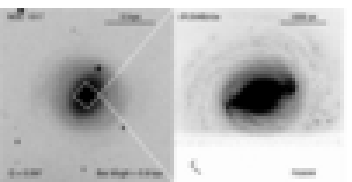}
\includegraphics[width=0.48\textwidth]{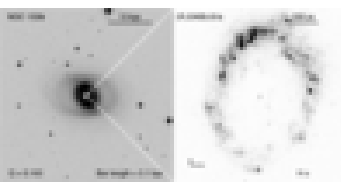}\\
\includegraphics[width=0.48\textwidth]{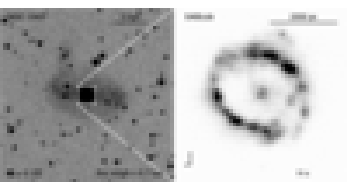}
\includegraphics[width=0.48\textwidth]{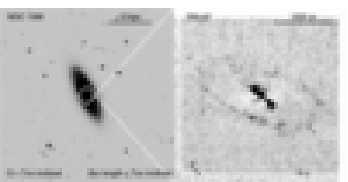}\\
\includegraphics[width=0.48\textwidth]{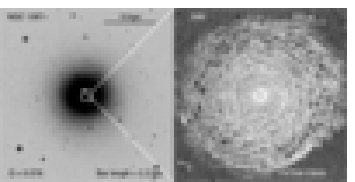}
\includegraphics[width=0.48\textwidth]{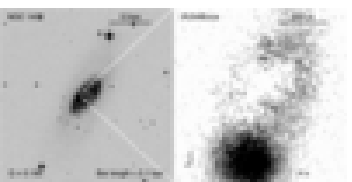}\\
\includegraphics[width=0.48\textwidth]{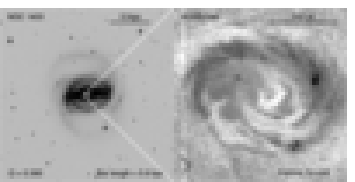}
\includegraphics[width=0.48\textwidth]{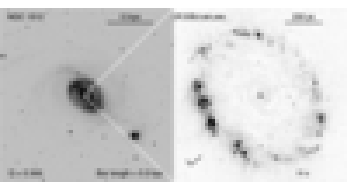}\\
\includegraphics[width=0.48\textwidth]{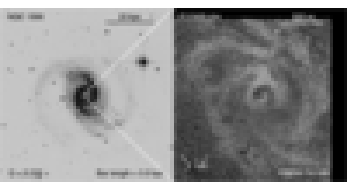}
\includegraphics[width=0.48\textwidth]{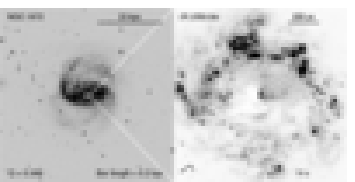}\\
\end{tabular}
\end{center}
{\bf Fig.~\ref{images}.} (continued)
\end{figure*}

\begin{figure*}
\begin{center}
\begin{tabular}{c}
\includegraphics[width=0.48\textwidth]{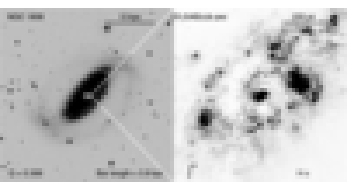}
\includegraphics[width=0.48\textwidth]{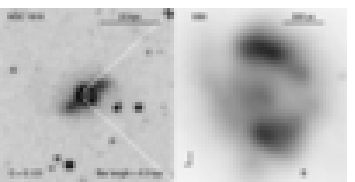}\\
\includegraphics[width=0.48\textwidth]{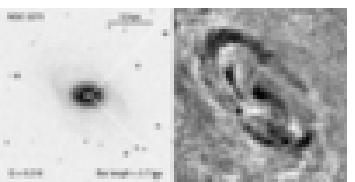}
\includegraphics[width=0.48\textwidth]{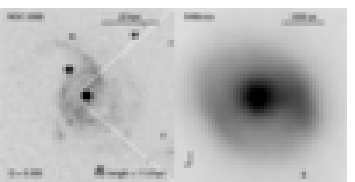}\\
\includegraphics[width=0.48\textwidth]{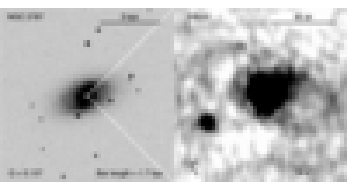}
\includegraphics[width=0.48\textwidth]{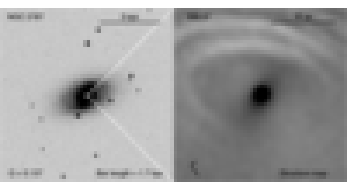}\\
\includegraphics[width=0.48\textwidth]{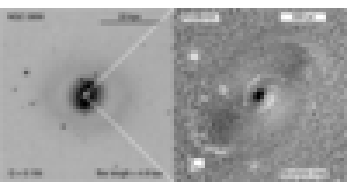}
\includegraphics[width=0.48\textwidth]{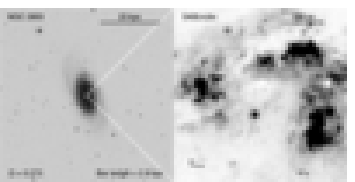}\\
\includegraphics[width=0.48\textwidth]{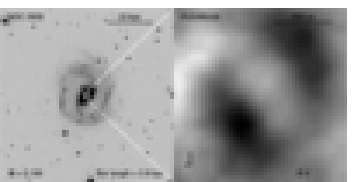}
\includegraphics[width=0.48\textwidth]{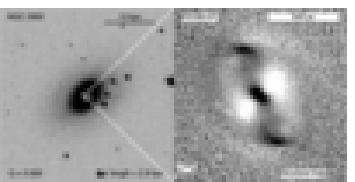}\\
\end{tabular}
\end{center}
{\bf Fig.~\ref{images}.} (continued)
\end{figure*}

\begin{figure*}
\begin{center}
\begin{tabular}{c}
\includegraphics[width=0.48\textwidth]{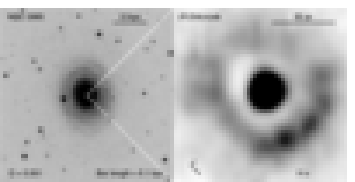}
\includegraphics[width=0.48\textwidth]{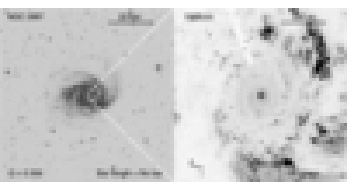}\\
\includegraphics[width=0.48\textwidth]{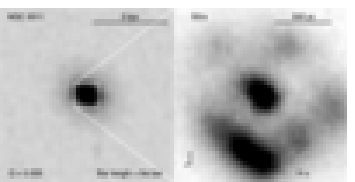}
\includegraphics[width=0.48\textwidth]{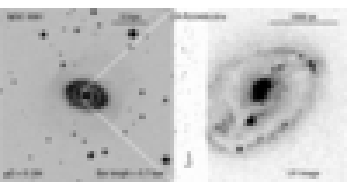}\\
\includegraphics[width=0.48\textwidth]{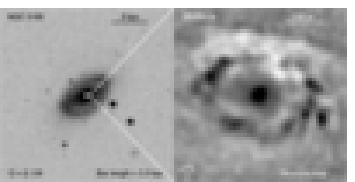}
\includegraphics[width=0.48\textwidth]{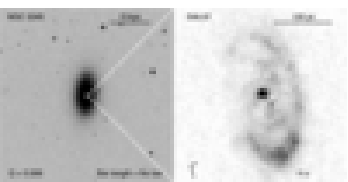}\\
\includegraphics[width=0.48\textwidth]{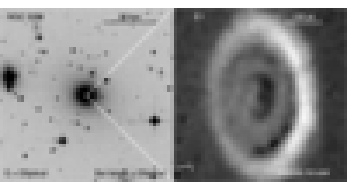}
\includegraphics[width=0.48\textwidth]{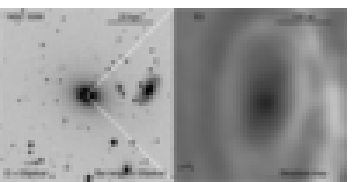}\\
\includegraphics[width=0.48\textwidth]{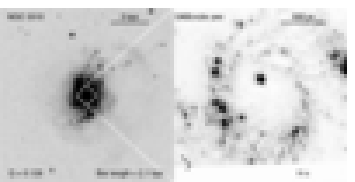}
\includegraphics[width=0.48\textwidth]{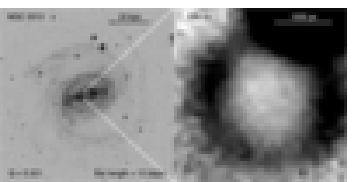}\\
\end{tabular}
\end{center}
{\bf Fig.~\ref{images}.} (continued)
\end{figure*}

\begin{figure*}
\begin{center}
\begin{tabular}{c}
\includegraphics[width=0.48\textwidth]{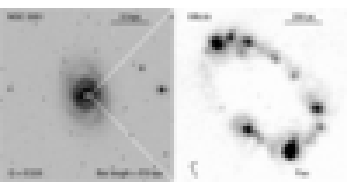}
\includegraphics[width=0.48\textwidth]{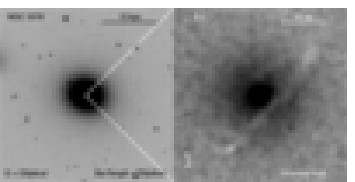}\\
\includegraphics[width=0.48\textwidth]{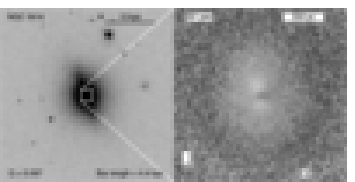}
\includegraphics[width=0.48\textwidth]{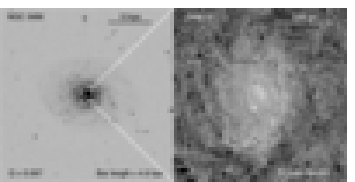}\\
\includegraphics[width=0.48\textwidth]{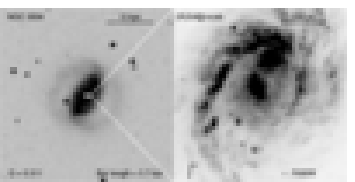}
\includegraphics[width=0.48\textwidth]{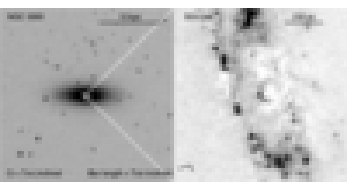}\\
\includegraphics[width=0.48\textwidth]{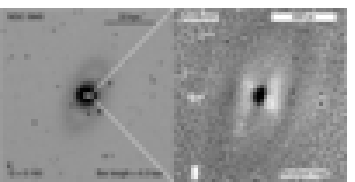}
\includegraphics[width=0.48\textwidth]{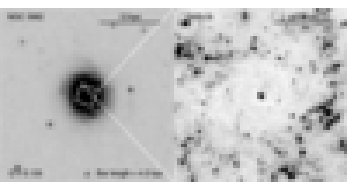}\\
\includegraphics[width=0.48\textwidth]{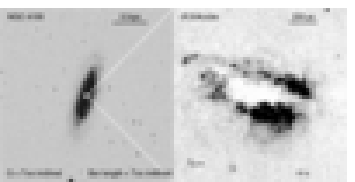}
\includegraphics[width=0.48\textwidth]{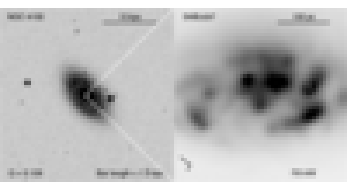}\\
\end{tabular}
\end{center}
{\bf Fig.~\ref{images}.} (continued)
\end{figure*}

\begin{figure*}
\begin{center}
\begin{tabular}{c}
\includegraphics[width=0.48\textwidth]{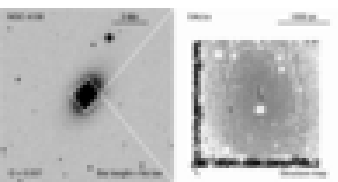}
\includegraphics[width=0.48\textwidth]{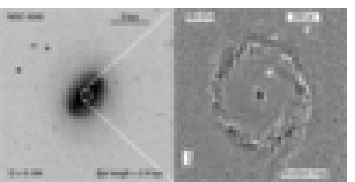}\\
\includegraphics[width=0.48\textwidth]{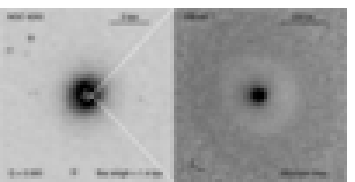}
\includegraphics[width=0.48\textwidth]{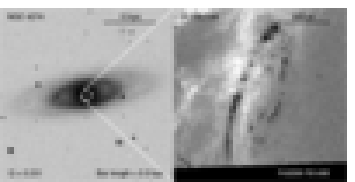}\\
\includegraphics[width=0.48\textwidth]{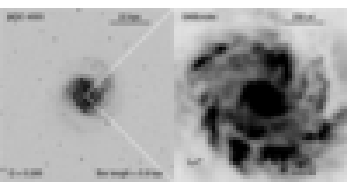}
\includegraphics[width=0.48\textwidth]{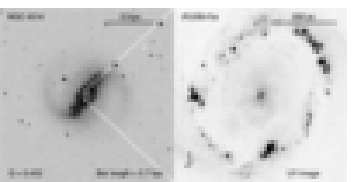}\\
\includegraphics[width=0.48\textwidth]{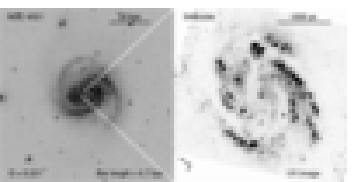}
\includegraphics[width=0.48\textwidth]{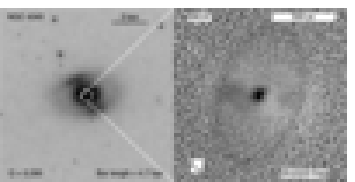}\\
\includegraphics[width=0.48\textwidth]{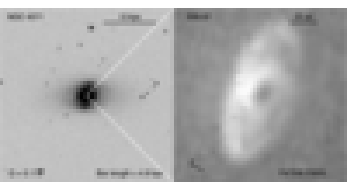}
\includegraphics[width=0.48\textwidth]{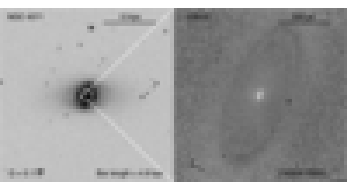}\\
\end{tabular}
\end{center}
{\bf Fig.~\ref{images}.} (continued)
\end{figure*}

\begin{figure*}
\begin{center}
\begin{tabular}{c}
\includegraphics[width=0.48\textwidth]{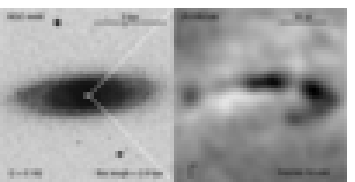}
\includegraphics[width=0.48\textwidth]{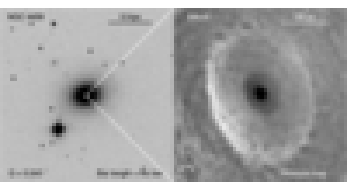}\\
\includegraphics[width=0.48\textwidth]{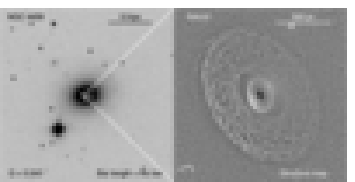}
\includegraphics[width=0.48\textwidth]{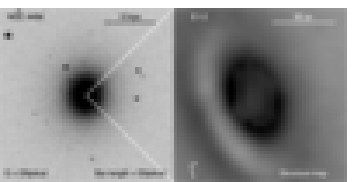}\\
\includegraphics[width=0.48\textwidth]{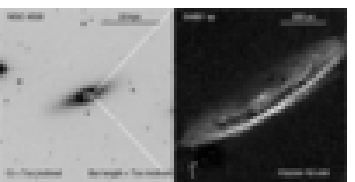}
\includegraphics[width=0.48\textwidth]{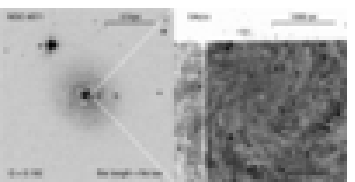}\\
\includegraphics[width=0.48\textwidth]{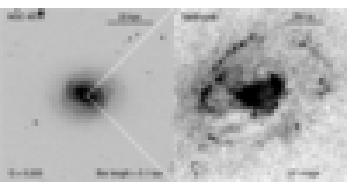}
\includegraphics[width=0.48\textwidth]{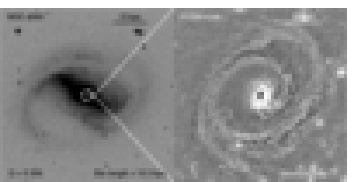}\\
\includegraphics[width=0.48\textwidth]{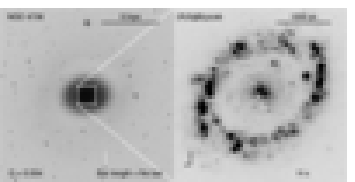}
\includegraphics[width=0.48\textwidth]{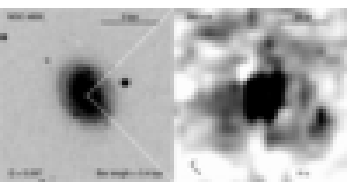}\\
\end{tabular}
\end{center}
{\bf Fig.~\ref{images}.} (continued)
\end{figure*}

\begin{figure*}
\begin{center}
\begin{tabular}{c}
\includegraphics[width=0.48\textwidth]{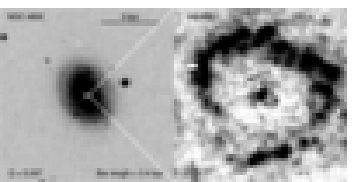}
\includegraphics[width=0.48\textwidth]{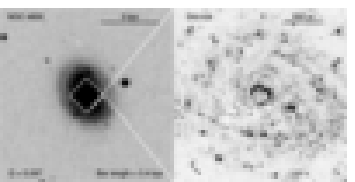}\\
\includegraphics[width=0.48\textwidth]{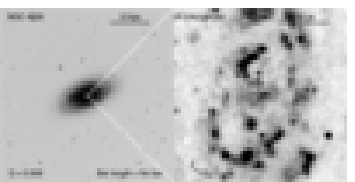}
\includegraphics[width=0.48\textwidth]{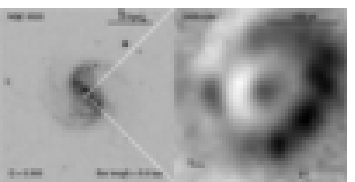}\\
\includegraphics[width=0.48\textwidth]{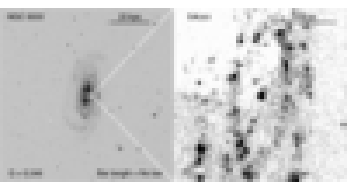}
\includegraphics[width=0.48\textwidth]{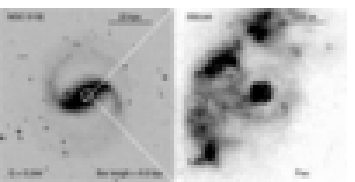}\\
\includegraphics[width=0.48\textwidth]{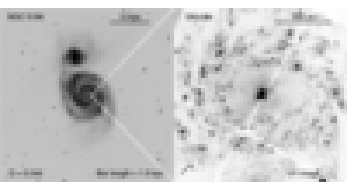}
\includegraphics[width=0.48\textwidth]{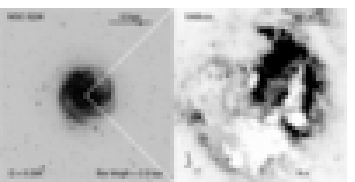}\\
\includegraphics[width=0.48\textwidth]{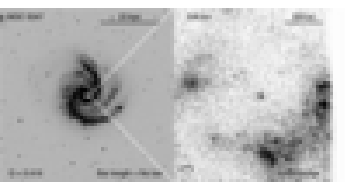}
\includegraphics[width=0.48\textwidth]{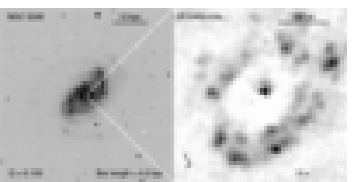}\\
\end{tabular}
\end{center}
{\bf Fig.~\ref{images}.} (continued)
\end{figure*}

\begin{figure*}
\begin{center}
\begin{tabular}{c}
\includegraphics[width=0.48\textwidth]{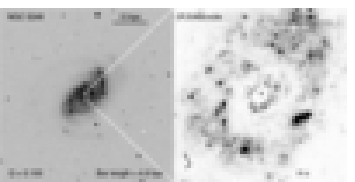}
\includegraphics[width=0.48\textwidth]{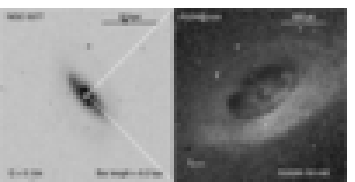}\\
\includegraphics[width=0.48\textwidth]{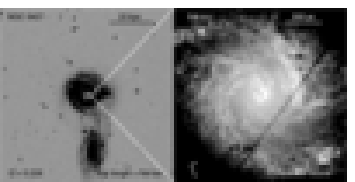}
\includegraphics[width=0.48\textwidth]{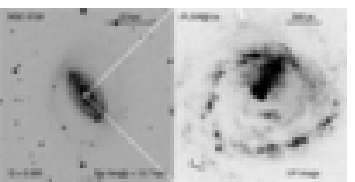}\\
\includegraphics[width=0.48\textwidth]{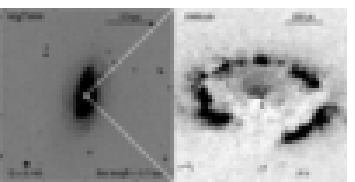}
\includegraphics[width=0.48\textwidth]{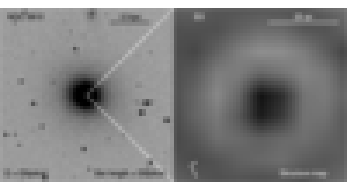}\\
\includegraphics[width=0.48\textwidth]{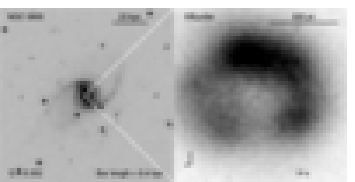}
\includegraphics[width=0.48\textwidth]{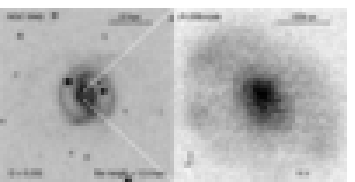}\\
\includegraphics[width=0.48\textwidth]{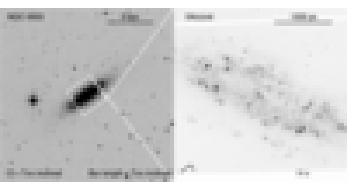}
\includegraphics[width=0.48\textwidth]{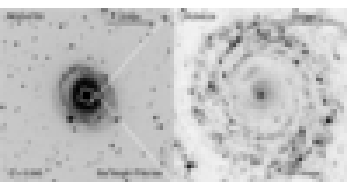}\\
\end{tabular}
\end{center}
{\bf Fig.~\ref{images}.} (continued)
\end{figure*}

\begin{figure*}
\begin{center}
\begin{tabular}{c}
\includegraphics[width=0.48\textwidth]{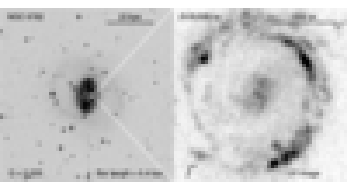}
\includegraphics[width=0.48\textwidth]{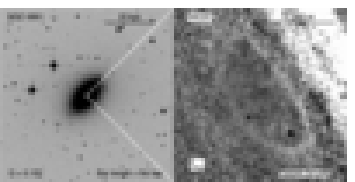}\\
\includegraphics[width=0.48\textwidth]{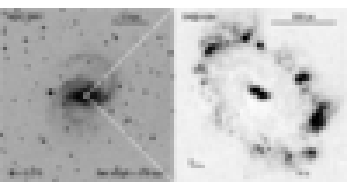}
\includegraphics[width=0.48\textwidth]{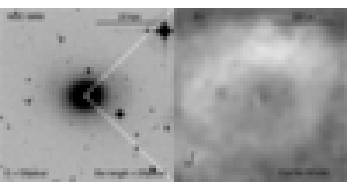}\\
\includegraphics[width=0.48\textwidth]{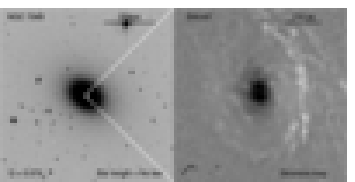}
\includegraphics[width=0.48\textwidth]{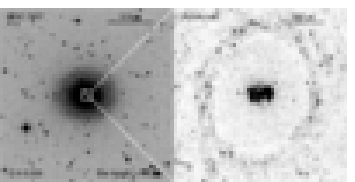}\\
\includegraphics[width=0.48\textwidth]{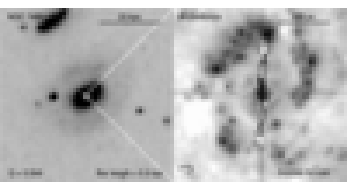}
\includegraphics[width=0.48\textwidth]{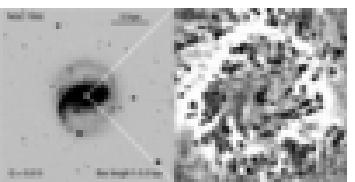}\\
\includegraphics[width=0.48\textwidth]{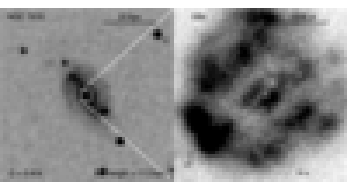}
\includegraphics[width=0.48\textwidth]{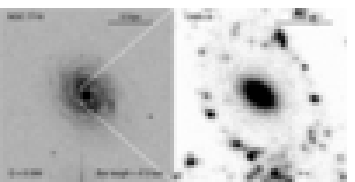}\\
\end{tabular}
\end{center}
{\bf Fig.~\ref{images}.} (continued)
\end{figure*}

\begin{figure*}
\begin{center}
\begin{tabular}{c}
\includegraphics[width=0.48\textwidth]{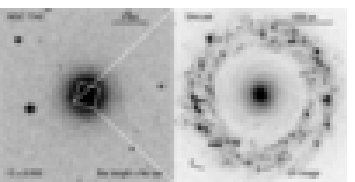}
\includegraphics[width=0.48\textwidth]{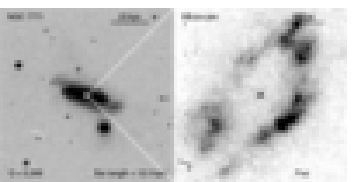}\\
\includegraphics[width=0.48\textwidth]{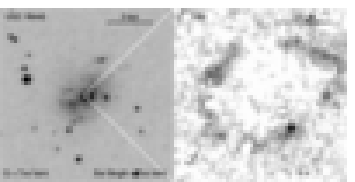}
\end{tabular}
\end{center}
{\bf Fig.~\ref{images}.} (continued)
\end{figure*}

\label{lastpage}

\end{document}